\documentclass[prd,preprint,preprintnumbers,nofootinbib,eqsecnum,superscriptaddress]{revtex4}

 \usepackage[dvips,final]{graphicx}
  \usepackage{amssymb}
   \usepackage{amsmath}
    \usepackage{amsfonts}
     \usepackage{epsfig}
      \usepackage{bm}

\usepackage{mathpazo}

\usepackage[section]{placeins}

\usepackage{multirow}
\usepackage{ctable}
\usepackage{booktabs}
\usepackage{array}
\usepackage{tabularx}
\usepackage{xcolor}
\usepackage{pstricks}
\definecolor{fred}{rgb}{0.90053, 0.00369, 0.00159}  

\begin{document}

\author{Rafa{\l} Maciu{\l}a}
\email{rafal.maciula@ifj.edu.pl}
\affiliation{Institute of Nuclear
Physics, Polish Academy of Sciences, ul. Radzikowskiego 152, PL-31-342 Krak{\'o}w, Poland}

\author{Antoni Szczurek\footnote{also at University of Rzesz\'ow, PL-35-959 Rzesz\'ow, Poland}}
\email{antoni.szczurek@ifj.edu.pl} 
\affiliation{Institute of Nuclear
Physics, Polish Academy of Sciences, ul. Radzikowskiego 152, PL-31-342 Krak{\'o}w, Poland}

\author{Jakub Zaremba}
\email{jakub.zaremba@ifj.edu.pl}
\affiliation{Institute of Nuclear
Physics, Polish Academy of Sciences, ul. Radzikowskiego 152, PL-31-342 Krak{\'o}w, Poland}

\author{Izabela Babiarz}
\email{izabela.babiarz@ifj.edu.pl}
\affiliation{Institute of Nuclear
Physics, Polish Academy of Sciences, ul. Radzikowskiego 152, PL-31-342 Krak{\'o}w, Poland}
 
\title{
Production asymmetry of \bm{$\nu_{\tau}$} neutrinos and \bm{${\overline \nu}_{\tau}$} antineutrinos \\from a fixed target experiment SHiP}

\begin{abstract}
We discuss how to calculate cross sections as well as rapidity, transverse
momentum and energy distributions of 
$\nu_{\tau}$ and ${\overline \nu}_{\tau}$ produced
from the direct $D_s^{\pm} \to \nu_{\tau}/{\overline \nu}_{\tau}$ and chain $D_s^{\pm} \to \tau^+/\tau^- \to \nu_{\tau}/{\overline \nu}_{\tau}$ decays in $p\!+^{96}\!\mathrm{Mo}$ scattering with proton beam $E_{\mathrm{lab}}$ = 400 GeV \textit{i.e.} at $\sqrt{s}_{NN}$ = 27.4 GeV. The $\tau$ decays are simulated with 
the help of the \textsc{Tauola} code and include multiple decay channels of $\tau$ in amounts proportional to their branching ratios.
In our calculations we include $D_s^{\pm}$ from charm fragmentation
$c \to D_s^{+}$ and $\bar c \to D_s^-$ as well as those from
subleading fragmentation of strange quarks/antiquarks $s \to D_s^-$ and $\bar s \to D_s^+$.
The $s \ne \bar s$ asymmetry of the strange quark content of proton is included.
The different contributions to $D_s^{\pm}$ and
$\nu_{\tau} / {\overline \nu}_{\tau}$ are shown explicitly.
We discuss and quantify a not discussed so far effect of asymmetries 
for production of $\nu_{\tau}$ and ${\overline \nu}_{\tau}$ caused by 
subleading fragmentation mechanism and discuss
related uncertainties.
A potential measurement of the asymmetry is discussed. 
Estimates of a number of observed 
$\nu_{\tau} / \overline{\nu}_{\tau}$ in 
the $\nu_{\tau} / \overline{\nu}_{\tau} +^{208}\!\mathrm{Pb}$ reaction, 
with 2m long target are given with the help of the NuWro program.
We refer also to the production of the high-energy
(anti)neutrinos in the atmosphere.
\end{abstract} 

\maketitle

\section{Introduction}

The $\nu_{\tau}$ and $\overline{\nu}_{\tau}$ particles were ones of last ingredients
of the Standard Model discovered experimentally \cite{DONUT1}.
So far only a few $\nu_\tau/\overline{\nu}_{\tau}$ neutrinos/antineutrinos were observed
experimentally \cite{DONUT2,OPERA}. Recently the IceCube experiment 
observed 2 cases of the $\tau$ neutrinos/antineutrinos \cite{IceCube_tau}.

The proposed SHiP (Search for Hidden Particles) experiment \cite{SHiP1,SHiP2}
may change the situation \cite{SHiP3}.
It was roughly estimated that about $300-1000$ neutrinos ($\nu_{\tau}+\overline{\nu}_{\tau}$)
will be observed by the SHiP experiment \cite{SHiP3,BR2018}. This will 
considerably improve our knowledge in this weakly tested corner of the Standard Model.

The $\nu_\tau/\overline{\nu}_{\tau}$ neutrinos/antineutrinos are known to be primarily produced
from $D_s^{\pm}$ decays. The corresponding branching fraction is
relatively well known \cite{PDG} and is 
BR$(D_s^{\pm} \to \tau^{\pm})$ = 0.0548.
The $D_s$ mesons are abundantly produced in proton-proton collisions.
They were measured \textit{e.g.} at the LHC by the ALICE
\cite{Abelev:2012tca} and the LHCb experiments \cite{Aaij:2013mga}.
The LHCb experiment in the collider-mode has observed even a small
asymmetry in the production of $D_s^+$ and $D_s^-$ \cite{Aaij:2018afd}. So far the asymmetry
is not fully understood from first principles. In Ref.~\cite{Goncalves:2018zzf}
two of us proposed a possible explanation of the fact in terms
of subleading $s \to D_s^-$ or $\bar s \to D_s^+$ fragmentations.
However, the corresponding fragmentation functions are not well known. 

Here we wish to investigate possible consequences for forward production
of $D_s$ mesons and forward production of $\nu_\tau$
neutrinos and $\overline{\nu}_{\tau}$ antineutrinos.
In our model $D_s^{\pm}$ mesons can be produced from both, charm and strange quark/antiquark
fragmentation, with a similar probability of the transition. The $s \to
D_s$ mechanism is expected to be especially
important at large rapidities (or large Feynman $x_F$) \cite{Goncalves:2018zzf}.  
Does it has consequences
for forward production of neutrinos/antineutrinos for the SHiP experiment?
We shall analyze this issue in the present paper.
In short, we wish to make as realistic as possible predictions of the cross section
for production of $\nu_\tau/\overline{\nu}_{\tau}$ neutrinos/antineutrinos.
Here we will also discuss interactions of the neutrinos/antineutrinos 
with the matter (Pb target was proposed for identifying 
neutrinos/antineutrinos). This was discussed already in the literature 
(see \textit{e.g.} Ref.~\cite{JR2010} and references therein).

\section{Some details of the approach}

Here we discuss in short mechanisms of production of $D_s$ mesons,
weak decays of $D_s$ mesons to $\nu_\tau/\overline{\nu}_{\tau}$ neutrinos/antineutrinos 
and interactions of $\nu_\tau/\overline{\nu}_{\tau}$ neutrinos/antineutrinos with nuclear targets.

\subsection{$D_s$ meson production}

\begin{figure}[!h]
\begin{minipage}{1.0\textwidth}
  \centerline{\includegraphics[width=1.0\textwidth]{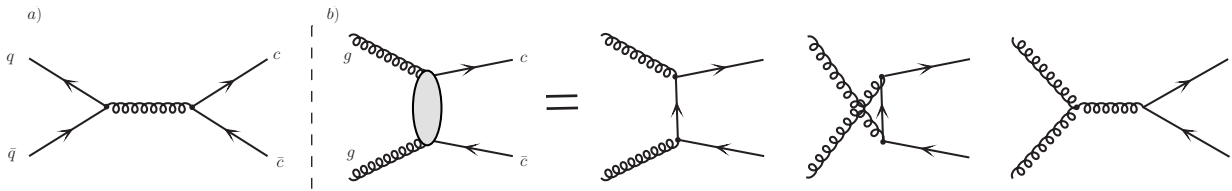}}
\end{minipage}
  \caption{
\small Dominant mechanisms of charm quark production at leading-order: $q\bar q$-annihilation (diagram a) and $gg$-fusion (diagrams b).
These partonic processes lead to leading (standard) fragmentation component 
       of $D_s$ production.
}
\label{fig:diagrams_leading}
\end{figure}

\begin{figure}[!h]
\begin{minipage}{1.0\textwidth}
  \centerline{\includegraphics[width=1.0\textwidth]{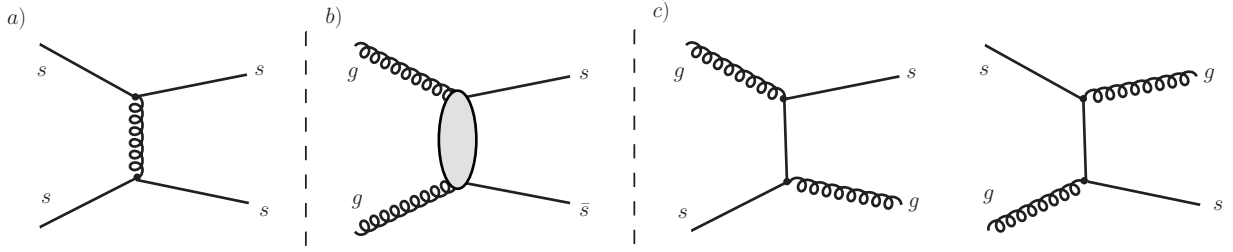}}
\end{minipage}
  \caption{
\small An example of strange quark (or antiquark) production mechanisms at leading-order: $ss \to ss$ (diagram a), $gg \to s\bar s$ (diagram b), $gs \to gs$ and $sg \to sg$ (diagrams c). These partonic processes lead to subleading (unfavored) fragmentation component 
       of $D_s$ production.}
\label{fig:diagrams_subleading}
\end{figure}

In the present paper we discuss two mechanisms of $D_s$ meson
production:
\vskip-2mm
\begin{itemize}
\item $c \to D_s^+$, $\bar c \to D_s^-$, called leading fragmentation,
\item $\bar s \to D_s^+$, $s \to D_s^-$, called subleading fragmentation.
\end{itemize}
The underlying leading-order pQCD partonic mechanisms for charm and strange quark production are shown schematically in 
Fig.~\ref{fig:diagrams_leading} and Fig.~\ref{fig:diagrams_subleading}, 
respectively. 
At high energies, for charm quark production higher-order (NLO and even NNLO) 
corrections are very important, especially when considering differential
distributions, such as quark transverse momentum distribution 
or quark-antiquark correlation observables (see \textit{e.g.}~Refs.~\cite{Cacciari:2012ny,Maciula:2019izq}).

The $c$ and $\bar c$ cross sections are calculated in the collinear NLO 
approximation using the \textsc{Fonll} framework \cite{FONLL}
or in the $k_t$-factorization approach \cite{kTfactorization}. 
The latter calculations are done within both, the standard
scheme with $2\to2$ hard subprocesses as well as within a new scheme 
with higher-order ($2\to3$ and $2 \to 4$) mechanisms
included at the tree level\footnote{We have checked numerically, that both prescriptions almost
coincide for the KMR uPDF also at the rather low c.m.s. collision energy considered here.} \cite{Maciula:2019izq}. Here, both 
the $gg$-fusion and $q\bar q$-annihilation production mechanisms for 
$c\bar c$-pairs with off-shell initial state partons are taken into 
consideration.

Not all charm hadrons must be created from the $c /{\bar c}$ fragmentation.
An extra hidden associated production of $c$ and $\bar c$
can occur in a complicated hadronization process. 
In principle, $c$ and $\bar c$ partons
can also hadronize into light mesons (e.g. kaons) with 
non-negligible fragmentation fraction (see e.g. Ref.~\cite{Epele:2018ewr}).
Similarly, fragmentation of light partons into heavy mesons may be well 
possible \cite{Kneesch:2007ey}.
In the present study we will discuss also results of 
\textsc{Pythia} hadronization 
to $D_s$ mesons in this context as well as our simple model of 
subleading fragmentation $s \to D_s^-$ and $\bar s \to D_s^+$ \cite{Goncalves:2018zzf}. 

The $s$ and $\bar s$ distributions are calculated here in the
  leading-order 
(LO) collinear factorization approach with on-shell initial state partons and with a special treatment of minijets at low transverse momenta, as adopted \textit{e.g.} in \textsc{Pythia}, 
by multiplying standard cross section by a somewhat arbitrary
suppression factor \cite{Sjostrand:2014zea}
\begin{equation}
F_{sup}(p_t) = \frac{p_t^4}{((p_{t}^{0})^{2} + p_t^2)^2}  \; .
\end{equation}
Within this framework the cross section of course strongly depends on the free parameter $p_{t}^{0}$ which could be, in principle, fitted to low energy charm experimental data \cite{Maciula:2017wov}. Here, we use rather conservative value $p_{t}^{0} = 1.5$ GeV. We use three different sets of the collinear parton distribution functions (PDFs): the MMHT2014 \cite{Harland-Lang:2014zoa}, the NNPDF30 \cite{Ball:2014uwa} and the JR14NLO08FF \cite{Jimenez-Delgado:2014twa} parametrizations. All of them provide an asymmetric strange sea quark distributions in the proton with $s(x)\neq \bar{s}(x)$. The dominant partonic mechanisms are $gs \to gs$, $g\bar s \to g\bar s$ (and their symmetric counterparts) and $gg \to s\bar s$.
In some numerical calculations we take into account also other $2\to 2$ diagrams with $s(\bar s)$-quarks in the final state, however, their contributions are found to be almost negligible.

The transition from quarks to hadrons in our calculations is done within the independent parton fragmentation picture.
Here, we follow the assumptions relevant for the case of low c.m.s. collision energies and/or small transverse momenta of hadrons,
as discussed in our recent analysis \cite{Maciula:2019iak}, and we assume that the hadron $H$ is emitted in the direction of parent
quark/antiquark $q$, i.e. $\eta_H = \eta_q$ (the same pseudorapidities or
polar angles). Within this approach we set the light-cone $z$-scaling, \textit{i.e.} we define $p_H^+ = z p_q^+$, where $p^{+} = E+p$.
In the numerical calculations we also include ``energy conservation'' conditions: $E_H > m_H$ and $E_{H} \leq E_{q}$.
If we take the parton as the only reservoir of energy (independent parton fragmentation) these conditions (especially the latter one)
may be strongly broken in the standard fragmentation framework with
constant rapidity $y_q = y_H$ scenario, especially, when discussing
small transverse momenta of hadrons. The light-cone scaling prescription reproduces the standard approach in the limit: $m_{q}, m_{H} \to 0$.

For $c/\bar c \to D_s^{\pm}$ fragmentation we take the traditional Peterson fragmentation function with $\varepsilon$ = 0.05.
In contrast to the standard mechanism, the fragmentation function for
$s/{\bar s} \to D_s^{\mp}$ transition is completely unknown which makes
the situation more difficult. For the case of light-to-light (light parton to light meson) transition rather softer fragmentation functions
(peaked at smaller $z$-values) are supported by phenomenological studies
\cite{Bertone:2017tyb}.
However, the light-to-heavy fragmentation should not be significantly different than for the heavy-to-heavy case.
The shape of the fragmentation function depends on mass of the hadron rather than on the mass of parton (see \textit{e.g.} Ref.~\cite{Kneesch:2007ey}). 
Therefore, here we take the same fragmentation function for the $s/{\bar s} \to D_s^{\mp}$ as for the
$c/\bar c \to D_s^{\pm}$. Besides the shape of the $s/{\bar s} \to D_s^{\mp}$ fragmentation function the relevant fragmentation fraction
is also unknown. The transition probability $P = P_{s \to D_s}$ can be treated as a free parameter and needs to be extracted
from experimental data. First attempt was done very recently in Ref.~\cite{Goncalves:2018zzf}, where $D^{+}_{s}/D^{-}_{s}$ production asymmetry was studied.

\begin{table}[tb]%
\caption{Cross sections for charm and strangeness production
in $pp$-collisions for $\sqrt{s}$ = 27.4 GeV. For charm mesons $P_{c \to D_s}=0.08$ and $P_{s \to D_s}=0.05$ are used.}

\label{tab:low_energies}
\centering %
\begin{tabularx}{0.9\linewidth}{c c c}
\\[-3.5ex] 
\toprule[0.1em] %

\multirow{2}{7.5cm}{Framework/mechanism} & \multicolumn{2}{c}{\multirow{1}{6.cm}{Total cross section   [$\mu$b]}}  \\ [-0.ex]
\multirow{2}{7.5cm}{}                                         & \multirow{1}{4.cm}{partonic} & \multirow{1}{4.cm}{$D_{s}^{+}$ or $D_{s}^{-}$}\\ [+0.1ex]

\bottomrule[0.1em]
\multirow{1}{7.5cm}{FONLL: all processes $\to c/\bar{c} \to D_{s}^{+}/D_{s}^{-} $ } & \multirow{2}{3.cm}{} & \multirow{2}{3.cm}{}\\ [-1.0ex]
\multirow{1}{7.5cm}{MMHT2014nlo} & \multirow{1}{3.cm}{12.568} & \multirow{1}{3.cm}{0.510}\\ [-1.0ex]
\multirow{1}{7.5cm}{CT14nlo} & \multirow{1}{3.cm}{10.751} & \multirow{1}{3.cm}{0.445}\\ [-1.0ex]
\multirow{1}{7.5cm}{JR14NLO08FF} & \multirow{1}{3.cm}{7.806} & \multirow{1}{3.cm}{0.277}\\ [-1.0ex]
\multirow{1}{7.5cm}{NNPDF30 NLO} & \multirow{1}{3.cm}{4.955} & \multirow{1}{3.cm}{0.200}\\ [+0.1ex]
\bottomrule[0.1em]
\multirow{1}{7.5cm}{$k_{T}$-fact. + KMR MMHT2014lo uPDF} & \multirow{1}{3.cm}{} & \multirow{1}{3.cm}{}\\ [-1.0ex]
\multirow{1}{7.5cm}{$g^*g^* \to c\bar c\;\;\;\;\;\;$  ($c/\bar{c} \to D_{s}^{+}/D_{s}^{-} $)} & \multirow{1}{3.cm}{3.191} & \multirow{1}{3.cm}{0.142}\\ [-1.0ex]
\multirow{1}{7.5cm}{$q^*\bar{q}^* \to c\bar c\;\;\;\;\;\;$  ($c/\bar{c} \to D_{s}^{+}/D_{s}^{-}$)} & \multirow{1}{3.cm}{0.164} & \multirow{1}{3.cm}{0.007}\\ [+0.1ex]
\hline
\multirow{1}{7.5cm}{$k_{T}$-fact. + KMR CT14lo uPDF} & \multirow{1}{3.cm}{} & \multirow{1}{3.cm}{}\\ [-1.0ex]
\multirow{1}{7.5cm}{$g^*g^* \to c\bar c\;\;\;\;\;\;$  ($c/\bar{c} \to D_{s}^{+}/D_{s}^{-} $)} & \multirow{1}{3.cm}{4.642} & \multirow{1}{3.cm}{0.241}\\ [-1.0ex]
\multirow{1}{7.5cm}{$q^*\bar{q}^* \to c\bar c\;\;\;\;\;\;$  ($c/\bar{c} \to D_{s}^{+}/D_{s}^{-}$)} & \multirow{1}{3.cm}{1.069} & \multirow{1}{3.cm}{0.050}\\ [+0.1ex]
\hline
\multirow{1}{7.5cm}{$k_{T}$-fact. + PB-NLO-set1 uPDF} & \multirow{1}{3.cm}{} & \multirow{1}{3.cm}{}\\ [-1.0ex]
\multirow{1}{7.5cm}{$g^*g^* \to c\bar c\;\;\;\;\;\;$  ($c/\bar{c} \to D_{s}^{+}/D_{s}^{-} $)} & \multirow{1}{3.cm}{2.254} & \multirow{1}{3.cm}{0.073}\\ [-1.0ex]
\multirow{1}{7.5cm}{$q^*\bar{q}^* \to c\bar c\;\;\;\;\;\;$  ($c/\bar{c} \to D_{s}^{+}/D_{s}^{-}$)} & \multirow{1}{3.cm}{1.286} & \multirow{1}{3.cm}{0.055}\\ [+0.1ex]
\bottomrule[0.1em]
\multirow{1}{7.5cm}{LO coll. + MMHT2014lo PDF} & \multirow{1}{3.cm}{} & \multirow{1}{3.cm}{}\\ [-1.0ex]
\multirow{1}{7.5cm}{$gg \to s\bar s\;\;\;\;\;\;\;\;\;$  ($s/\bar{s} \to D_{s}^{-}/D_{s}^{+} $)} & \multirow{1}{3.cm}{1.603} & \multirow{1}{3.cm}{0.032}\\ [-1.0ex]
\multirow{1}{7.5cm}{$is \to is + si \to si\;\;\;\;\;\;$  ($s \to D_{s}^{-}$)} & \multirow{1}{3.cm}{4.789 $\times 2$} & \multirow{1}{3.cm}{0.149}\\ [-1.0ex]
\multirow{1}{7.5cm}{$i\bar{s} \to i\bar{s} + \bar{s}i \to \bar{s}i\;\;\;\;\;\;$  ($\bar s \to D_{s}^{+}$)} & \multirow{1}{3.cm}{3.769 $\times 2$} & \multirow{1}{3.cm}{0.098}\\ [+0.1ex]
\hline
\multirow{1}{7.5cm}{LO coll. + NNPDF30 LO PDF} & \multirow{1}{3.cm}{} & \multirow{1}{3.cm}{}\\ [-1.0ex]
\multirow{1}{7.5cm}{$gg \to s\bar s\;\;\;\;\;\;\;\;\;$  ($s/\bar{s} \to D_{s}^{-}/D_{s}^{+} $)} & \multirow{1}{3.cm}{0.947} & \multirow{1}{3.cm}{0.016}\\ [-1.0ex]
\multirow{1}{7.5cm}{$is \to is + si \to si\;\;\;\;\;\;$  ($s \to D_{s}^{-}$)} & \multirow{1}{3.cm}{1.960 $\times 2$} & \multirow{1}{3.cm}{0.114}\\ [-1.0ex]
\multirow{1}{7.5cm}{$i\bar{s} \to i\bar{s} + \bar{s}i \to \bar{s}i\;\;\;\;\;\;$  ($\bar s \to D_{s}^{+}$)} & \multirow{1}{3.cm}{0.988 $\times 2$} & \multirow{1}{3.cm}{0.047}\\ [+0.1ex]
\hline
\multirow{1}{7.5cm}{LO coll. + JR14NLO08FF PDF} & \multirow{1}{3.cm}{} & \multirow{1}{3.cm}{}\\ [-1.0ex]
\multirow{1}{7.5cm}{$gg \to s\bar s\;\;\;\;\;\;\;\;\;$  ($s/\bar{s} \to D_{s}^{-}/D_{s}^{+} $)} & \multirow{1}{3.cm}{0.733} & \multirow{1}{3.cm}{0.010}\\ [-1.0ex]
\multirow{1}{7.5cm}{$is \to is + si \to si\;\;\;\;\;\;$  ($s \to D_{s}^{-}$)} & \multirow{1}{3.cm}{2.616 $\times 2$} & \multirow{1}{3.cm}{0.086}\\ [-1.0ex]
\multirow{1}{7.5cm}{$i\bar{s} \to i\bar{s} + \bar{s}i \to \bar{s}i\;\;\;\;\;\;$  ($\bar s \to D_{s}^{+}$)} & \multirow{1}{3.cm}{2.413 $\times 2$} & \multirow{1}{3.cm}{0.082}\\ [+0.1ex]

\bottomrule[0.1em]
\end{tabularx}
\end{table}

For further discussions in Table 1 we have collected total cross sections for 
different contributions to charm and strange quark production as well as to subsequent production of $D_{s}^{\pm}$ mesons
in proton-proton scattering at $\sqrt{s}$ = 27.4 GeV. For the leading fragmentation mechanism here we compare results for $c\bar c$-pair production
calculated in the $k_{T}$-factorization approach and in the
\textsc{Fonll} framework. The $k_{T}$-factorization approach leads to a
slightly smaller cross sections than in the case of \textsc{Fonll}. At
the rather low
energy considered here the dominant production mechanism is still the $gg$-fusion, however, the $q\bar q$-annihilation is found to be also important.
This statement is true for calculations with both, on-shell and off-shell partons. For the calculations with off-shell partons we use three different sets of uPDFs: two sets of the Kimber-Martin-Ryskin (KMR) model \cite{Watt:2003vf} based on MMHT2014lo \cite{Harland-Lang:2014zoa} and CT14lo \cite{Dulat:2015mca} collinear PDFs and one set of parton-branching model PB-NLO-set1 \cite{Martinez:2018jxt}. In general, several uPDFs lead to a quite similar results, especially for $g^*g^* \to c \bar c$ mechanism. In the case of $q^*\bar{q}^* \to c \bar c$ channel results of the KMR-CT14lo and PB-NLO-set1 almost coincide, however, we found a significant difference between these two results and the result obtained with the KMR-MMHT2014lo. The discrepancy between KMR-CT14lo and KMR-MMHT2014lo results comes from a significant
differences of up and down quark distributions at very small-$Q^2$ and large-$x$ incorporated in the MMHT2014lo and CT14lo collinear PDFs\footnote{For gluon and strange quark production both models provide rather similar distributions, the differences are much smaller than in the case of up and down quarks.}.
The major part of the cross section for $q^*\bar{q}^* \to c \bar c$
mechanism at $\sqrt{s}=27.4$ GeV comes from the $x \approx 10^{-1}$ and
$k_{t} \approx 1$ GeV kinematical regime. In the KMR procedure,
transverse momentum of the incoming parton $k_{t}$ plays a role of the scale $Q$ in the collinear PDF
which is an input for the calculation of the uPDF. In fact, for calculations with the KMR uPDFs here, one is very sensitive to the rather poorly constrained region of the collinear PDFs.

Switching to collinear approximation, we do not approach these problematic regions
since the lower limit of the factorization scale is then set to be equal to the charm quark mass, so the minimal scale is $Q^{2}_{\textrm{min}}=2.25$ GeV$^{2}$.         
For the \textsc{Fonll} predicitions different collinear PDFs were used. Various PDF parametrizations lead to very different results.
The total cross sections are very sensitive to the low-$p_{t}$ region
which is very uncertain at this low energy. The predicted cross sections
 strongly depend on the low-$Q^2$
parametrizations of the collinear PDFs which are not under full theoretical control. This uncertainty may be crucial for the further predictions of neutrino production at the SHiP experiment which, as will be shown in the next sections, is mostly driven by this problematic kinematical region. As already mentioned, the situation may be even more complicated in the case of the $k_{T}$-factorization approach where less-known objects, i.e. transverse-momentum-dependent uPDFs, are used. Besides the PDF uncertainties, at very low charm quark transverse momenta \textsc{Fonll} predictions (in general any pQCD calculations) are very sensitive to the choice of the renormalization/factorization scale and the charm quark mass (see e.g. Ref.~\cite{Cacciari:2012ny}). These uncertainties
may be crucial especially in the case of charm flavour production at low energies but were discussed many times in the literature. 

For strange quark and/or antiquark production we consider all the dominant partonic $2\to 2$ processes. Here we show separately results for
$gg$-fusion and for other mechanisms with $s$ or $\bar s$ quark or antiquark in the initial and final state, denoted as $is \to is$ or $i\bar{s} \to i\bar{s}$
where $i =u,d,s,g, \bar{u}, \bar{d}, \bar{s}$. The cross sections for strange quark/antiquark production are of the same order of magnitude as in the case of charm production.
According to the obtained partonic cross sections, both fragmentation mechanisms - leading and subleading, are predicted to contribute to the $D_{s}^{\pm}$-meson cross section at the similar level. Also here, we show results for different collinear PDFs. We have intentionally chosen the PDF parametrizations that lead to an asymmetry in production of $s$ and $\bar s$. Within these models the $s$-quarks are produced more frequently than the $\bar s$-antiquarks and the largest production asymmetry is obtained for the NNPDF30 PDF. 

\begin{table}[tb]%
\caption{Number of $D_s$ mesons per $10^6$ generations of hard processes,
fraction of a given mechanism and respective cross sections in nanobarns from \textsc{Pythia} Monte Carlo generator. Here we collected numbers
for $D_s^+ + D_s^-$ mesons.}

\label{tab:pythia}
\centering %
\begin{tabularx}{0.9\linewidth}{c c c c}
\\[-3.5ex] 
\toprule[0.1em] %

\multirow{1}{5.cm}{\textsc{Pythia}: $10^6$ generations} & \multirow{1}{3.cm}{N($D_{s}^{+}$)$+$N($D_{s}^{-}$)} & \multirow{1}{3.cm}{fraction [$\%$]} & \multirow{1}{3.cm}{cross section [nb]} \\
\bottomrule[0.1em]
\multirow{1}{5.cm}{total}                   & \multirow{1}{3.cm}{1467+1771} &  \multirow{1}{3.cm}{100.0}  &  \multirow{1}{3.cm}{1156} \\
\hline
\multirow{1}{5.cm}{$g g \to c \bar c$}  &  \multirow{1}{3.cm}{1099}     & \multirow{1}{3.cm}{33.9}  &   \multirow{1}{3.cm}{392} \\
\multirow{1}{5.cm}{$q \bar q \to c \bar c$} &   \multirow{1}{3.cm}{163}     & \multirow{1}{3.cm}{5.0}  &    \multirow{1}{3.cm}{58} \\
\multirow{1}{5.cm}{$g g \to g g$}           &   \multirow{1}{3.cm}{174}     & \multirow{1}{3.cm}{5.4}  &    \multirow{1}{3.cm}{62} \\
\multirow{1}{5.cm}{$q/{\bar q} g$}          &  \multirow{1}{3.cm}{1088}     & \multirow{1}{3.cm}{33.6}  &   \multirow{1}{3.cm}{388} \\
\multirow{1}{5.cm}{$q q'$, etc.}        &   \multirow{1}{3.cm}{713}  & \multirow{1}{3.cm}{22.0}  &   \multirow{1}{3.cm}{255} \\

\bottomrule[0.1em]
\end{tabularx}
\end{table}

The overall picture for $D_{s}^{\pm}$-meson production based on the independent parton fragmentation framework with leading and subleading fragmentation
components seem to be similar to the picture present in the
\textsc{Pythia} Monte Carlo generator. In Table \ref{tab:pythia} we show
the number of $D_{s}^{\pm}$ mesons per $10^6$ generations of hard
processes, fraction of a given mechanism and respective cross sections
in nanobarns obtained from the \textsc{Pythia} generator. The partonic structure of the $D_s^{\pm}$ meson
production in PYTHIA is rather similar to the structure obtained in our model. The dominant mechanisms here are $gg\to c\bar c$ (our leading component) and 
$q(\bar{q})g\to q(\bar{q})g$ and $qq' \to qq'$ (our subleading
components). Both models lead to a very similar results for the leading
component. For the subleading contributions, our model slightly
underestimates the \textsc{Pythia} predictions. Also here we got a clear production asymmetry $N(D_s^+)$ = 1467 and $N(D_s^-)$ = 1771, which is important in the context of the production asymmetry of neutrinos/antineutrinos.



\begin{figure}[!htbp]
\begin{minipage}{0.47\textwidth}
 \centerline{\includegraphics[width=1.0\textwidth]{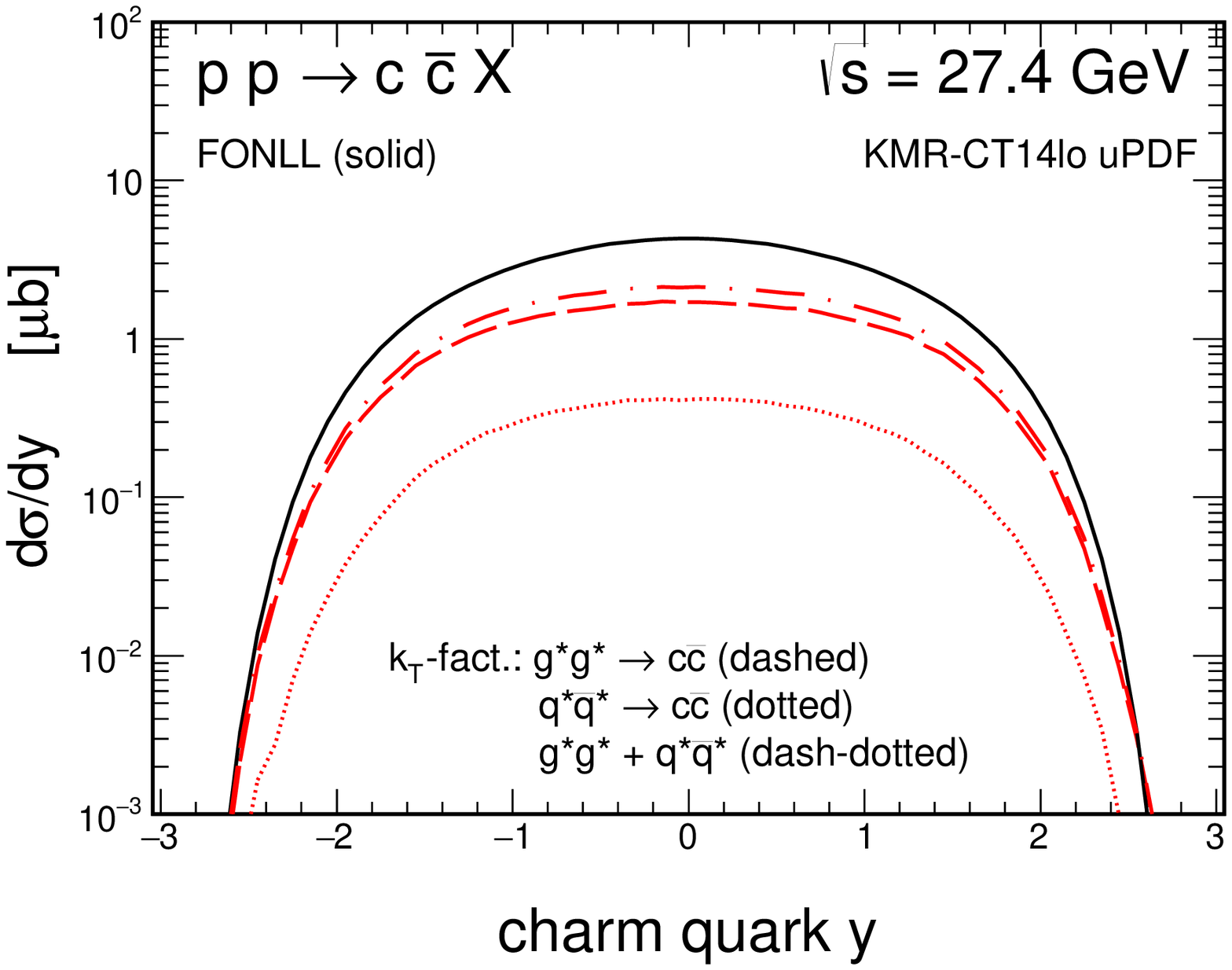}}
\end{minipage}
\hspace{0.5cm}
\begin{minipage}{0.47\textwidth}
 \centerline{\includegraphics[width=1.0\textwidth]{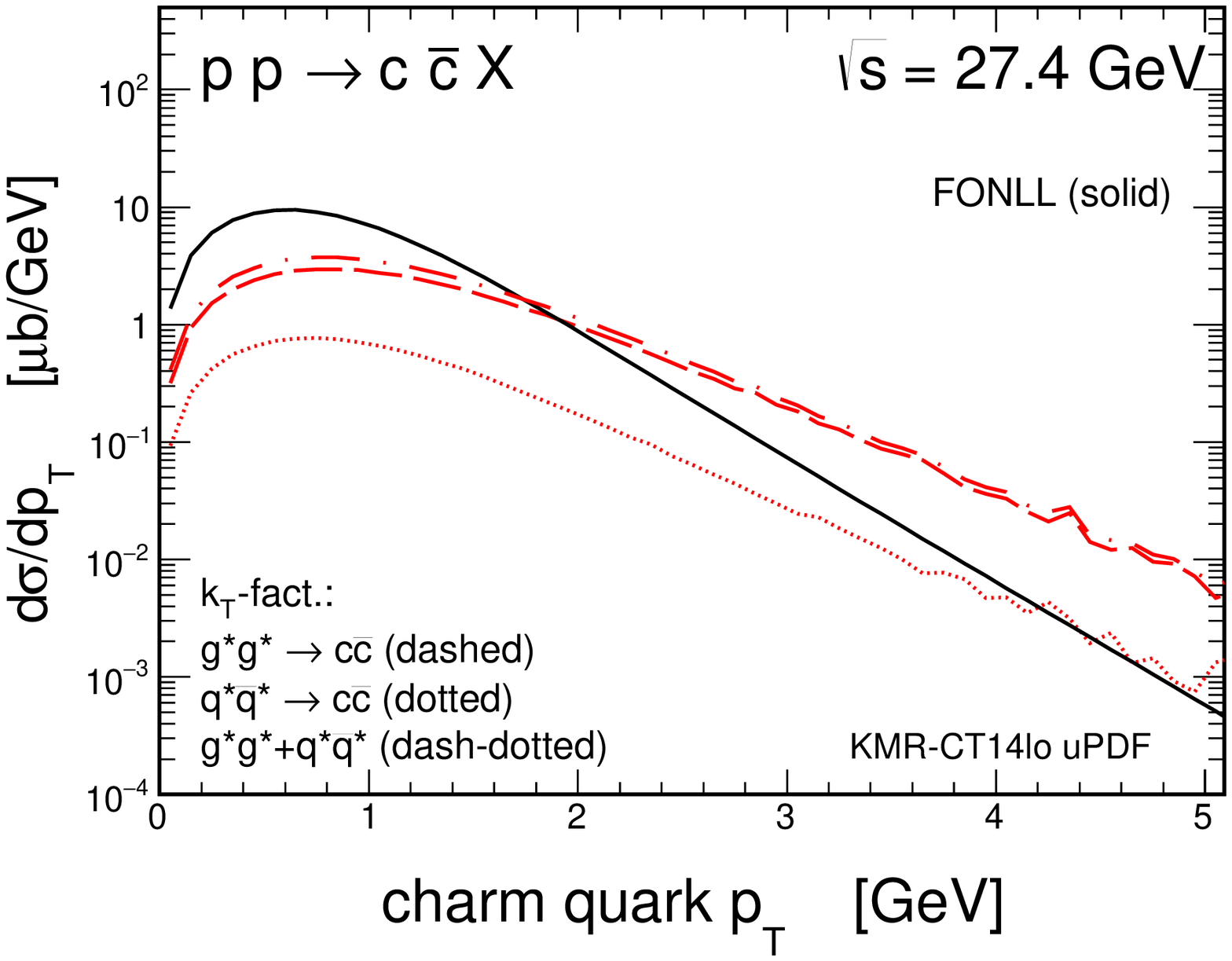}}
\end{minipage}\\
\begin{minipage}{0.47\textwidth}
 \centerline{\includegraphics[width=1.0\textwidth]{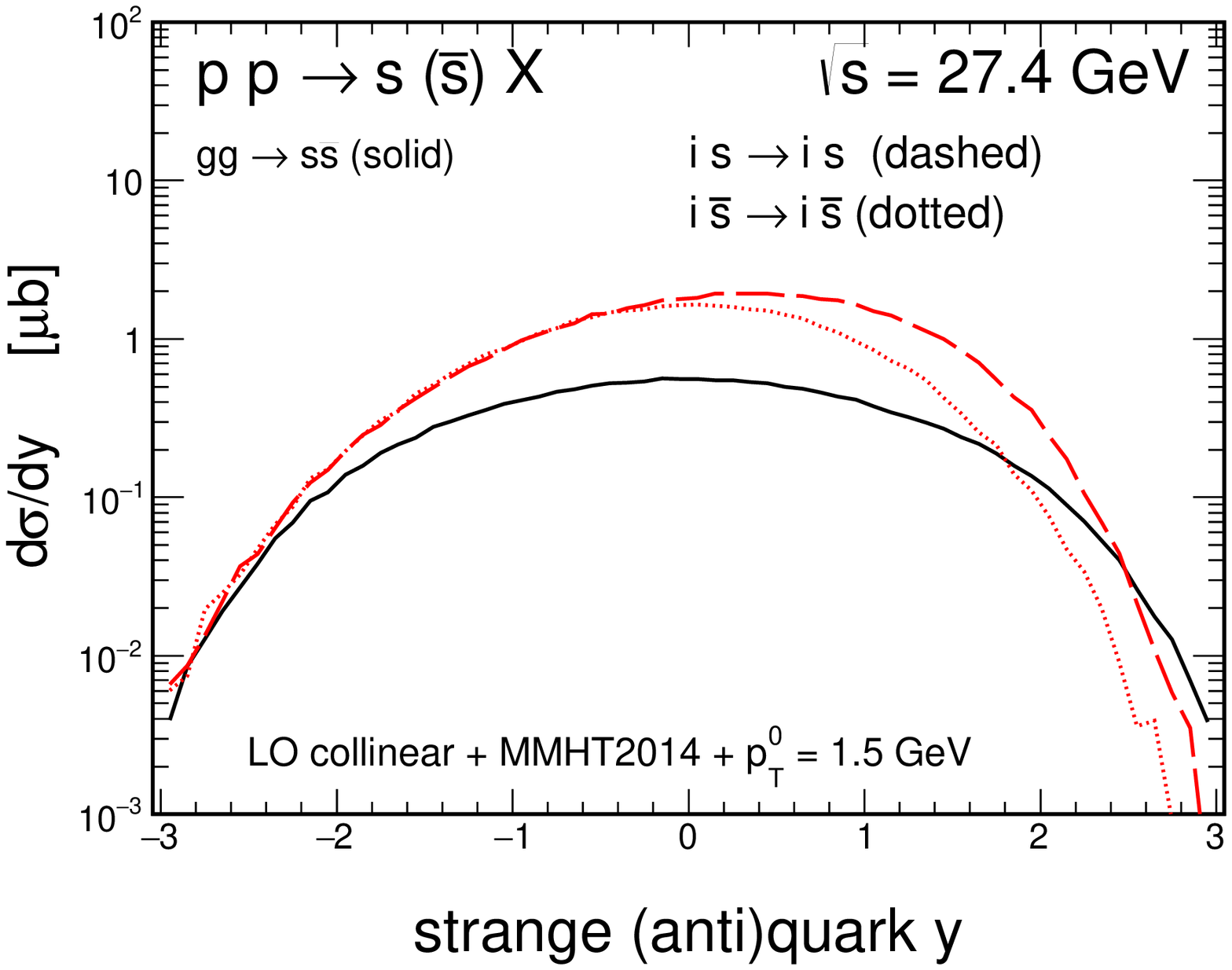}}
\end{minipage}
\hspace{0.5cm}
\begin{minipage}{0.47\textwidth}
 \centerline{\includegraphics[width=1.0\textwidth]{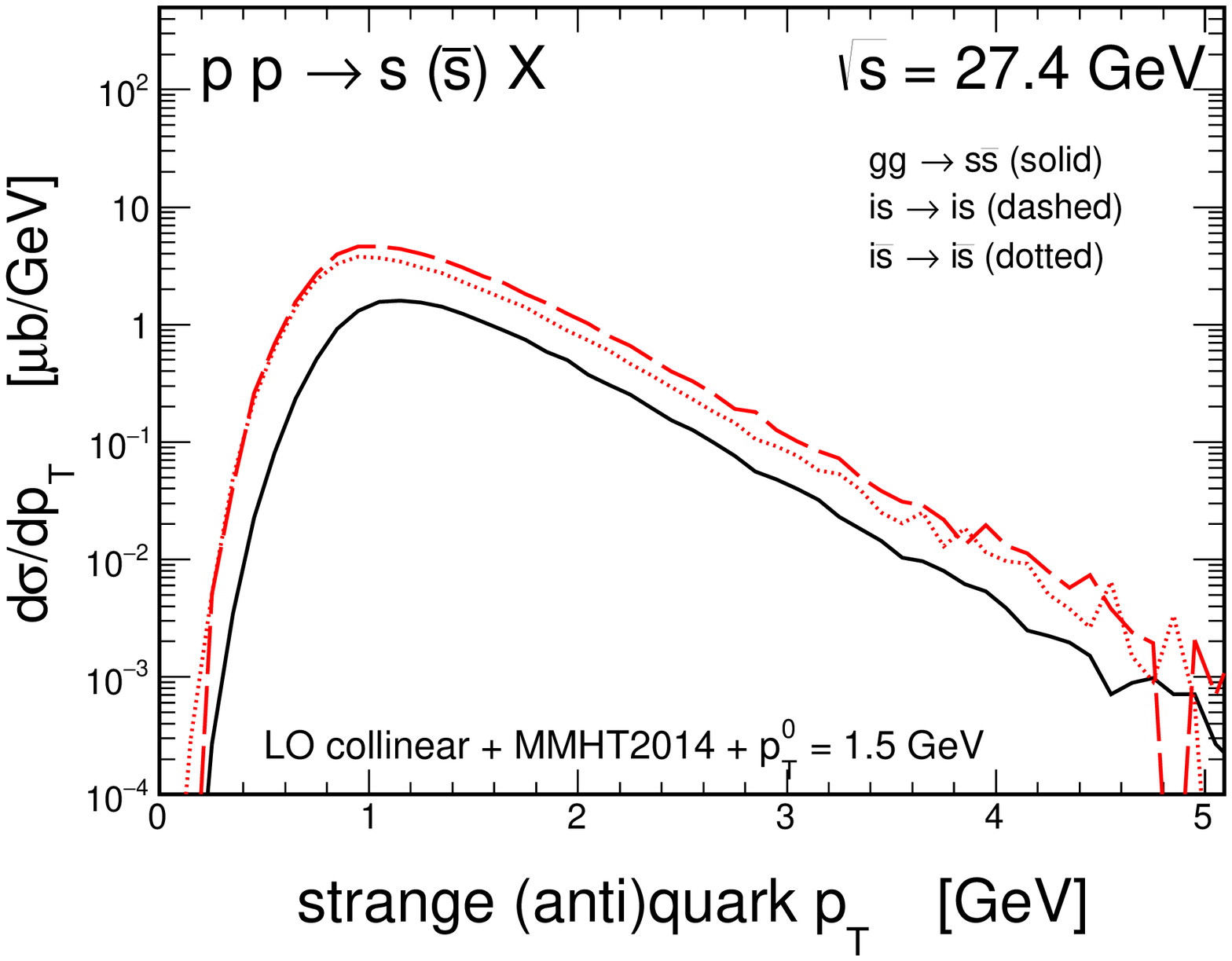}}
\end{minipage}
\caption{
\small Rapidity (left) and transverse momentum (right) distributions of charm (top) and strange (bottom) quarks.
Contributions of different mechanisms are shown separately. 
For strange quarks only contributions of $i s \to is$ or $i \bar s \to i \bar s$ 
($i = g, q, \bar q$) are shown. Similar contributions for
$s i \to s i$ or $\bar s i \to \bar s i$ can be obtained by the
$y \to -y$ symmetry operation.
Other details are specified in the figure.
}
 \label{fig:partons}
\end{figure}

In Fig.~\ref{fig:partons} we compare distributions of $c/\bar c$ and $s/\bar s$ quarks/antiquarks (top and bottom panels)
produced in proton-proton collisions at $\sqrt{s}$ = 27.4 GeV. For charm
quarks we get a very similar rapidity ditributions within both the
\textsc{Fonll} (solid lines) and the $k_{T}$-factorization (dash-dotted
lines) frameworks. For the latter case we show separately $g^*g^*$
(dashed lines) and $q^*\bar{q}^*$ (dotted lines) components. For the
quark transverse momentum distribution we obtain some differences
between both approaches. At very small transverse momenta the
\textsc{Fonll} code leads to a larger cross section than that obtained in the $k_{T}$-factorization. However, at larger $p_{T}$'s the situation reverses and the $k_{T}$-factorization result now become larger. This may be a combined result of several effects, \textit{i.e.} the effect of keeping exact kinematics from the very beginning in the $k_{T}$-factorization, the effect of the off-shellness of the incident partons and in some limited amount also the effect of the beyond NLO contributions effectively included at the tree-level in the $k_{T}$-factorization approach \cite{Maciula:2019izq} which are missing in the \textsc{Fonll} framework.

In general, the cross section for $s/\bar s$ quarks/antiquarks are of similar order of
magnitude as that for $c \bar c$ production (see top and bottom panels of Fig.~\ref{fig:partons}). For strange quarks we show separately the two dominant channels (or classes of channels): $gg$-fusion (solid lines) and $is \to is$ or $i\bar{s} \to i\bar{s}$ (dashed and dotted lines). For the latter mechanisms we obtain a clear asymmetry between production of $s$-quark and $\bar s$-antiqark which is a direct consequence of the $s(x)\neq \bar{s}(x)$ asymmetry in the MMHT2014lo PDFs.

\begin{figure}[!htbp]
\begin{minipage}{0.47\textwidth}
 \centerline{\includegraphics[width=1.0\textwidth]{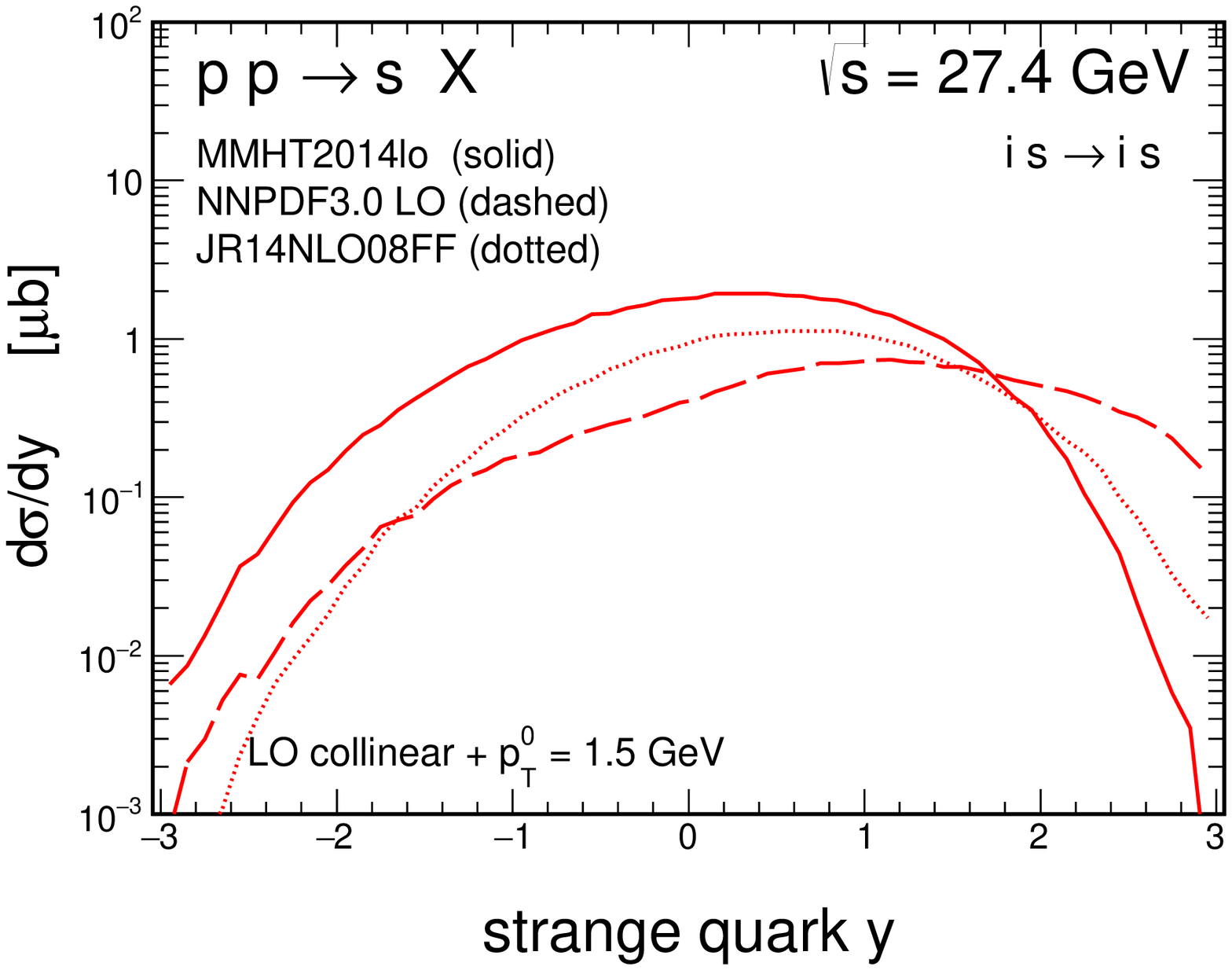}}
\end{minipage}
\hspace{0.5cm}
\begin{minipage}{0.47\textwidth}
 \centerline{\includegraphics[width=1.0\textwidth]{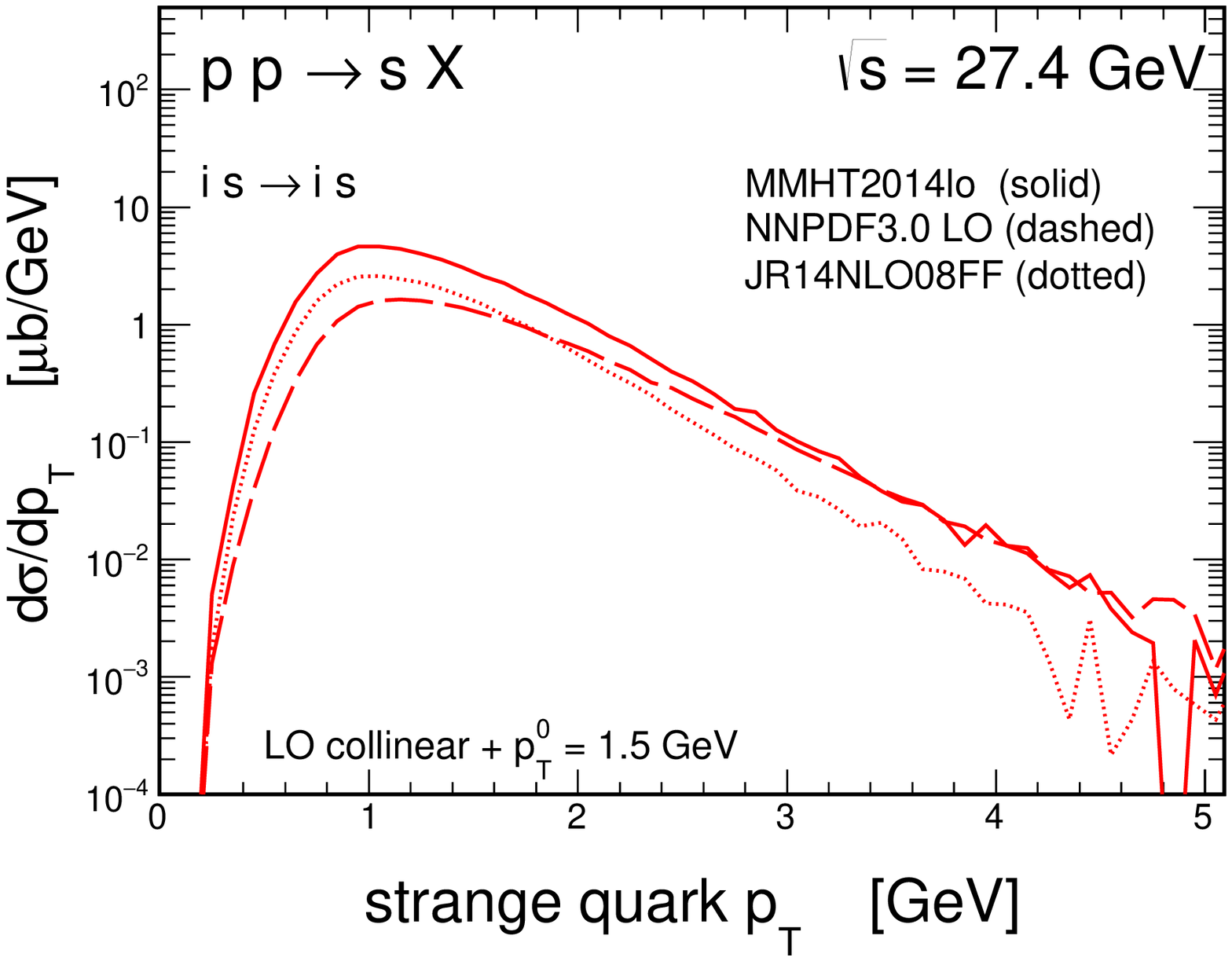}}
\end{minipage}
\caption{
\small Rapidity (left) and transverse momentum (right) distributions of strange quarks
       from $is\to is$ mechanisms for different sets of collinear PDFs. Details are specified in the figure.
}
 \label{fig:partons_pdfs}
\end{figure}

Of course, different PDFs may lead to a different distributions of $s$-quark and $\bar s$-antiquark and also to different size of their production asymmetry. In Fig.~\ref{fig:partons_pdfs} we present rapidity (left panel) and transverse momentum (right panel) distributions of $s$-quark from the $is \to is$ class of processes for
different PDFs from the literature. Quite different rapidity
distributions are obtained from the different PDFs, especially in the (very)forward region where the differences are really large.
In the consequence this will generate uncertainties for far-forward (very large rapidities) production of $D_s$ meson
and $\nu_{\tau} / \overline{\nu}_{\tau}$ neutrinos/antineutrinos as well.

\begin{figure}[!htbp]
\begin{minipage}{0.47\textwidth}
 \centerline{\includegraphics[width=1.0\textwidth]{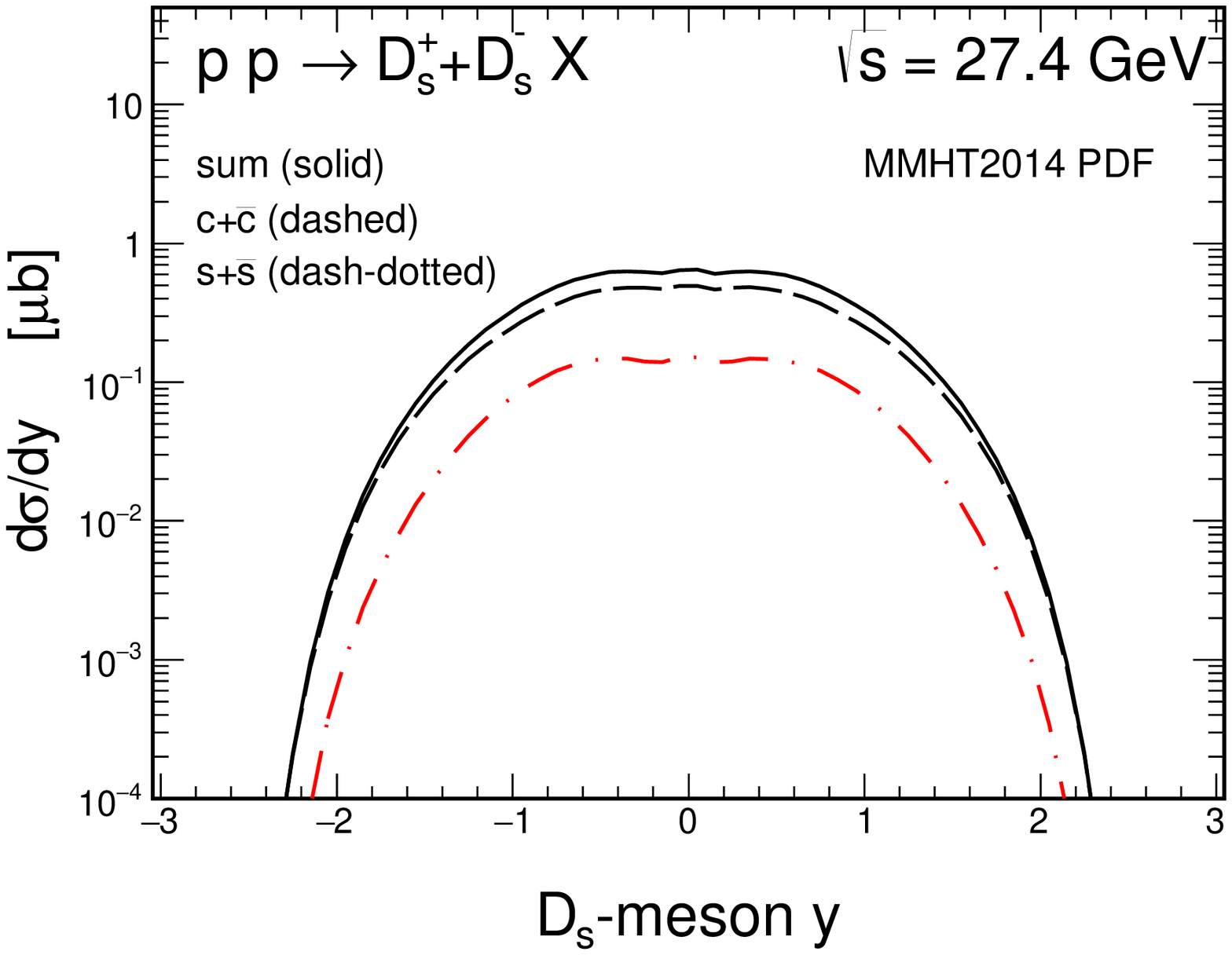}}
\end{minipage}
\hspace{0.5cm}
\begin{minipage}{0.47\textwidth}
 \centerline{\includegraphics[width=1.0\textwidth]{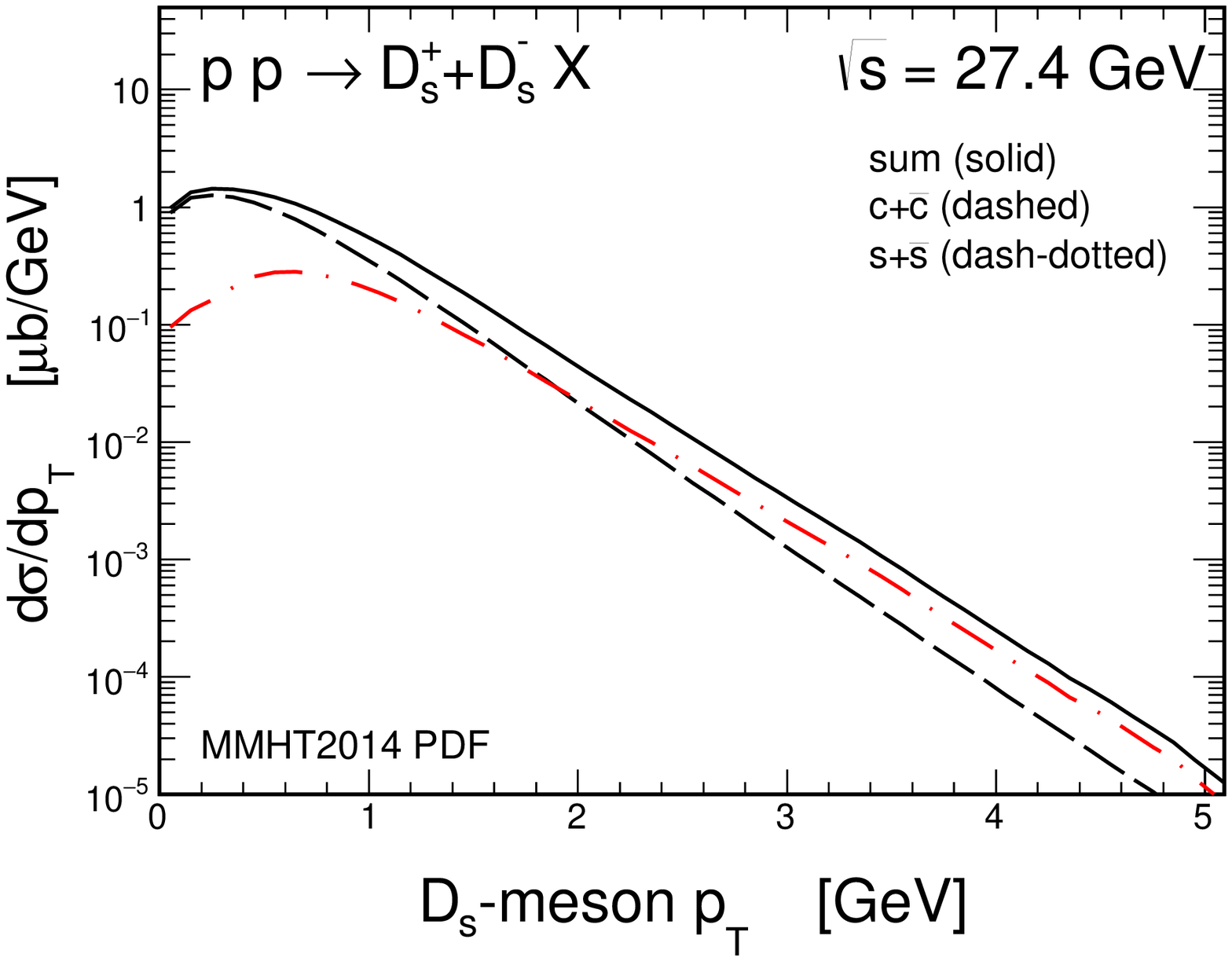}}
\end{minipage}\\
\begin{minipage}{0.47\textwidth}
 \centerline{\includegraphics[width=1.0\textwidth]{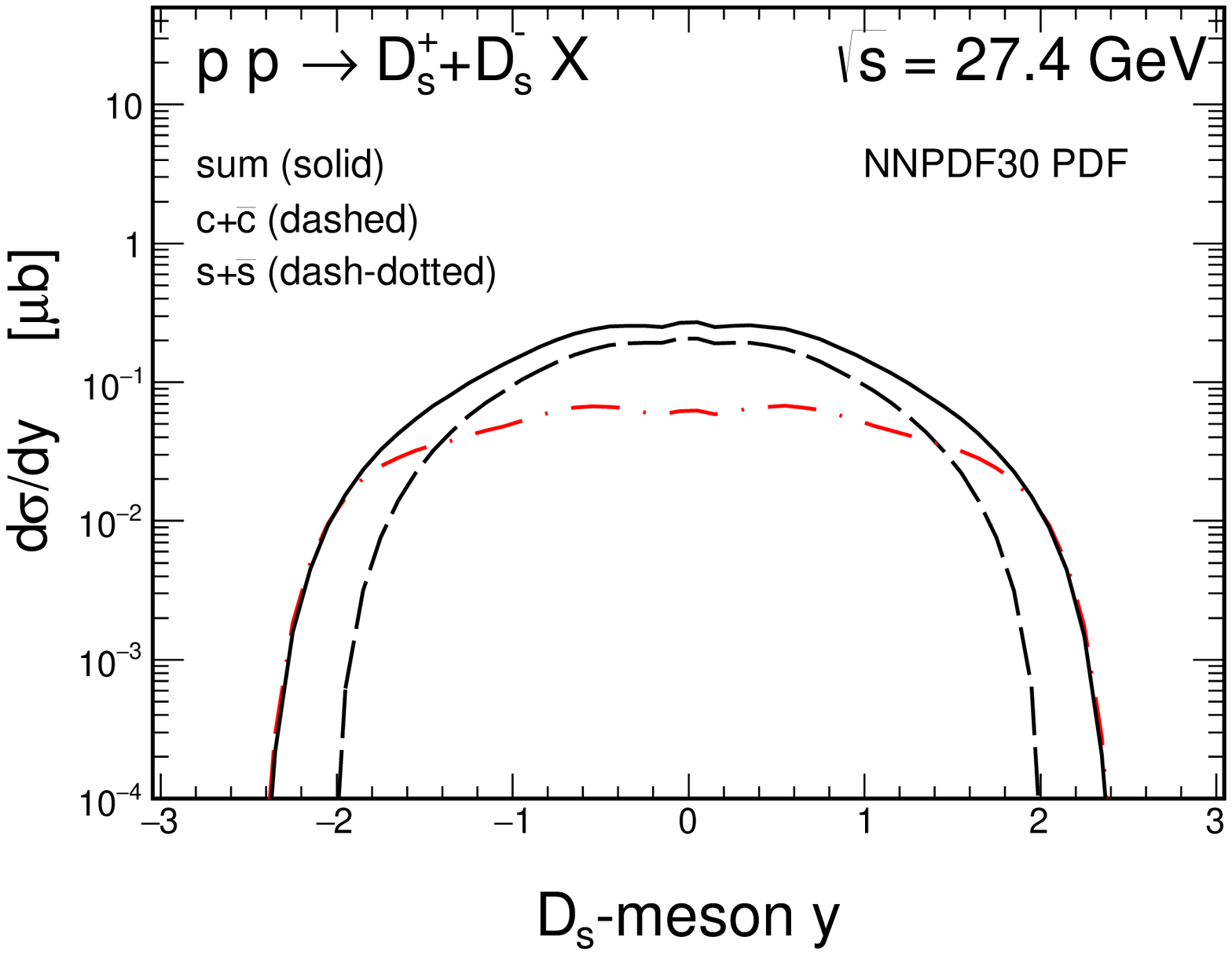}}
\end{minipage}
\hspace{0.5cm}
\begin{minipage}{0.47\textwidth}
 \centerline{\includegraphics[width=1.0\textwidth]{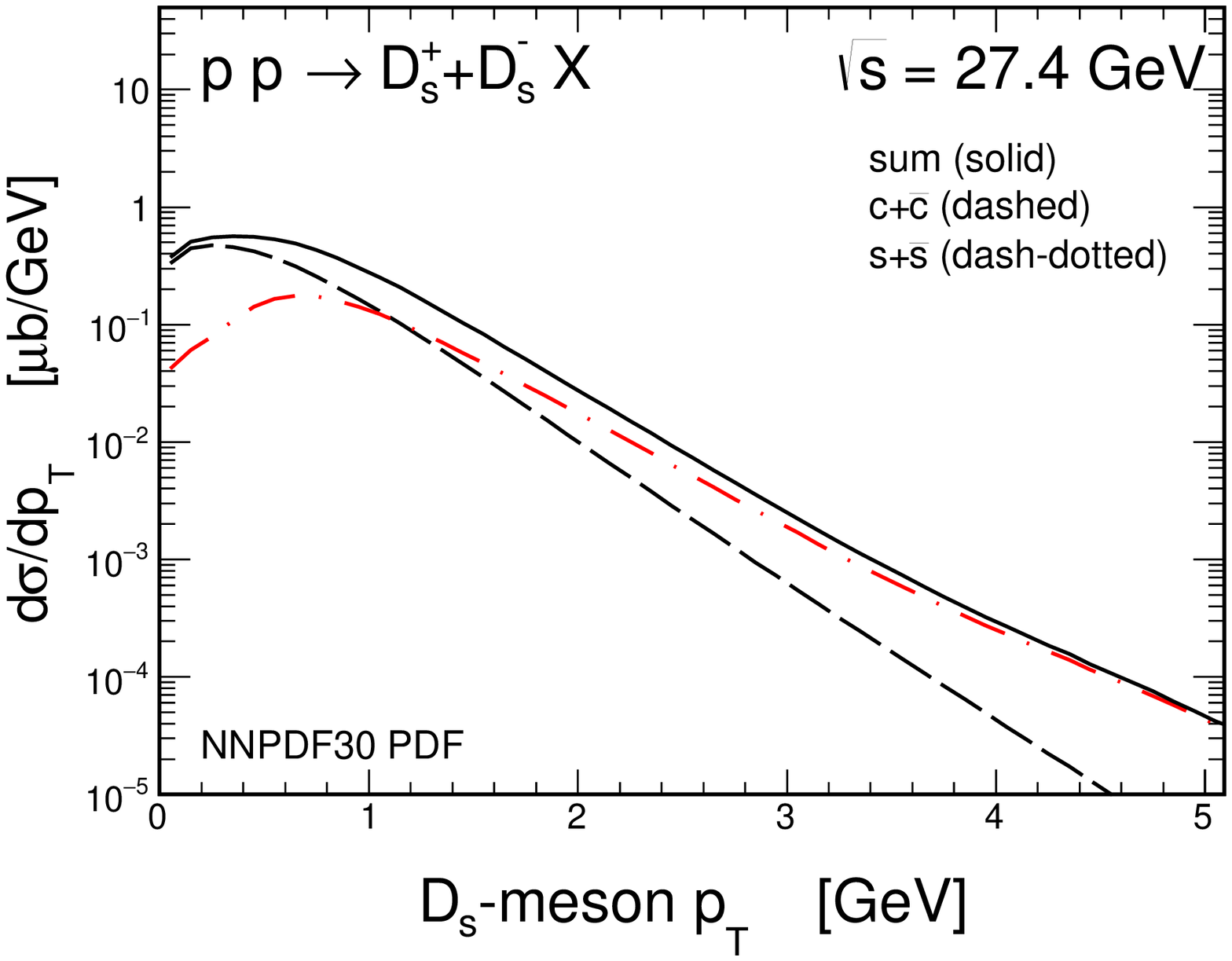}}
\end{minipage}
\caption{
\small Rapidity (left) and transverse momentum (right) distributions of
$D_{s}$ mesons for the MMHT2014 (top) and NNPDF30 (bottom) sets of collinear PDFs. Contributions from charm and strange quark fragmentation are shown separately. Details are specified in the figure.
}
 \label{fig:mesons}
\end{figure}

In Fig.~\ref{fig:mesons} we show the resulting rapidity (left panel) and transverse momentum (right panel)
distributions of $D_s$ mesons from proton-proton scattering at $\sqrt{s}=27.4$ GeV.
We compare contributions of the leading ($c/\bar c \to D_s^{\pm}$) and the
subleading ($s/\bar s \to D_s^{\mp}$) mechanisms, calculated in the \textsc{Fonll} and in the LO collinear approach, respectively.
In this calculation $P_{c \to D_s}=0.08$ and $P_{s \to D_s}=0.05$ were used.
Top and bottom panels show results for different collinear PDF sets from the literature. While for the MHHT2014 PDF the subleading
contribution is always smaller than the leading one, for the NNPDF30 PDF
it is not the case and the subleading contribution wins above $|y|>$ 2.
The subleading contribution also wins at larger meson transverse momenta and changes the slope of the distribution in a visible way.
Again the effect is stronger for the calculations with the NNPDF30 PDF which leads to a smaller leading contribution than in the case of the MMHT2014 PDF.
This demonstrates uncertainties related to the production mechanism.
Related consequences for the production of $\nu_{\tau}/\overline{\nu}_{\tau}$
will be discussed in section \ref{sec:nutau_observed}.

\begin{figure}[!htbp]
\begin{minipage}{0.47\textwidth}
 \centerline{\includegraphics[width=1.0\textwidth]{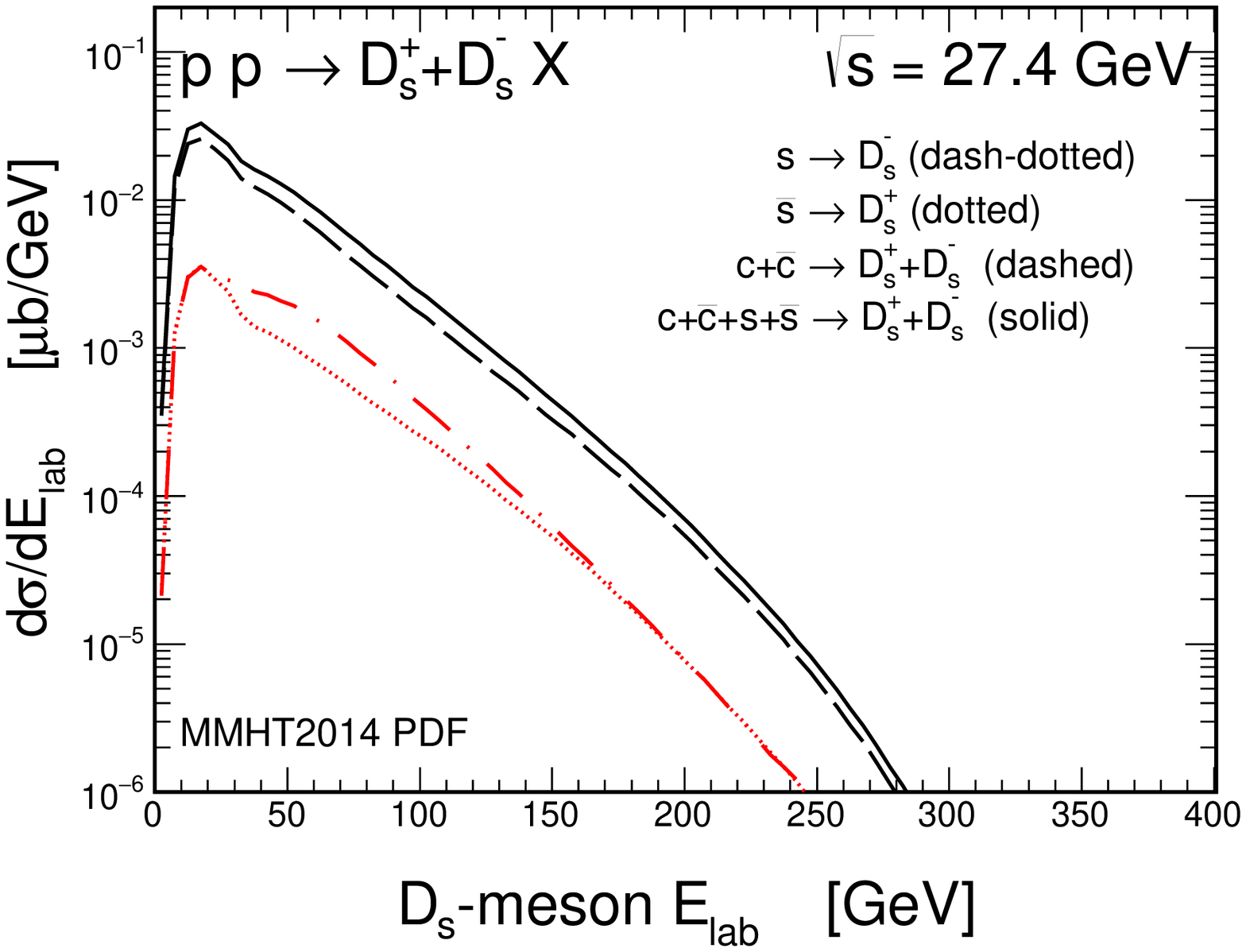}}
\end{minipage}
\hspace{0.5cm}
\begin{minipage}{0.47\textwidth}
 \centerline{\includegraphics[width=1.0\textwidth]{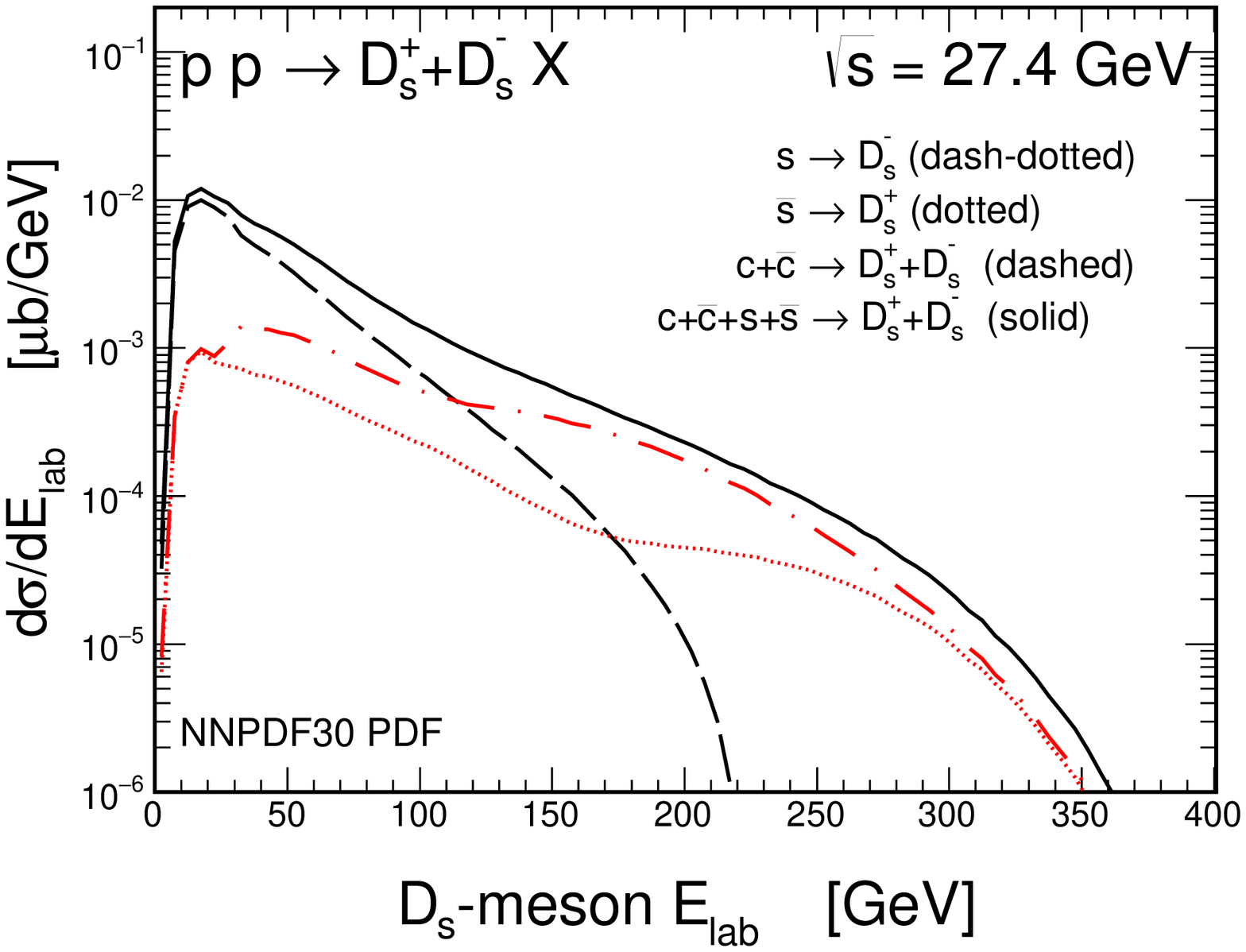}}
\end{minipage}
\caption{
\small Energy distributions of $D_{s}$ mesons in the laboratory frame for the MMHT2014 (left) and the NNPDF30 (right) sets of collinear PDFs. Contributions from charm and strange quark fragmentation are shown separately. Details are specified in the figure.
}
 \label{fig:mesons_Elab}
\end{figure}

Finally, in Fig.~\ref{fig:mesons_Elab}, we show how these PDF uncertainties discussed above 
affect predictions for the energy distribution of $D_{s}$ mesons in the laboratory frame. Here we show separately the leading $c + \bar c \to D_{s}^{+} + D_{s}^{-}$ (dashed lines) and two subleading $s\to D_{s}^{-}$ (dash-dotted lines) and $s\to D_{s}^{-}$ (dotted lines) contributions as well as their sum $c + \bar c + s + \bar s \to D_{s}^{+} + D_{s}^{-}$ (solid lines). The left and right panels correspond to the MMHT2014 and the NNPDF30 PDFs, respectively.
Again a pretty much different results are obtained for the two different PDF sets, especially for large meson energies. Depending on the collinear PDFs used
our model leads to a rather small (the MMHT2014 PDF) or a
fairly significant (the NNPDF30 PDF) contribution to the $D_{s}$ meson production at large energies which comes from the $s/\bar s$-quark fragmentation.   

Summarizing this part we see big uncertainties in our predictions for the production of $D_{s}$ mesons at the low $\sqrt{s}=27.4$ GeV energy.
A future measurement of $D_{s}$ mesons at low energies would definitely help to better understand underlying mechanism and in the consequence 
improve predictions for $\nu_{\tau}/\overline{\nu}_{\tau}$ production
for the SHiP experiment.

\subsection{Direct decay of $\bm{D_{s}^{\pm}}$ mesons}

The considered here decay channels: $D_s^+ \to \tau^+ \nu_{\tau}$ and $D_s^- \to \tau^- {\overline \nu}_{\tau}$, which are the sources of the direct neutrinos,
are analogous to the standard text book cases of $\pi^+ \to \mu^+
\nu_{\mu}$ and $\pi^- \to \mu^- {\overline \nu}_{\mu}$ decays, discussed in detail in the past (see e.g. Ref~\cite{Renton}). The same formalism used for the pion decay applies also to the $D_s$ meson decays.
Since pion has spin zero it decays isotropically in its rest frame. However, the produced muons are polarized in its direction of motion
which is due to the structure of weak interaction in the Standard Model. The same is true for $D_s^{\pm}$ decays and polarization of
$\tau^{\pm}$ leptons.

\begin{figure}[!htbp]
\begin{minipage}{0.47\textwidth}
 \centerline{\includegraphics[width=1.0\textwidth]{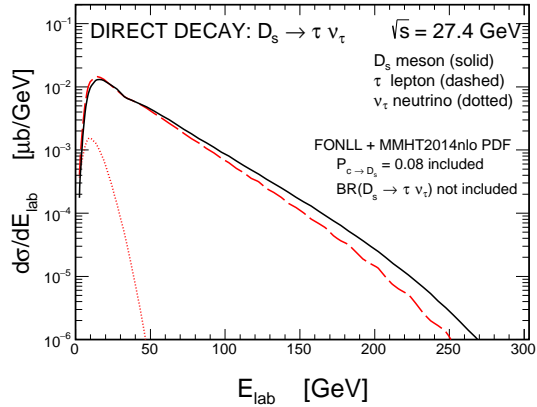}}
\end{minipage}
\hspace{0.5cm}
\caption{
\small Laboratory energy distributions of $D_{s}$ mesons (solid), $\tau$ leptons
(dashed) and $\nu_{\tau}$ neutrinos (dotted) from the direct decay
$D_{s} \to \tau \nu_{\tau}$. Here we show only
the leading contribution to $D_{s}$ meson production from charm quarks
calculated with the~\textsc{Fonll} code. The decay branching fraction is
not included here for easier comparison. 
}
 \label{fig:direct_Elab}
\end{figure}

Therefore the $\tau$ decay must be carefully considered.
In such decays the $\tau$ particles are strongly polarized with
$P_{\tau^+} = -P_{\tau^-}$.
In the following we assume that in the rest frame of $D_s$ meson:
\begin{center}
$P_{\tau^-}$ = 1 and $P_{\tau^+}$ = -1 \; .
\end{center}
This is also very good approximation in the rest frame of $\tau^{\pm}$.

To calculate cross section for $\nu_{\tau}/{\overline \nu}_{\tau}$ production
the $D_s^{\pm} \to \tau^{\pm} \nu_{\tau}/{\overline \nu}_{\tau}$ branching
fraction must be included.
The decay branching fraction is rather well known: 
BR$(D_s^{\pm} \to \tau^{\pm} \nu_{\tau}/{\overline \nu}_{\tau})$ = 0.0548$\pm$
0.0023 \cite{PDG}.

In Fig.~\ref{fig:direct_Elab} we show laboratory frame energy distribution of $D_s$
meson (solid line) and $\tau$ lepton (dashed line) and $\nu_{\tau}$
neutrino (dotted line) from the direct decay. It can be clearly seen
that the $\tau$ lepton takes almost whole energy of the mother $D_s$ meson.

\subsection{Neutrinos from chain decay of $\tau$ leptons}

The $\tau$ decays are rather complicated due to having many possible decay 
channels~\cite{PDG}. Nevertheless, all confirmed decays lead to production of 
$\nu_{\tau}$ (${\overline \nu}_{\tau}$). This means total amount of neutrinos/antineutrinos produced from $D_s$ decays into $\tau$ lepton 
is equal to the amount of antineutrinos/neutrinos produced in subsequent $\tau$ decay.
But, their energy distributions will be different due to $D_s$ 
production asymmetry in the case of the subleading fragmentation mechanism.

The purely leptonic channels (three-body decays), analogous to the
$\mu^{\pm} \to e^{\pm} ({\overline \nu}_{\mu}/\nu_{\mu})( \nu_e / {\overline \nu}_e)$
decay (discussed e.g. in Refs.~\cite{Renton,G1990}) cover only about 35\% of all 
$\tau$ lepton decays. Remaining 65\% are semi-leptonic decays. 
They differ quite drastically from each other and each gives 
slightly different energy distribution for $\nu_{\tau}$ (${\overline \nu}_{\tau}$).
In our model for the decay of $D_s$ mesons there
is almost full polarization of $\tau$ particles with respect to the direction of their motion.

Since $P_{\tau^+} = -P_{\tau^-}$ (see the previous subsection) and the angular distributions of polarized $\tau^{\pm}$ are
antisymmetric with respect to the spin axis the resulting distributions of $\nu_{\tau}$ and ${\overline \nu}_{\tau}$
from decays of $D_s^{\pm}$ are then identical, consistent with CP symmetry (see e.g. Ref.~\cite{Barr1988}).

The mass of the $\tau$ lepton ($1.777$ GeV) is very similar as the mass of the $D_s$
meson ($1.968$ GeV). Therefore, direct neutrino takes away only a small fraction of
energy/momentum of the mother $D_s$.
In this approximation: 
\begin{equation}
\vec{v}_{\tau} = \vec{v}_{D_s} \; , \;\;\; 
\vec{p}_{\tau} = \vec{p}_{D_s} 
\end{equation}
polarization of $\tau$ in its rest frame is 100 \%.
In reality polarization of $\tau^{\pm}$ is somewhat smaller.
In the approximate $Z$-moment method often used for production of
neutrinos/antineutrinos in the atmosphere
discussed e.g. in Ref.~\cite{G1990} the polarization is a function of
$E_{\tau}/E_{D_s}$ (see also Ref.~\cite{Pasquali:1998xf}).

Before we go to distribution of neutrinos/antineutrinos in the laboratory
system (fixed target $p +^{96}\!\mathrm{Mo}$ collisions) we shall present 
distributions of neutrinos/antineutrinos in the $\tau^{\pm}$ 
center-of-mass system, separately for different decay channels of
$\tau$.
In this calculation we use \textsc{Tauola} code \cite{TAUOLA}.

\begin{figure}[!h]
\begin{minipage}{0.47\textwidth}
  \centerline{\includegraphics[width=1.0\textwidth]{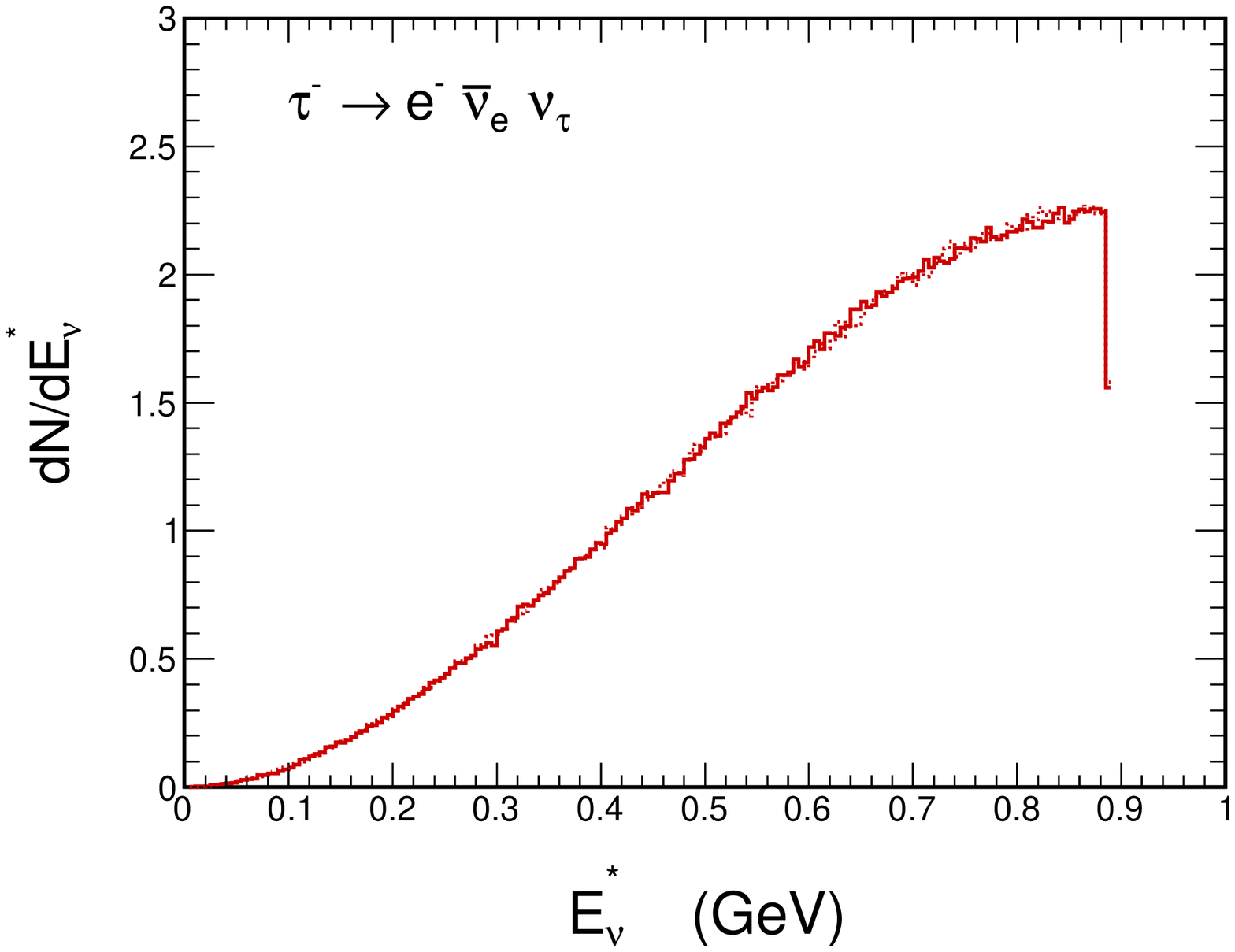}}
\end{minipage}
\begin{minipage}{0.47\textwidth}
  \centerline{\includegraphics[width=1.0\textwidth]{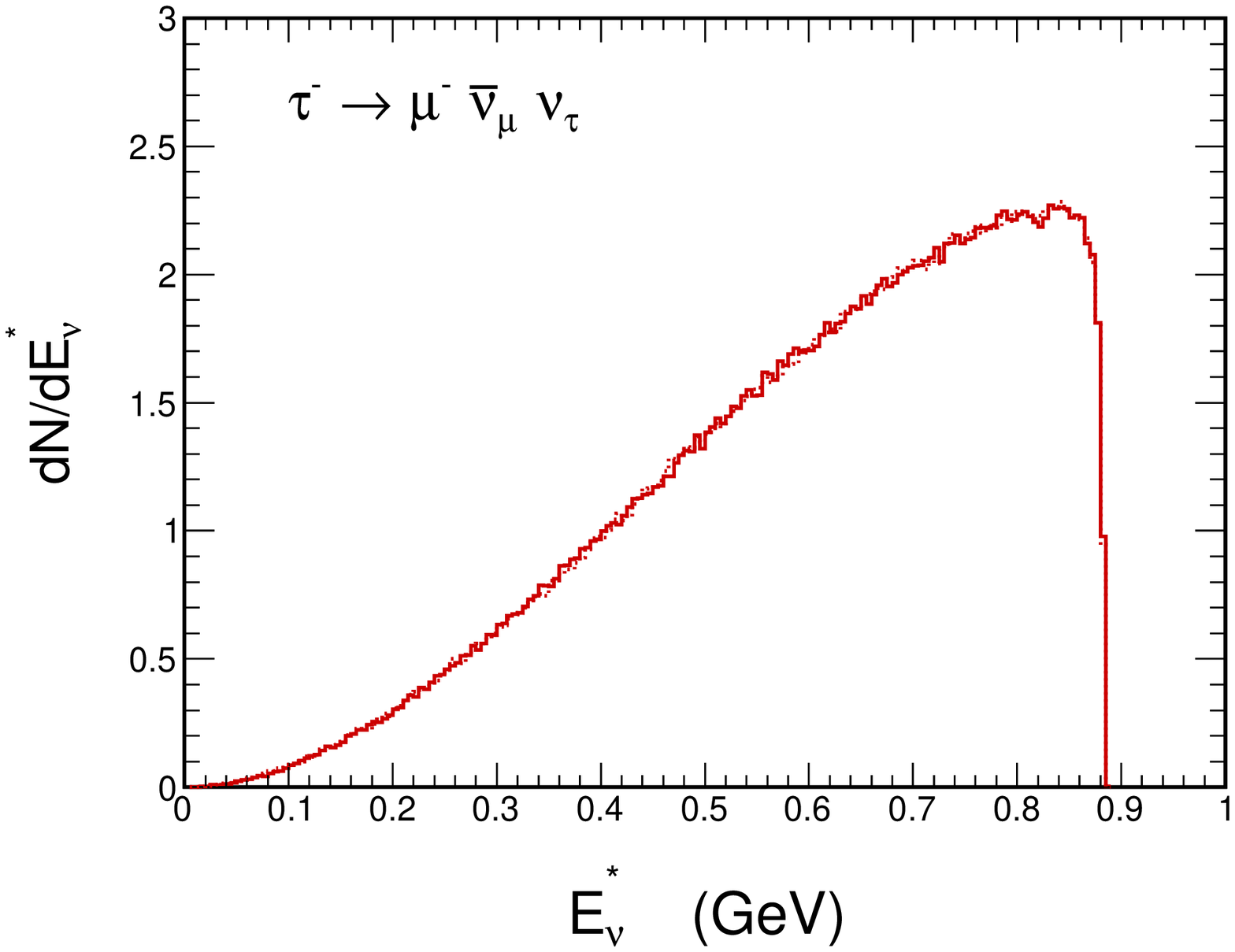}}
\end{minipage}\\
\begin{minipage}{0.47\textwidth}
  \centerline{\includegraphics[width=1.0\textwidth]{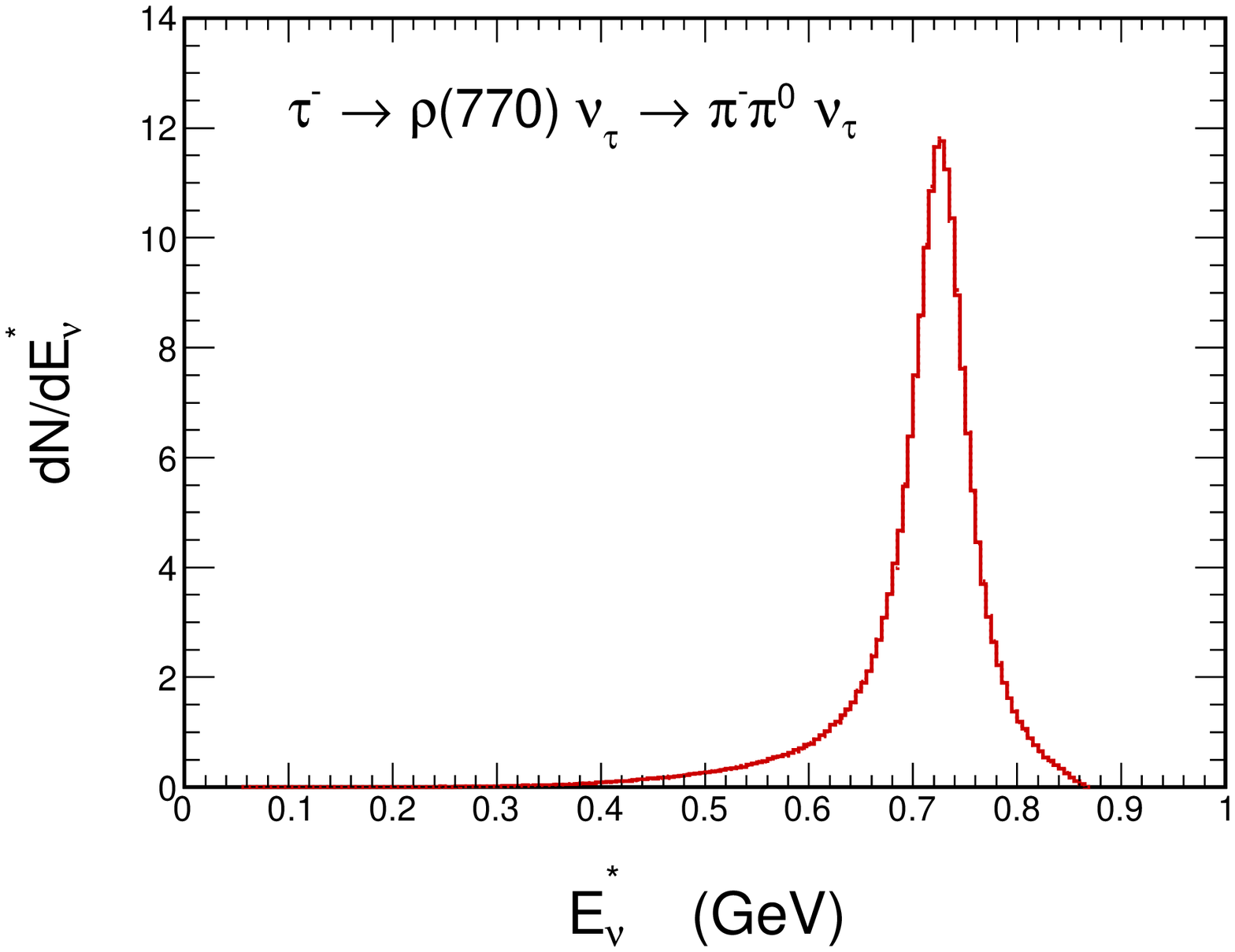}}
\end{minipage}
\begin{minipage}{0.47\textwidth}
  \centerline{\includegraphics[width=1.0\textwidth]{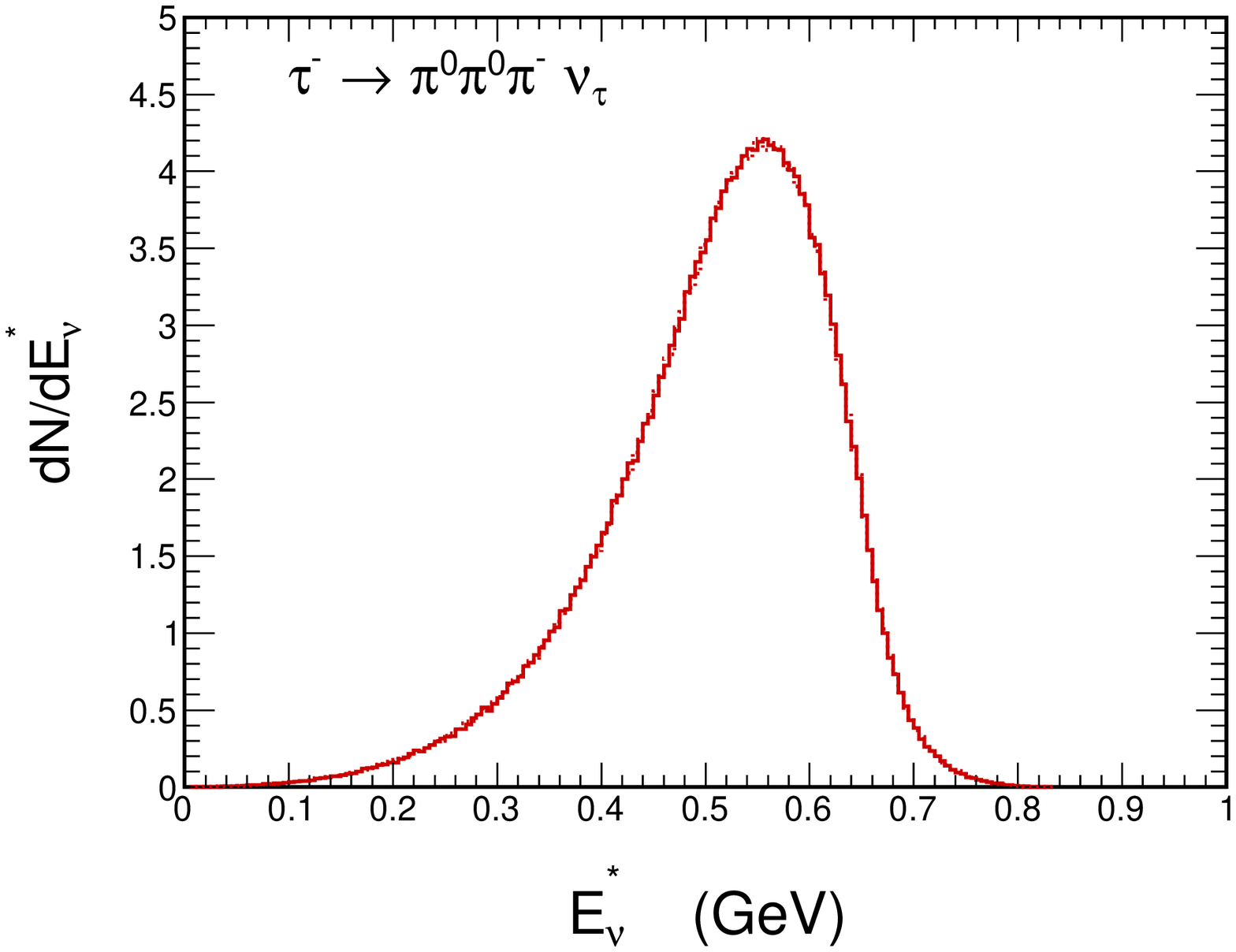}}
\end{minipage}
  \caption{
\small Distributions in energy of $\nu_{\tau}$
in the $\tau^-$ center-of-mass system, for selected decay channels. The counterpart distributions for $\overline{\nu}_{\tau}$ from the decay of $\tau^+$ are identical.
Note that, top plots, which are quite similar, cover about 35\% of all tau decays.
Therefore, dominant contribution comes from semi-leptonic decays, 
which lead to rather different distributions.
}
\label{fig:dN_dE*_channels}
\end{figure}

\begin{figure}[!h]
\begin{minipage}{0.47\textwidth}
  \centerline{\includegraphics[width=1.0\textwidth]{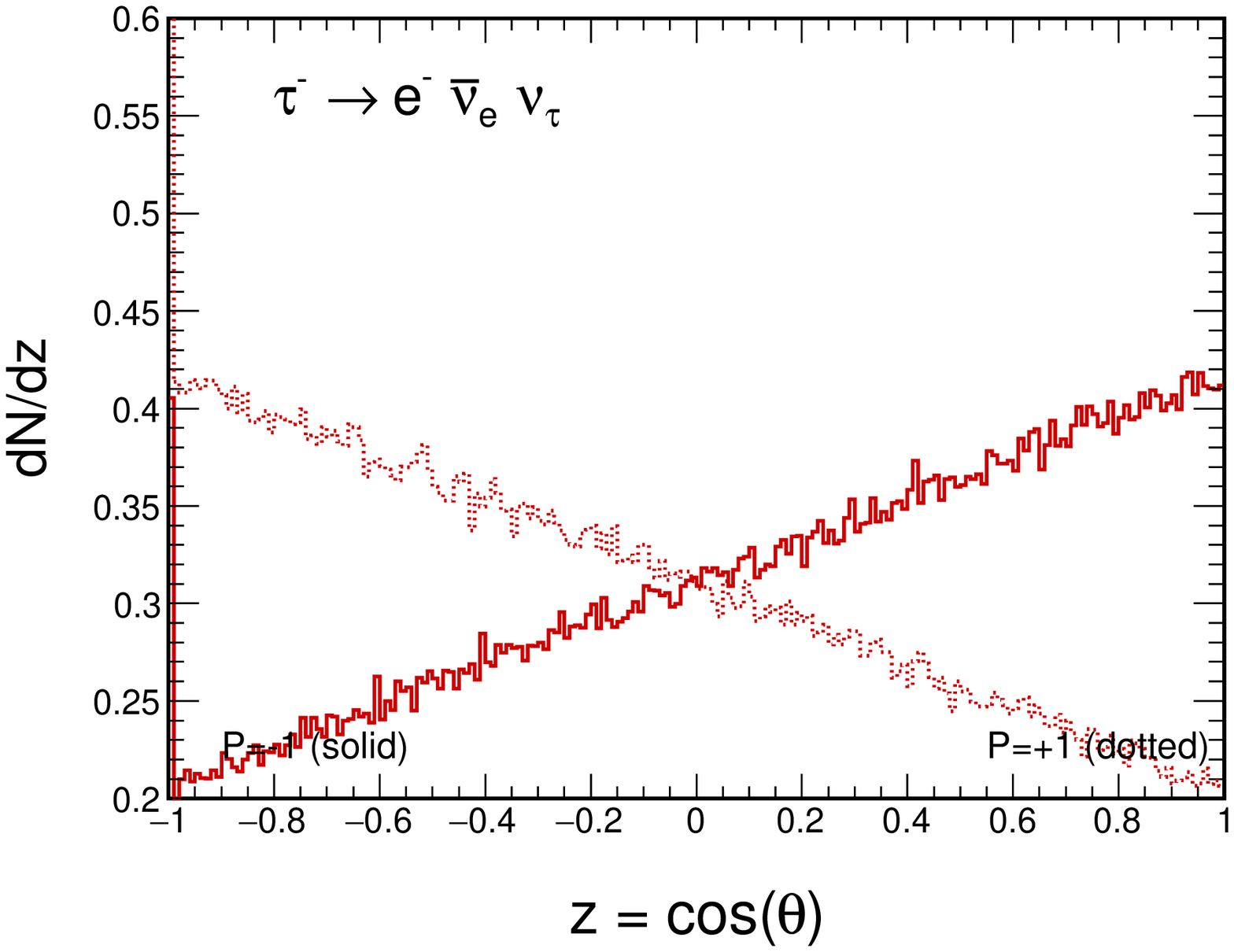}}
\end{minipage}
\begin{minipage}{0.47\textwidth}
  \centerline{\includegraphics[width=1.0\textwidth]{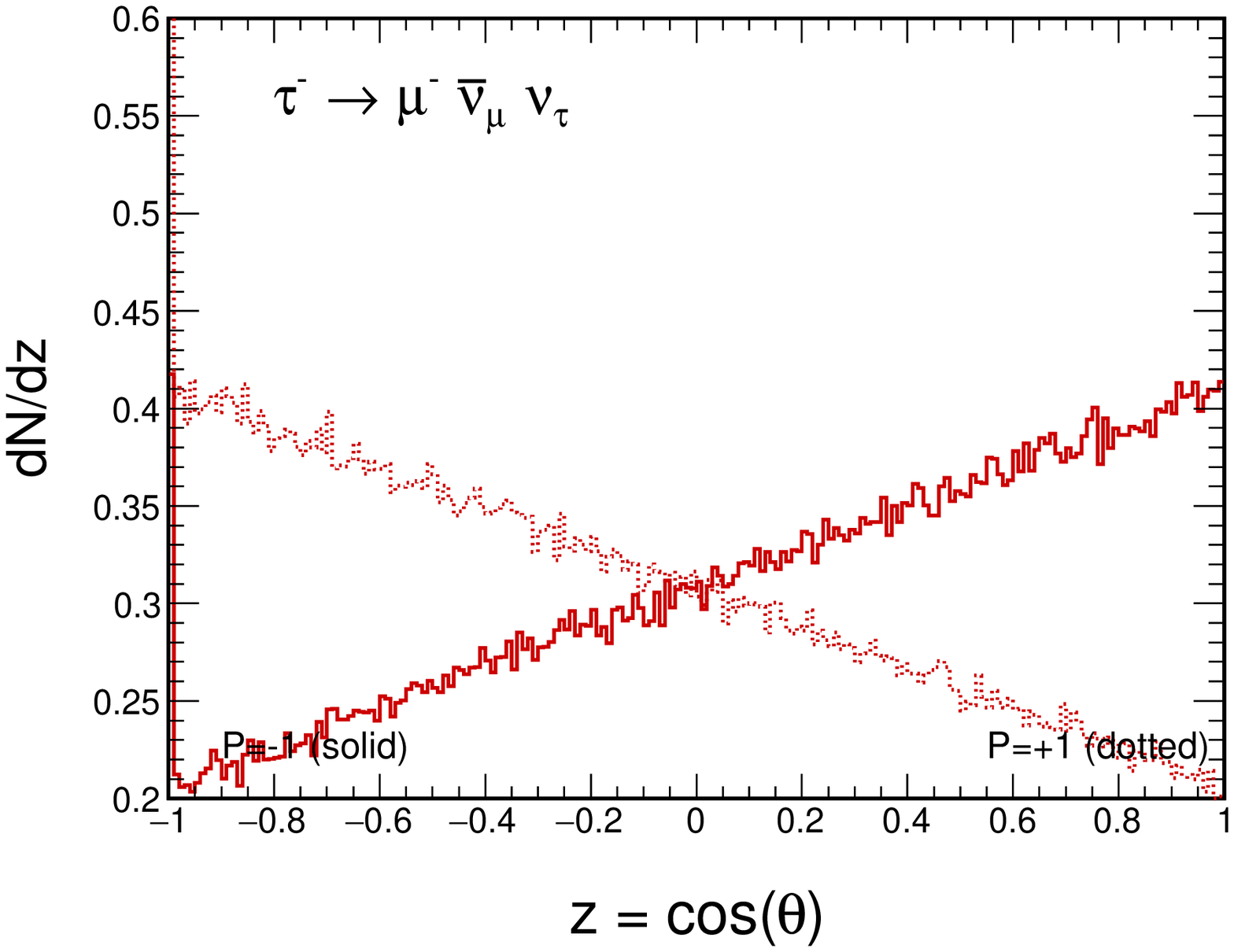}}
\end{minipage}\\
\begin{minipage}{0.47\textwidth}
  \centerline{\includegraphics[width=1.0\textwidth]{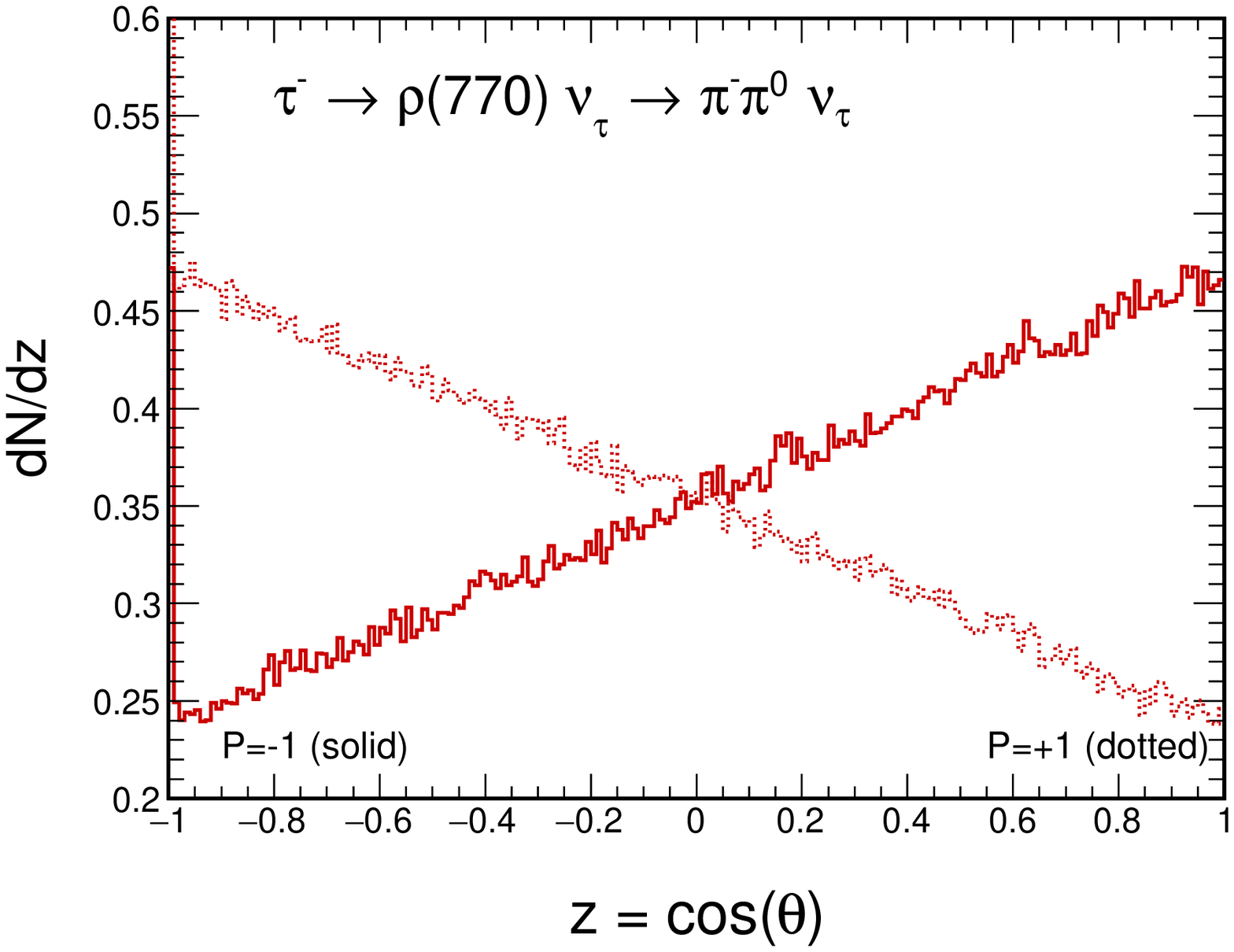}}
\end{minipage}
\begin{minipage}{0.47\textwidth}
  \centerline{\includegraphics[width=1.0\textwidth]{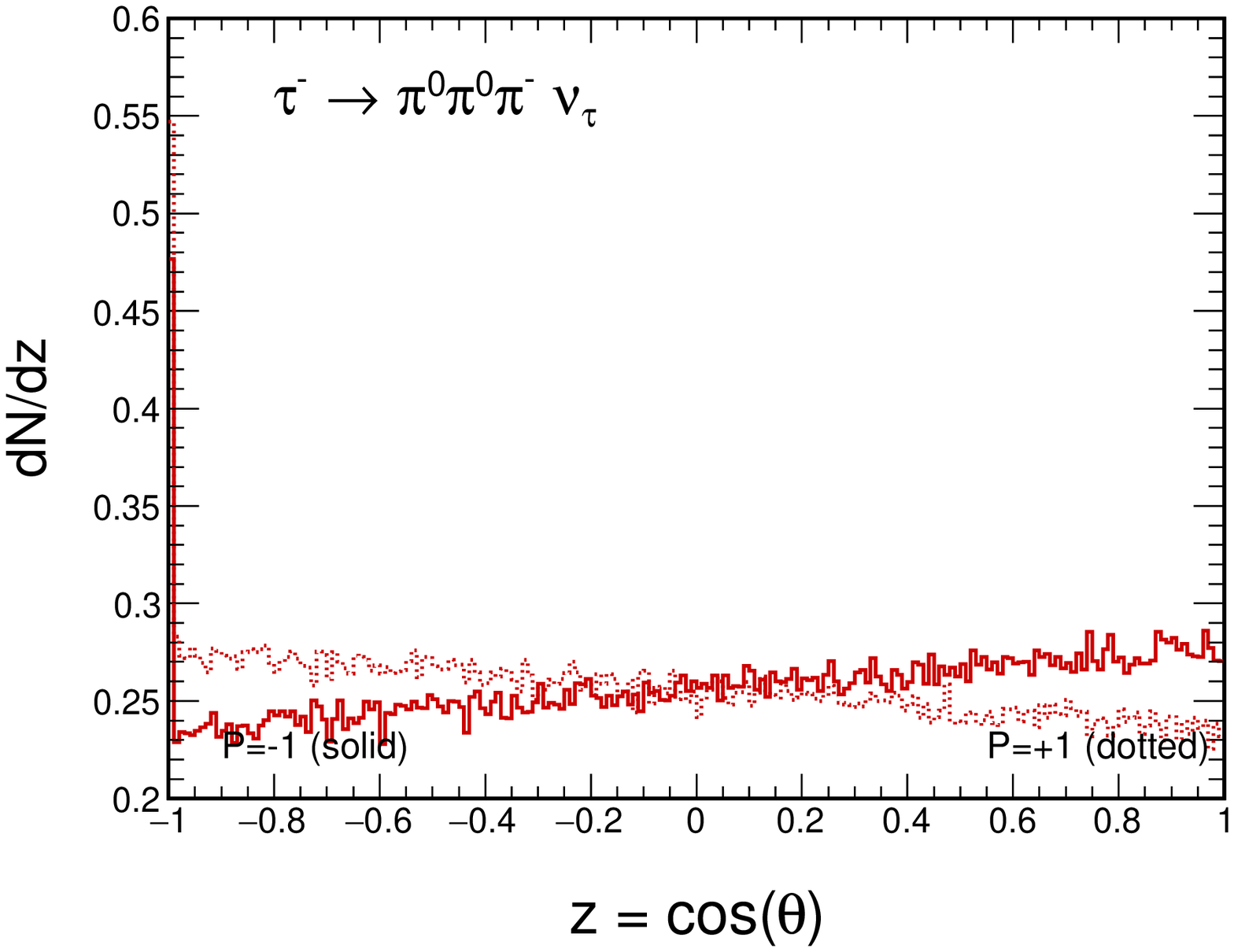}}
\end{minipage}
  \caption{
\small Distributions in $z^{*} = cos(\theta^*)$ of $\nu_{\tau}$ 
in the $\tau^{-}$ center-of-mass system for 
two polarizations of $\tau^{-}$ particles ($P = \pm$ 1). For charged
conjugate channels with $\overline{\nu}_{\tau}$ the distributions must be symmetrically inverted $P(\tau^{-}) =\pm 1 \to P(\tau^{+}) =\mp 1$.
}
\label{fig:dN_dz*_channels}
\end{figure}

In Fig.~\ref{fig:dN_dE*_channels} we show distribution in energy $E^*$ of
neutrinos in the $\tau$ center-of-mass system, for selected decay channels.
Quite different
distributions are obtained for different decay channels.
In Fig.~\ref{fig:dN_dz*_channels} we show distributions in 
$z^{*} = cos(\theta^*)$ of the neutrinos with respect to
$\tau$ spin direction, again separately for different decay channels. 
The distribution functions are linear in $z^*$ which could simplify calculations.

\subsection{$\bm{p +^{96}\!\mathrm{Mo}}$ collisions}

The differential cross section $d \sigma / d y dp_t$ for $D_s^{\pm}$ 
production in $p +^{96}\!\mathrm{Mo}$ collision is assumed to
approximately scale like
\begin{equation}
\frac{d \sigma_{p +\mathrm{Mo}}}{d y d p_t} = 
Z_{\mathrm{Mo}} \frac{d \sigma_{ p p}}{d y d p_t} +
(A_{\mathrm{Mo}} - Z_{\mathrm{Mo}})  \frac{d \sigma_{ p n}}{d y d p_t} \; .
\label{pMo_cs}
\end{equation}
It was shown in \cite{ALICE_RpA} that at much higher energies 
($\sqrt{s}_{NN}$ = 5.02 GeV) 
the nuclear modification factor for $D$ meson production in 
$p + \mathrm{Pb}$ is close to 1 in a broad range of rapidity and transverse momentum.
This is an approximation which is not easy to improve in a realistic
way. Therefore it is difficult at present to set uncertainties of such 
an approximation. In our calculation collision energy was fixed $\sqrt{s}_{NN}$ = 27.4 GeV.
For more precise calculation one has to consider calculating a cascade
which was attempted e.g. in \cite{SHiP_cascade}.

\subsection{Neutrino/antineutrino interactions
with the $\bm{\mathrm{Pb}}$ target}

How many neutrinos/antineutrinos will be observed
in the SHiP experiment depends on
the cross section for neutrino/antineutrino scattering of nuclei off 
the target.
In the case of the SHiP experiment a dedicated lead target was proposed.
At not too small energies ($\sqrt{s_{NN}} > 5$ GeV), the cross section for 
$\nu_{\tau} Pb$ and $\overline{\nu}_{\tau} Pb$ interactions
can be obtained from elementary cross sections as:
\begin{eqnarray}
\sigma(\nu_{\tau} Pb) &=& Z \sigma(\nu_{\tau} p) + (A-Z) 
\sigma(\nu_{\tau} n) \; ,  \\
\sigma(\overline{\nu}_{\tau} Pb) &=&
Z \sigma(\overline{\nu}_{\tau} p) + (A-Z) \sigma(\overline{\nu}_{\tau} n)
\; .
\label{Pb_cs}
\end{eqnarray}
Shadowing effects depend on $x$ variable 
(parton longitudinal momentum fraction),
i.e. on neutrino/antineutrino energy.
At not too high energies (not too small $x$) shadowing effects 
are rather small and can be neglected at present accuracy
having in mind other uncertainties. On the other hand for the $x$-ranges considered here
the antishadowing and/or EMC-effect may appear non-negligible but still rather small and shall not affect the numerical predictions presented here. The nuclear modifications of the PDFs goes beyond the scope of the present study and will be considered elsewhere.

Both elementary as well as nuclear cross sections strongly depend on
neutrino/antineutrino energy \cite{JR2010}. For $\tau$
neutrino/antineutrino interactions there is also an energy threshold
related to the mass of $\tau^{\pm}$ 
which reduces cross section compared to
$\nu_{\mu}/{\overline{\nu}_{\mu}}$ 
and practically cuts off contributions of nucleon resonances. Therefore 
one should include practically only deep-inelastic region.

The probability of interacting of neutrino/antineutrino with the lead
target can be calculated as:
\begin{equation}
P_{\nu_{\tau}/\overline{\nu}_{\tau}}^{\mathrm{target}}(E) = \int_0^{d} n_{\mathrm{cen}} \sigma_{\nu_{\tau} Pb}(E) dz 
= n_{\mathrm{cen}} \sigma_{\nu_{\tau} Pb}(E) d
\; ,
\label{probability_of_interaction}
\end{equation}
where $n_{\mathrm{cen}}$ is a number of scattering centers (lead nuclei) per volume 
element and the target thickness is $d \approx$ 2 m \cite{SHiP3}.
Using the \textsc{NuWro} Monte Carlo generator \cite{NuWro}, we obtain   
$\sigma(E)/E \sim 1.09 \times 10^{-38}$ cm$^2/$GeV for neutrino and
$0.41 \times 10^{-38}$ cm$^2/$GeV for antineutrino
for the $E = 100$ GeV. The number of scattering centers is
\begin{equation}
n_{\mathrm{cen}} = (11.340 / 207.2) N_A \; ,
\end{equation}
where $N_A$ = 6.02 $\times$ 10$^{23}$ is the Avogadro number.

The energy dependent flux of neutrinos can be written as:
\begin{equation}
\Phi_{\nu_{\tau}/\overline{\nu}_{\tau}}(E) =
\frac{N_p}{\sigma_{pA}} d\sigma_{pA \to \nu_{\tau}}(E)/dE \; ,
\label{neutrino_flux}
\end{equation}
where $N_p$ is integrated number of beam protons ($N_p = 2 \times 10^{20}$
according to the current SHiP project).
The $\sigma_{pA}$ in Eq.~(\ref{neutrino_flux}) is a crucial quantity which 
requires a short disscusion.
In Ref.~\cite{BR2018} it was taken as
$\sigma_{pA} = A \cdot \sigma_{pN}$ where $\sigma_{pN}$ = 10.7 was used. 
We do not know the origin of this number.
Naively $\sigma_{pN}$ should be the inelastic $pN$ cross section.

The formula above can be used to estimate number of
neutrinos/antineutrinos produced at the beam dump. For the decays of $D_{s}$ meson produced from charm quark fragmentation it reads:
\begin{equation}
N_{\nu_{\tau}} = \frac{N_p}{\sigma_{pA}} \sigma_{pA \to \nu_{\tau}X}
= \frac{N_p}{\sigma_{pN}} \sigma_{p p \to c \bar c} \;  
\mathrm{BR}(D_s \to \tau) \; \mathrm{P}(c \to D_s) \; .
\label{number_of_neutrinos}
\end{equation}
Taking $\mathrm{P}(c \to D_s)$ = 0.08, $\mathrm{BR}(D_s \to \tau)$ = 0.0548, 
$\sigma_{p p \to c \bar c X}$ = 10 $\mu$b and $\sigma_{pN}$ = 20 mb we get
$N_{\nu_{\tau}} = 0.66 \times 10^{15}$. Already this number is rather
uncertain mostly due to the choice of $\sigma_{pA}$ and $p p \to c \bar
c$ cross section. This is almost an order of magnitude lower than 
the corresponding number(s) in Table 2 of Ref.~\cite{SHiP3}. The numbers presented there and the reason of the discrepancy are not clear for us.

\begin{figure}[!h]
\begin{minipage}{0.47\textwidth}
  \centerline{\includegraphics[width=1.0\textwidth]{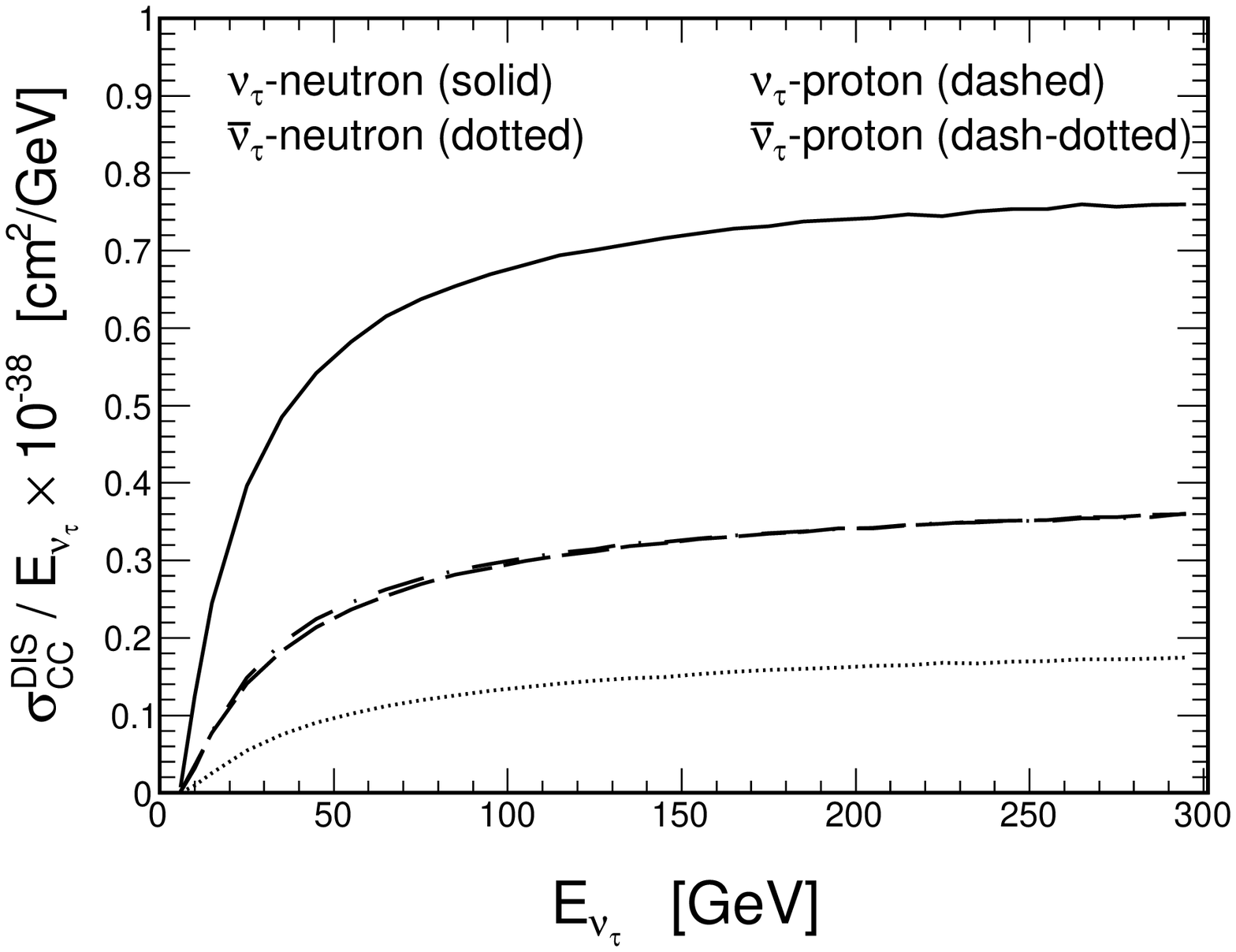}}
\end{minipage}
\begin{minipage}{0.47\textwidth}
  \centerline{\includegraphics[width=1.0\textwidth]{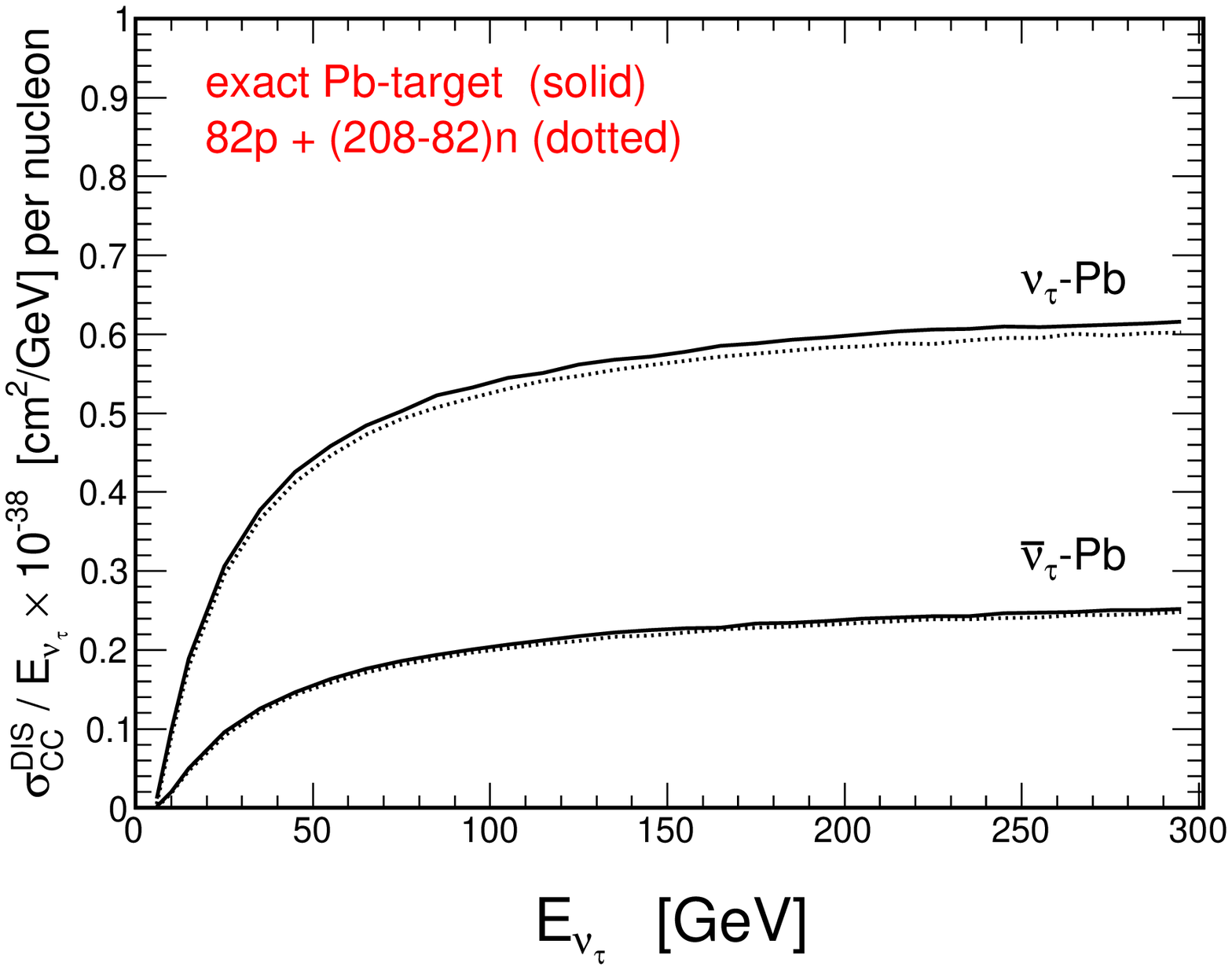}}
\end{minipage}
  \caption{
\small Left: The elementary cross sections $\sigma(\nu_{\tau}p)$, $\sigma(\nu_{\tau}n)$, $\sigma(\overline{\nu}_{\tau}p)$ and $\sigma(\overline{\nu}_{\tau}n)$ as a function of neutrino/antineutrino energy. Right: The $\sigma(\nu_{\tau}\textrm{Pb})$ and $\sigma(\overline{\nu}_{\tau}\textrm{Pb})$ cross sections per nucleon as a function of neutrino/antineutrino energy.
The results are obtained within the \textsc{NuWro} Monte Carlo generator. Details are specified in the figure.
}
\label{fig:nu_N}
\end{figure}

In the present paper the elementary cross sections $\sigma(\nu_{\tau}p)$, $\sigma(\nu_{\tau}n)$, $\sigma(\overline{\nu}_{\tau}p)$ and $\sigma(\overline{\nu}_{\tau}n)$ needed in Eq.(\ref{Pb_cs}) are calculated using the \textsc{NuWro} Monte Carlo generator.
In Fig.~\ref{fig:nu_N} we show the cross section for scattering neutrinos/antineutrinos on the protons and neutrons (left panel) and on the lead target (right panel) as a function of neutrino/antineutrino energy. The cross sections at larger energies are fully dominated by the charged current deep-inelastic scattering interactions (more than 90\% of the cross section for $E_{\nu} \ge 15$ GeV).
In the left panel we observe that all the cross sections strongly depend on neutrino energy.
While for the proton target the cross section for neutrino and antineutrino
is almost the same, for the neutron target they are quite different.
In the right panel we show the cross sections $\sigma(\nu_{\tau}\textrm{Pb})$ and $\sigma(\overline{\nu}_{\tau}\textrm{Pb})$.
Here we take only the dominant isotope $^{208}$Pb. Isotope admixture of 
$^{204}$Pb, $^{206}$Pb, $^{207}$Pb in the target for neutrino/antineutrino
observation makes only small corrections which is of academic value only.
We show separately results obtained by the elementary cross sections using Eq.(\ref{Pb_cs}) (dashed lines) and
those obtained directly for the lead target\footnote{obtained within the local Fermi gas model for the description of the nucleus as a target} including some nuclear effects (solid lines).
For our purpose the difference between the two results is rather marginal.
The cross sections will be used to estimate the number
of neutrino/antineutrino observations.

Finally the number of neutrinos/antineutrinos observed
in the $Pb$ target is calculated from the formula:
\begin{equation}
N_{\nu_{\tau}/\overline{\nu}_{\tau}}^{\mathrm{target}} =
\int dE \Phi_{\nu_{\tau}/\overline{\nu}_{\tau}}(E)
P_{\nu_{\tau}/\overline{\nu}_{\tau}}^{\mathrm{target}}(E) \; .
\label{number_of_observed_neutrinos}
\end{equation}
Here $\Phi_{\nu_{\tau}/\overline{\nu}_{\tau}}(E)$ is
calculated from different approaches to $D_s$ meson production including their subsequent decays
and $P_{\nu_{\tau}/\overline{\nu}_{\tau}}^{\mathrm{target}}(E)$ is obtained
using Eq.(\ref{probability_of_interaction}). The cross sections
for neutrino/antineutrino interactions with the lead target is shown 
in Fig.~\ref{fig:nu_N}.

\section{Numerical predictions for the SHiP experiment}

\subsection{Neutrino/antineutrino differential cross sections for $\bm{p+^{96}\!\mathrm{Mo}}$ at $\bm{\sqrt{s_{NN}}=27.4}$ GeV}

\begin{figure}[!h]
\begin{minipage}{0.47\textwidth}
  \centerline{\includegraphics[width=1.0\textwidth]{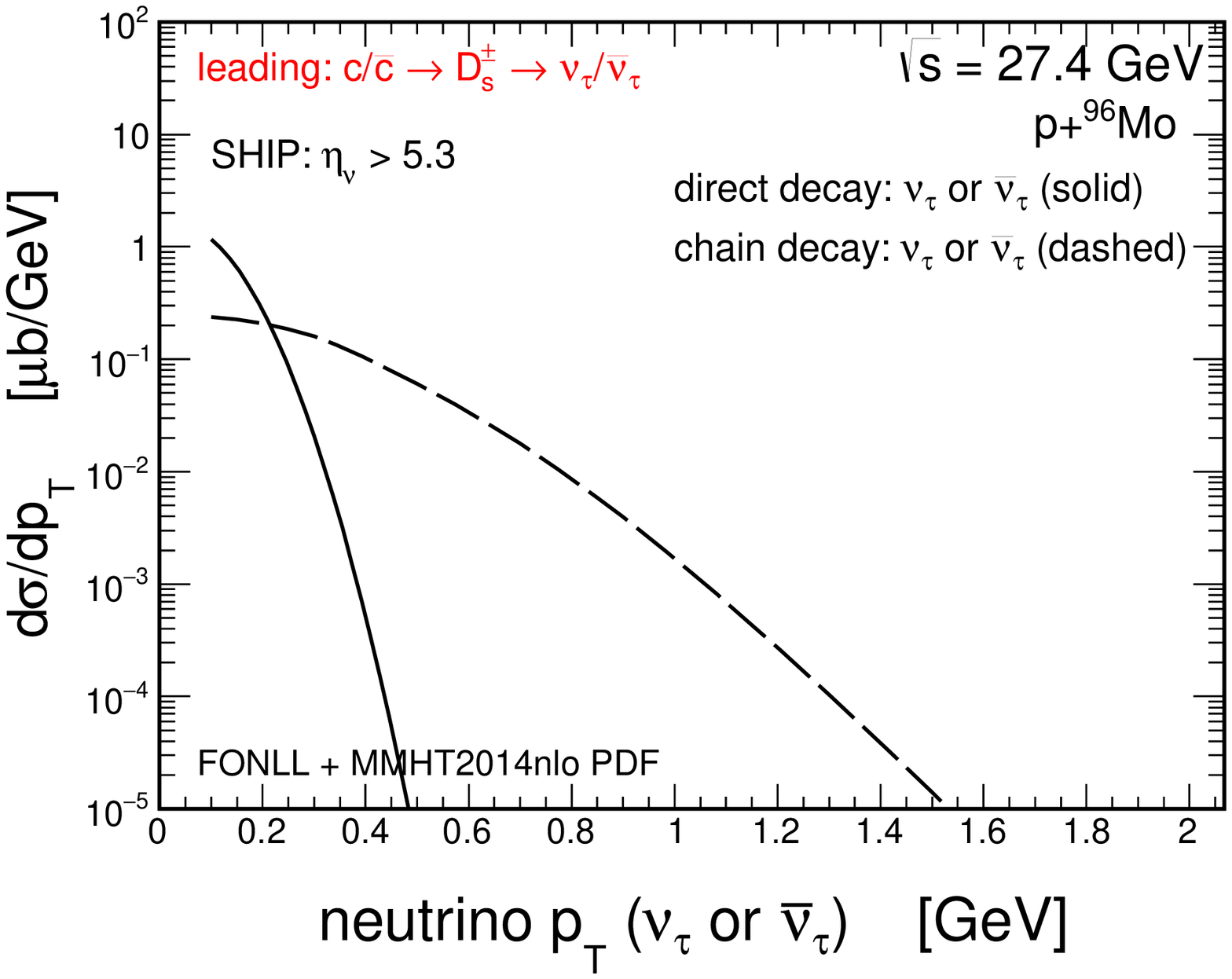}}
\end{minipage}
\begin{minipage}{0.47\textwidth}
  \centerline{\includegraphics[width=1.0\textwidth]{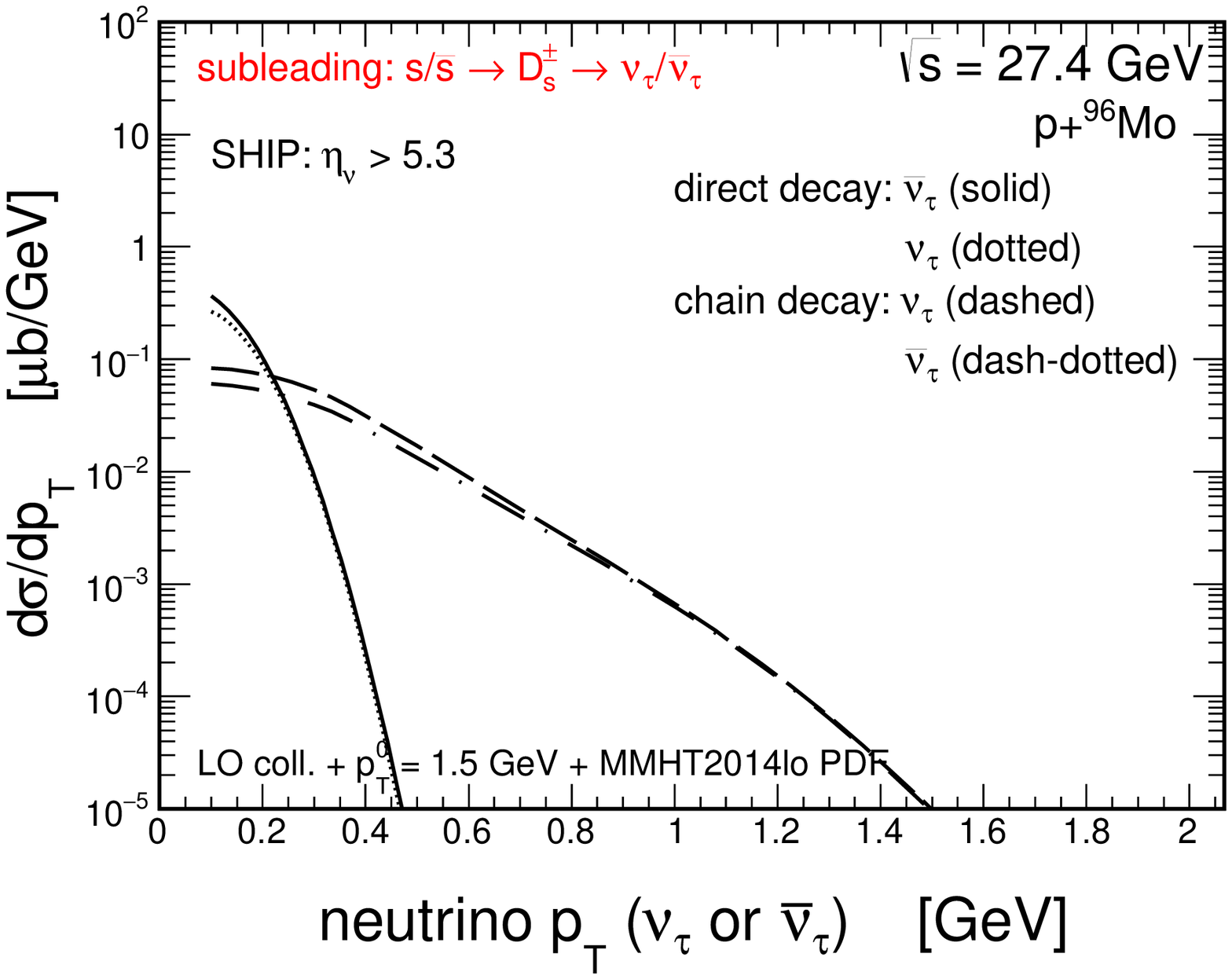}}
\end{minipage}\\
\begin{minipage}{0.47\textwidth}
  \centerline{\includegraphics[width=1.0\textwidth]{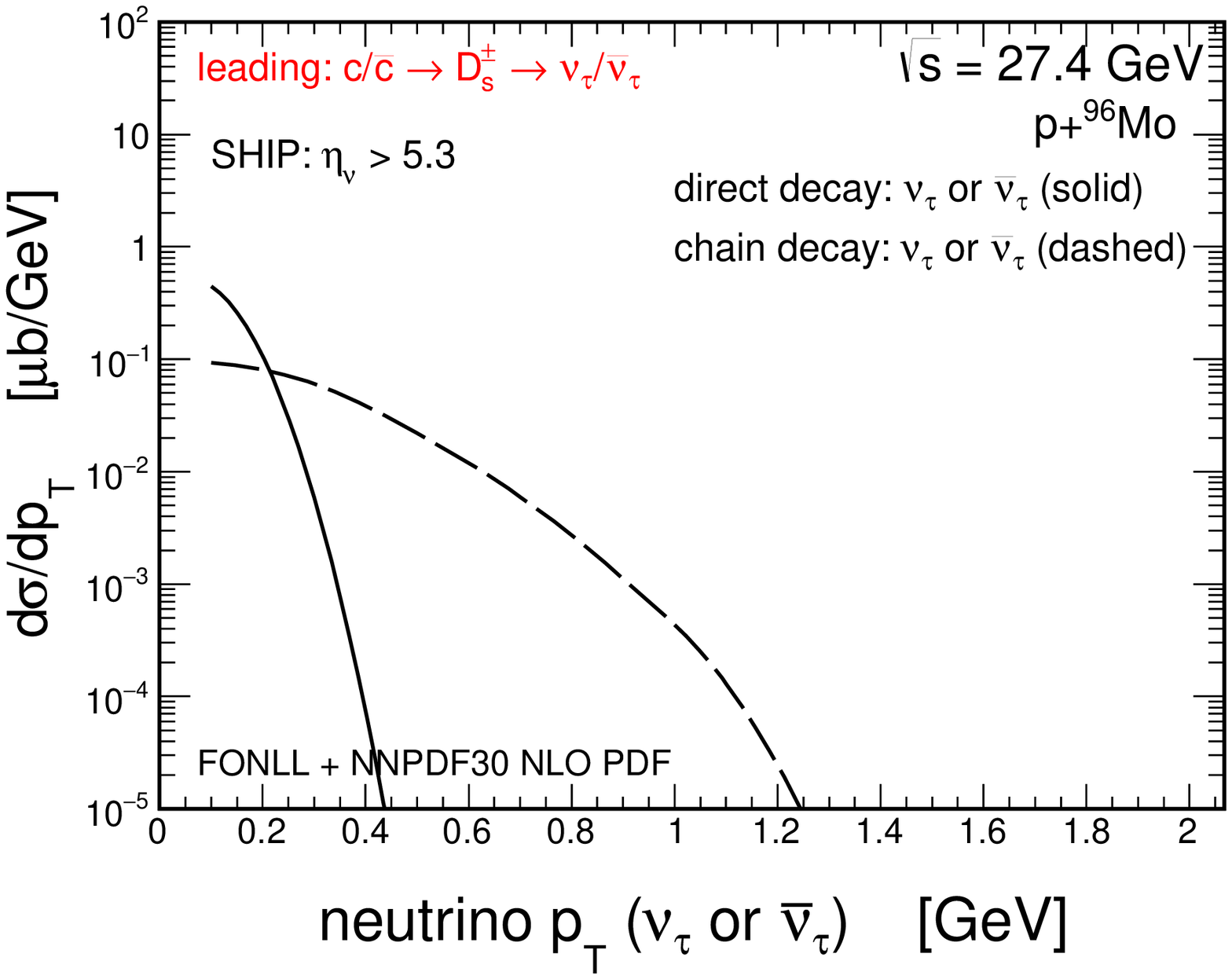}}
\end{minipage}
\begin{minipage}{0.47\textwidth}
  \centerline{\includegraphics[width=1.0\textwidth]{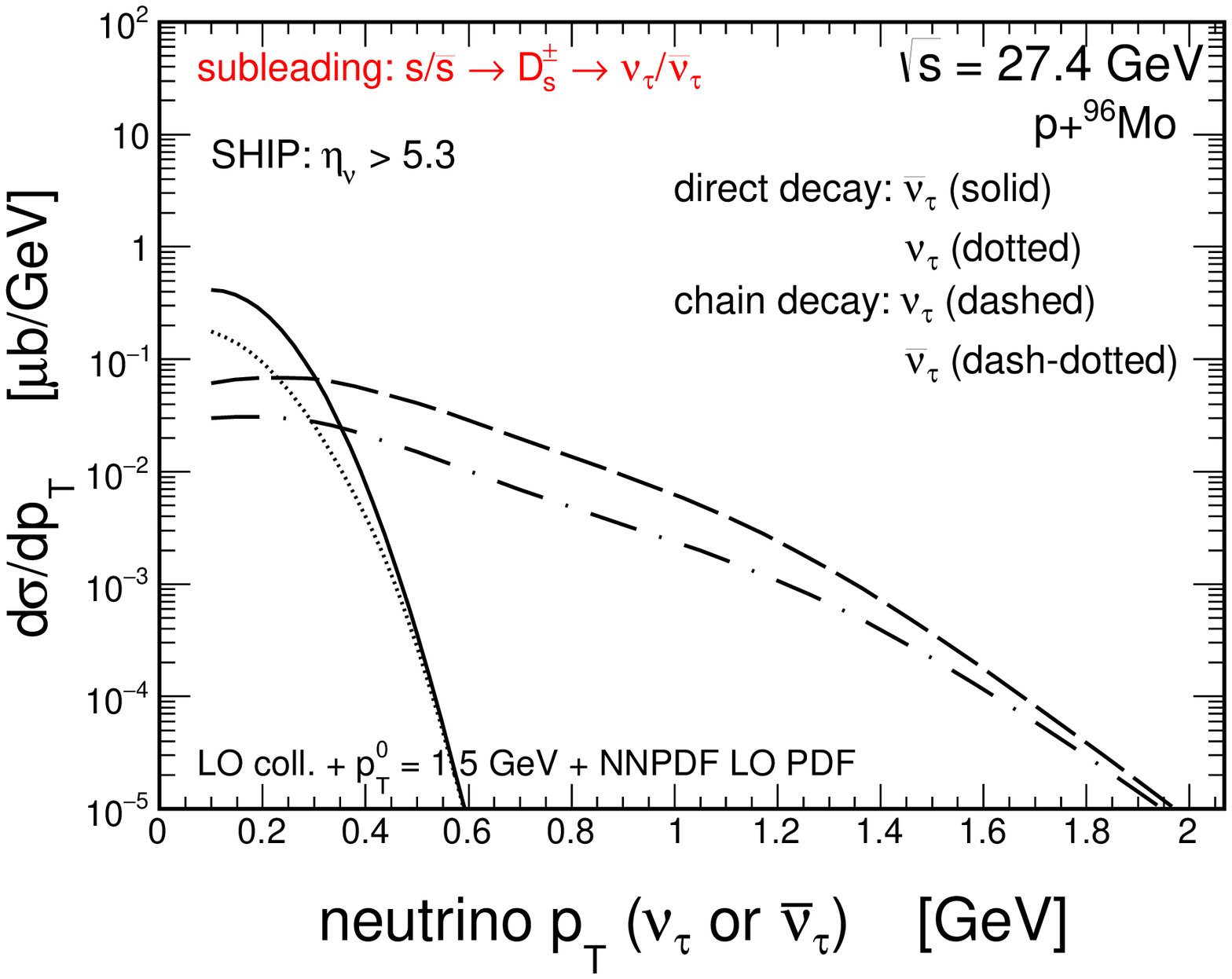}}
\end{minipage}
  \caption{
\small Transverse momentum distributions of $\nu_{\tau}$ (or
$\overline{\nu}_{\tau}$) neutrinos for the MMHT2014 (top) and the NNPDF30
(bottom) sets of collinear PDFs for the leading (left) and subleading
(right) $D_{s}$-meson production mechanisms. Contributions from the direct and chain decay modes are shown separately. Further details are specified in the figure.
}
\label{fig:nu_pT}
\end{figure}

We start presentation of our numerical results with the differential
cross sections for $\nu_{\tau}$ or $\overline{\nu}_{\tau}$ neutrino production.
In Fig.~\ref{fig:nu_pT} we show transverse momentum distributions of neutrinos/antineutrinos for $p+^{96}\!\mathrm{Mo}$ interactions at $\sqrt{s_{NN}}=27.4$ GeV with the $\eta_{\nu} > 5.3$ condition
relevant for the SHiP experiment. The predictions are done for both the leading (left panels) and for the subleading (right panels) $D_{s}$-meson production mechanisms
calculated with the MMHT2014 (top panels) and the NNPDF30 (bottom panels) PDFs. Here, we show separately distributions for $\nu_{\tau}$ and $\overline{\nu}_{\tau}$ 
from the direct and chain decay modes. The direct contributions are
concentrated in the region of extremely small transverse momenta 
while their chain counterparts have significantly longer tails in
$p_{T}$. Even for the chain decays the major parts of the cross sections
come from the region of small transverse momentum ($p_{T} < 2$ GeV) of neutrino. The
contributions for the subleading $s/\bar{s} \to D_{s}^{\pm}$ show a visible production asymmetry
for $\nu_{\tau}$ and $\overline{\nu}_{\tau}$ in contrast to the contributions for the standard leading $c/\bar{c} \to D_{s}^{\pm}$ mechanism. 
The $s(x)\neq \bar{s}(x)$ asymmetry in the parton distributions leads to the
neutrino/antineutrino production asymmetry for both, the direct and the
chain decay 
modes. The obtained cross sections for $s$-quark production are larger
than those for the $\bar{s}$-antiquark. For the direct decay mode this
leads to enhanced production of $\overline{\nu}_{\tau}$ 
antineutrinos with respect to the $\nu_{\tau}$ neutrinos. The effect is opposite in the case of the chain decay mode where $s \to \nu_{\tau}$ and $\bar{s} \to \overline{\nu}_{\tau}$.
The production asymmetry is larger when the NNPDF30 PDFs are used than in the case of the MMHT2014 PDFs. 

\begin{figure}[!h]
\begin{minipage}{0.47\textwidth}
  \centerline{\includegraphics[width=1.0\textwidth]{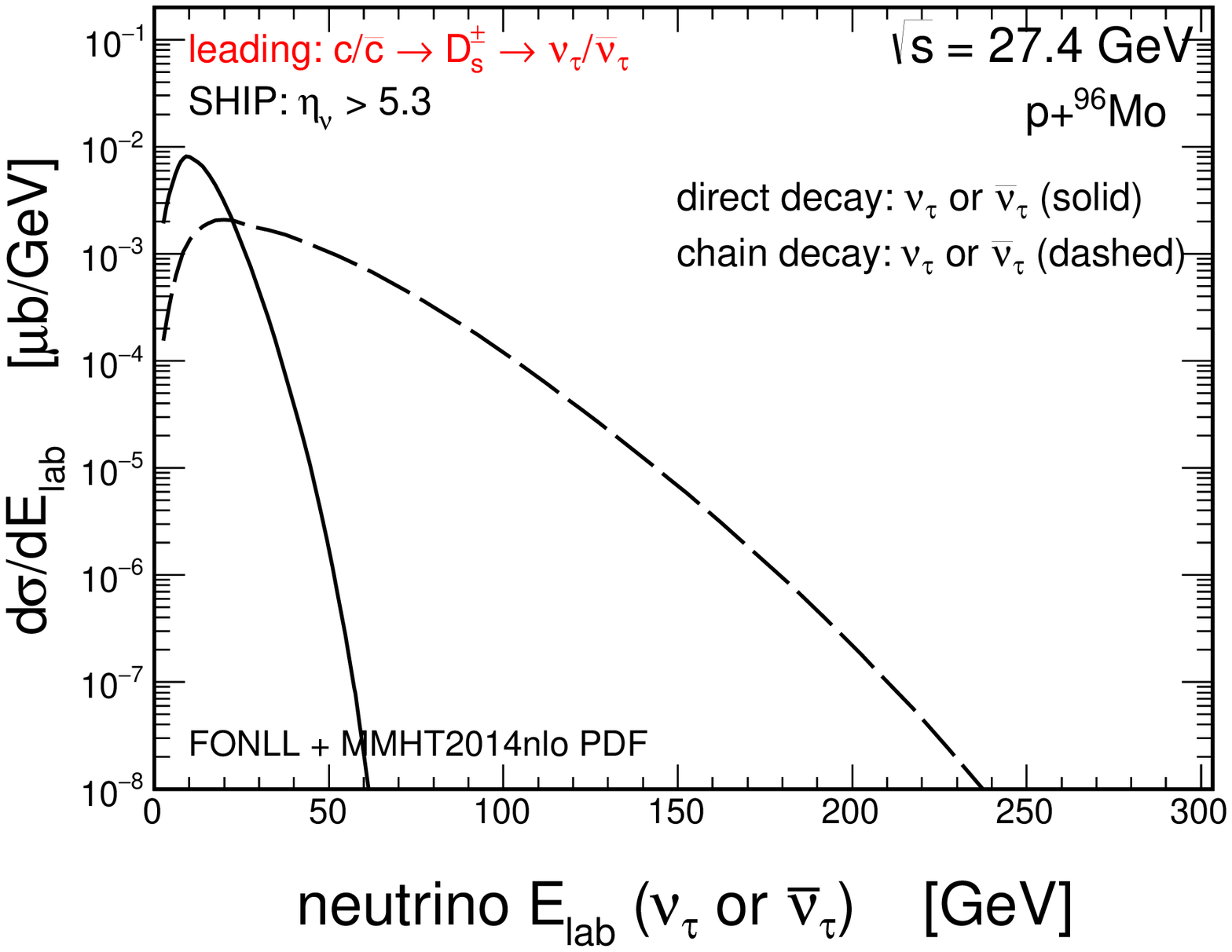}}
\end{minipage}
\begin{minipage}{0.47\textwidth}
  \centerline{\includegraphics[width=1.0\textwidth]{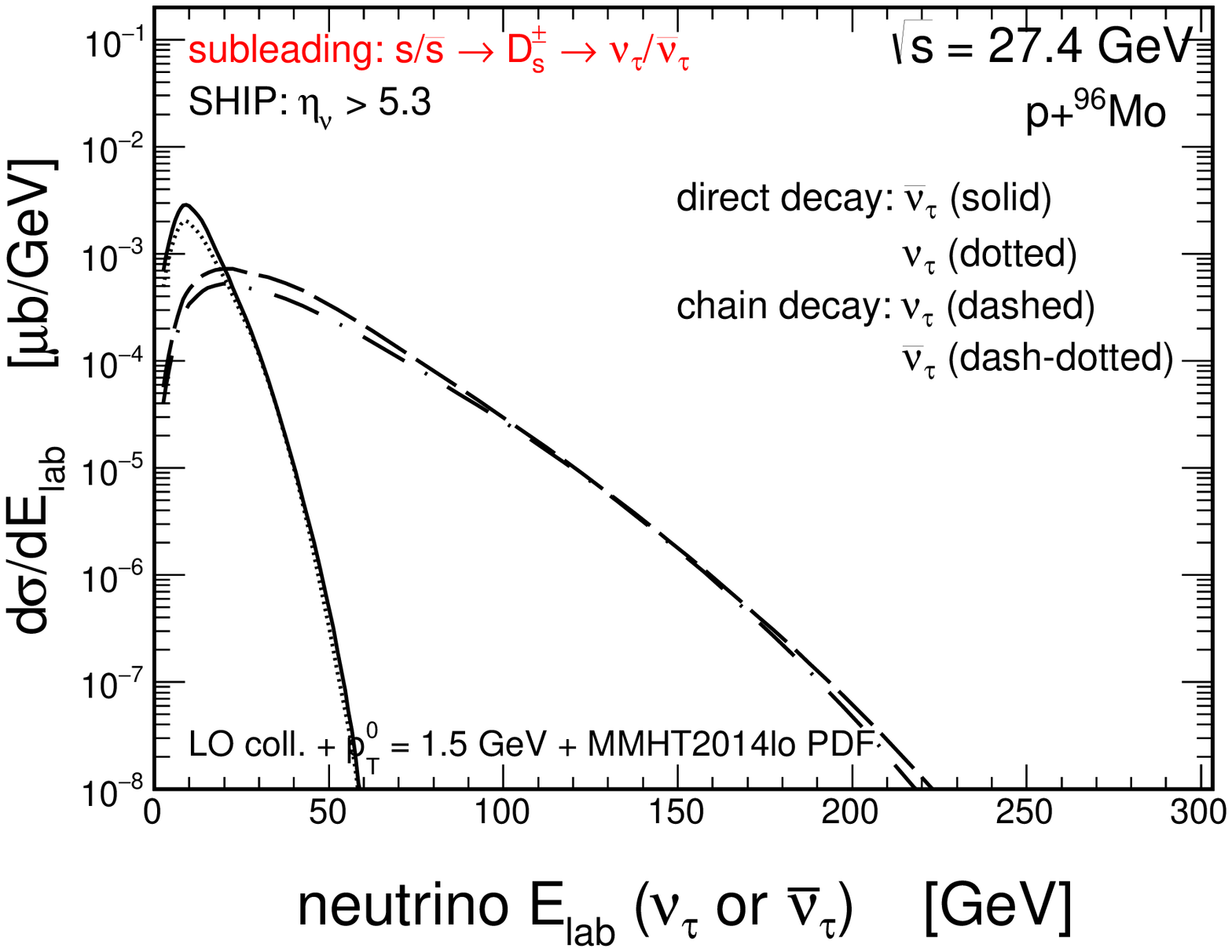}}
\end{minipage}\\
\begin{minipage}{0.47\textwidth}
  \centerline{\includegraphics[width=1.0\textwidth]{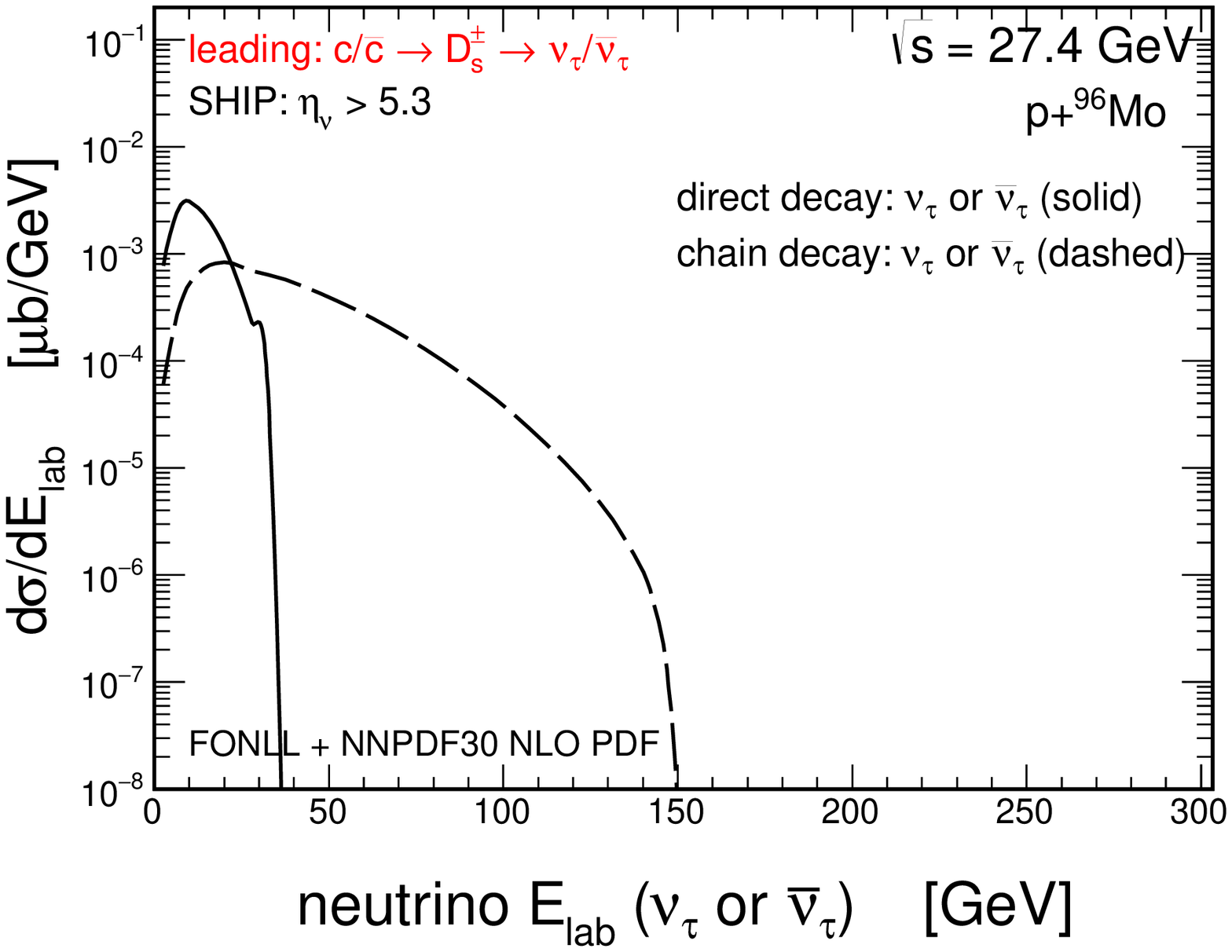}}
\end{minipage}
\begin{minipage}{0.47\textwidth}
  \centerline{\includegraphics[width=1.0\textwidth]{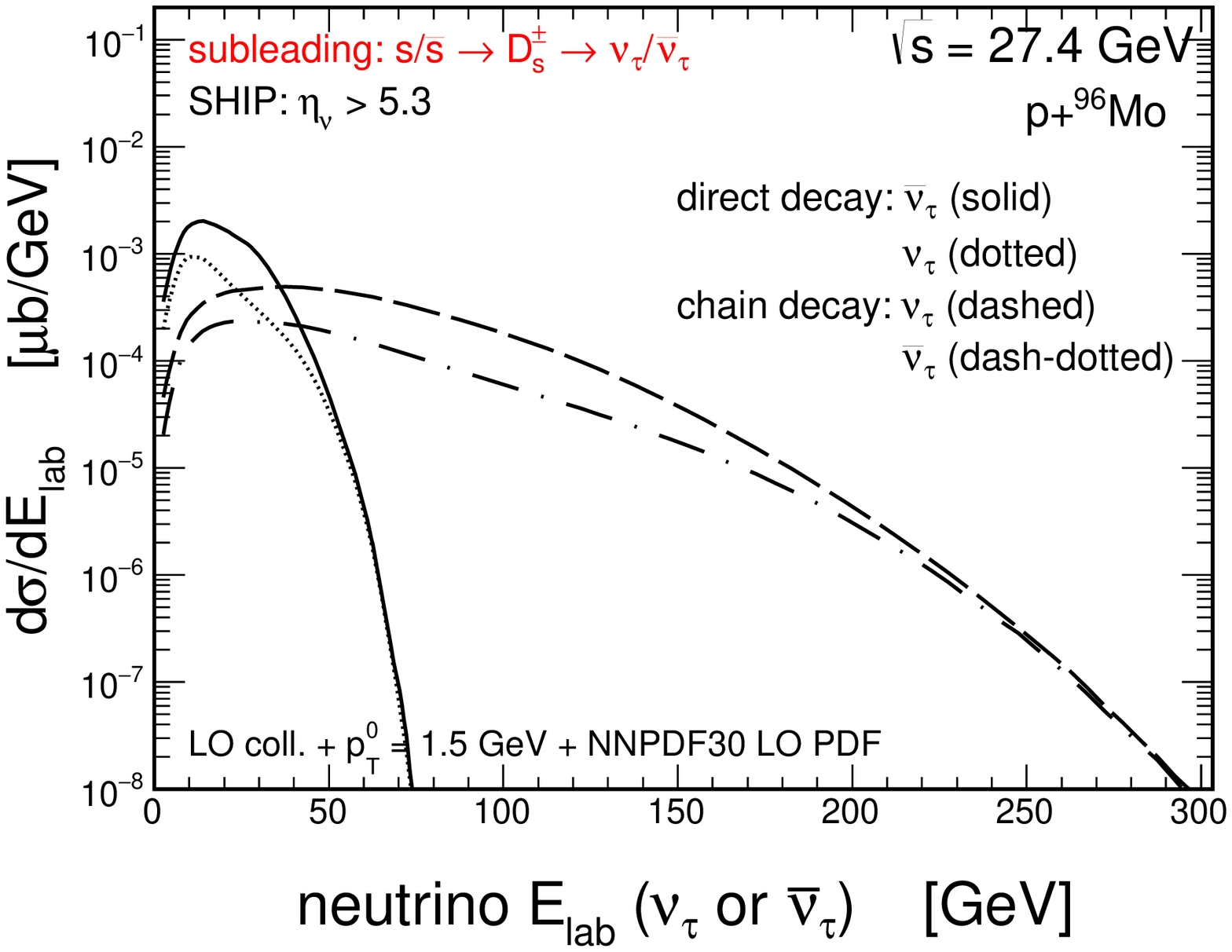}}
\end{minipage}
  \caption{
\small Laboratory frame energy distributions of $\nu_{\tau}$ (or
$\overline{\nu}_{\tau}$) neutrinos for the MMHT2014 (top) and NNPDF30
(bottom) sets of collinear PDFs for the leading (left) and subleading
(right) $D_{s}$-meson production mechanisms. Contributions from the
direct and the chain decay modes are shown separately. Further details are specified in the figure.
}
\label{fig:nu_Elab}
\end{figure}

Similar conclusions as above can be drawn from the analysis of the neutrino (antineutrino) laboratory frame energy distributions
shown in Fig.~\ref{fig:nu_Elab}. The direct decay mode dominates for smaller energies while the chain mode for larger energies. The crosspoint
is found to be between $20-40$ GeV and is slightly different for the leading and for the subleading contributions. The value for the leading contribution is consistent with the results reported in Ref.~\cite{BR2018}. The differences of the neutrino distributions obtained with the MMHT2014 and the NNPDF30 PDFs for the leading and the subleading mechanisms are driven by the respective differences of the $D_{s}$-meson distributions (see the discussion of Fig.~\ref{fig:mesons_Elab}).

\begin{figure}[!h]
\begin{minipage}{0.47\textwidth}
  \centerline{\includegraphics[width=1.0\textwidth]{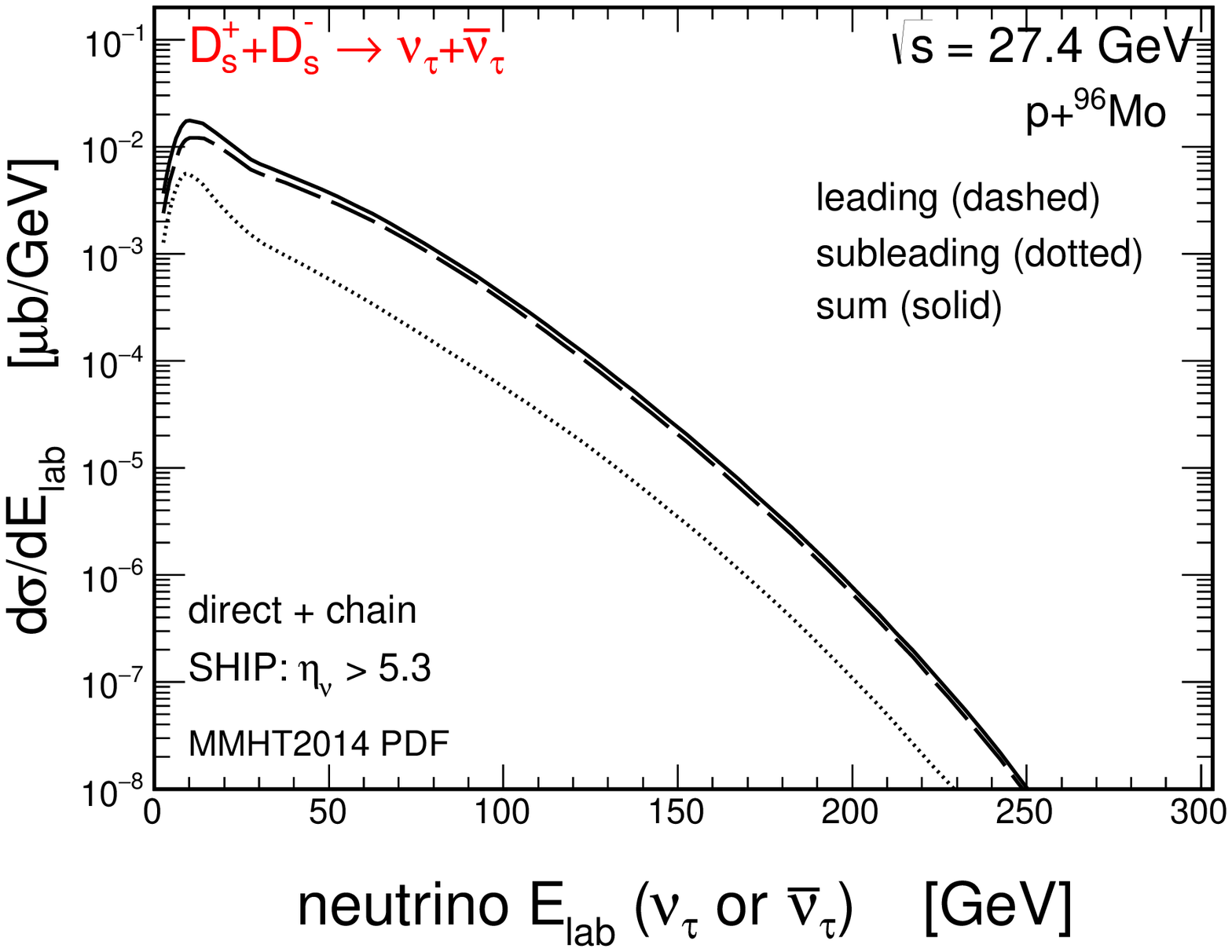}}
\end{minipage}
\begin{minipage}{0.47\textwidth}
  \centerline{\includegraphics[width=1.0\textwidth]{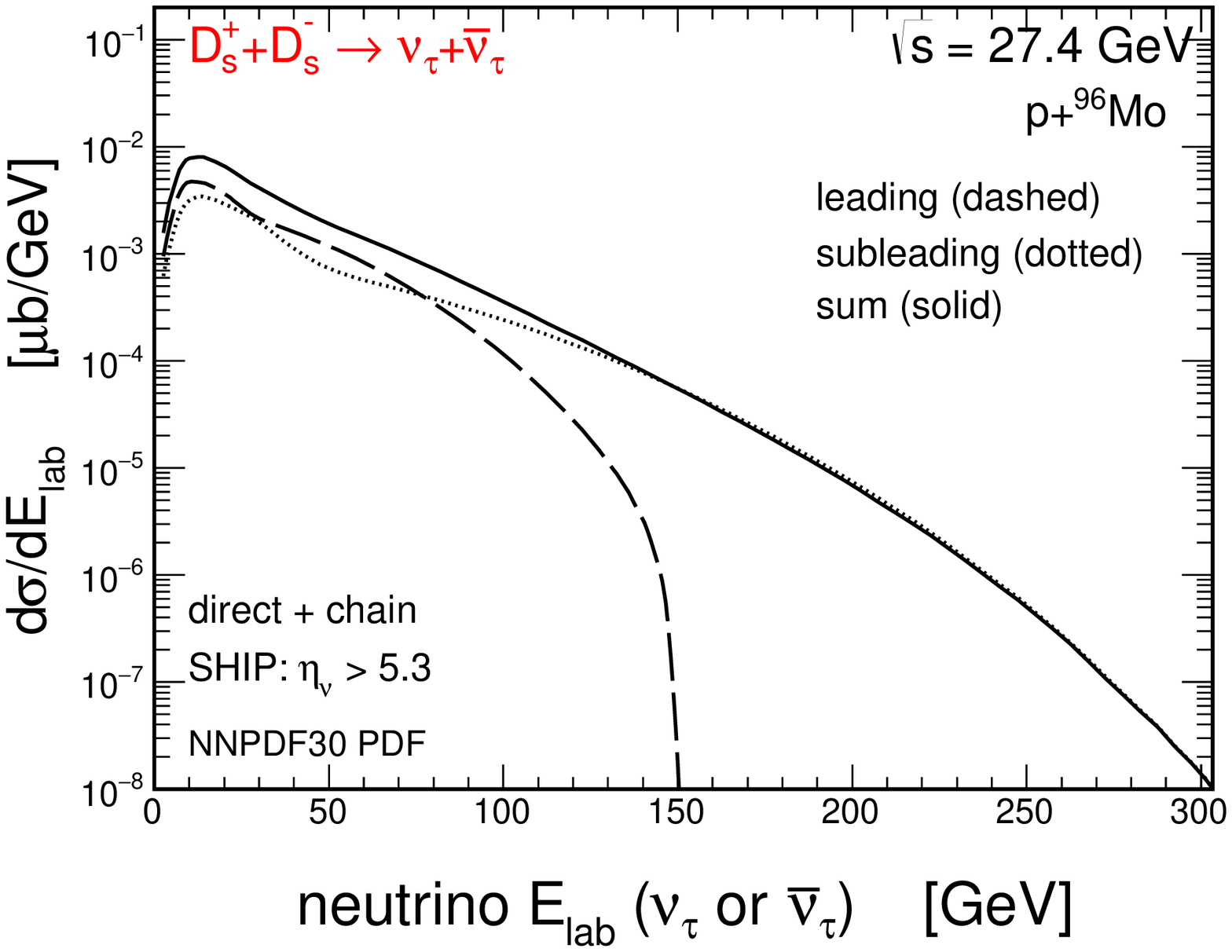}}
\end{minipage}
  \caption{
\small Laboratory frame energy distributions of $\nu_{\tau}$ (or $\overline{\nu}_{\tau}$) neutrinos for MMHT2014 (left) and NNPDF30 (right) sets of collinear PDFs
produced in $p+^{96}\!\mathrm{Mo}$ collisions.
Here we show in the same panel the leading and subleading contributions
as well as their sum. Contributions from the direct and chain decays 
are added together. Further details are specified in the figure.
}
\label{fig:nu_Elab_sum}
\end{figure}

In analogy to Fig.~\ref{fig:mesons_Elab}, where the laboratory frame energy distributions of the $D_{s}$ meson are shown,
here we wish to present similar distributions but for the neutrinos/antineutrinos. In Fig.~\ref{fig:nu_Elab_sum} we show the impact of the subleading contribution
for the predictions of $\nu_{\tau}$ and/or $\overline{\nu}_{\tau}$ energy distributions for the SHiP experiment. Again we obtain two different scenarios for the two different PDF sets.
The MMHT2014 PDFs set leads to an almost negligible subleading contribution in the whole energy range while the NNPDF30 PDFs set provides the subleading contribution to be dominant
at larger energies ($E_{\mathrm{lab}} > 100$ GeV). If such distributions could be measured by the SHiP experiment then 
they could be useful to constrain the PDFs in the purely known kinematical region.

\subsection{Number of neutrinos/antineutrinos observed for the $\bm{\mathrm{Pb}}$ target}
\label{sec:nutau_observed}

The calculated previously distributions are a bit theoretical. 
In this subsection we wish to make relations to what can be observed in the experiment.
As discussed previously the neutrino/antineutrino interaction with the matter
is strongly energy dependent.

\begin{figure}[!h]
\begin{minipage}{0.47\textwidth}
  \centerline{\includegraphics[width=1.0\textwidth]{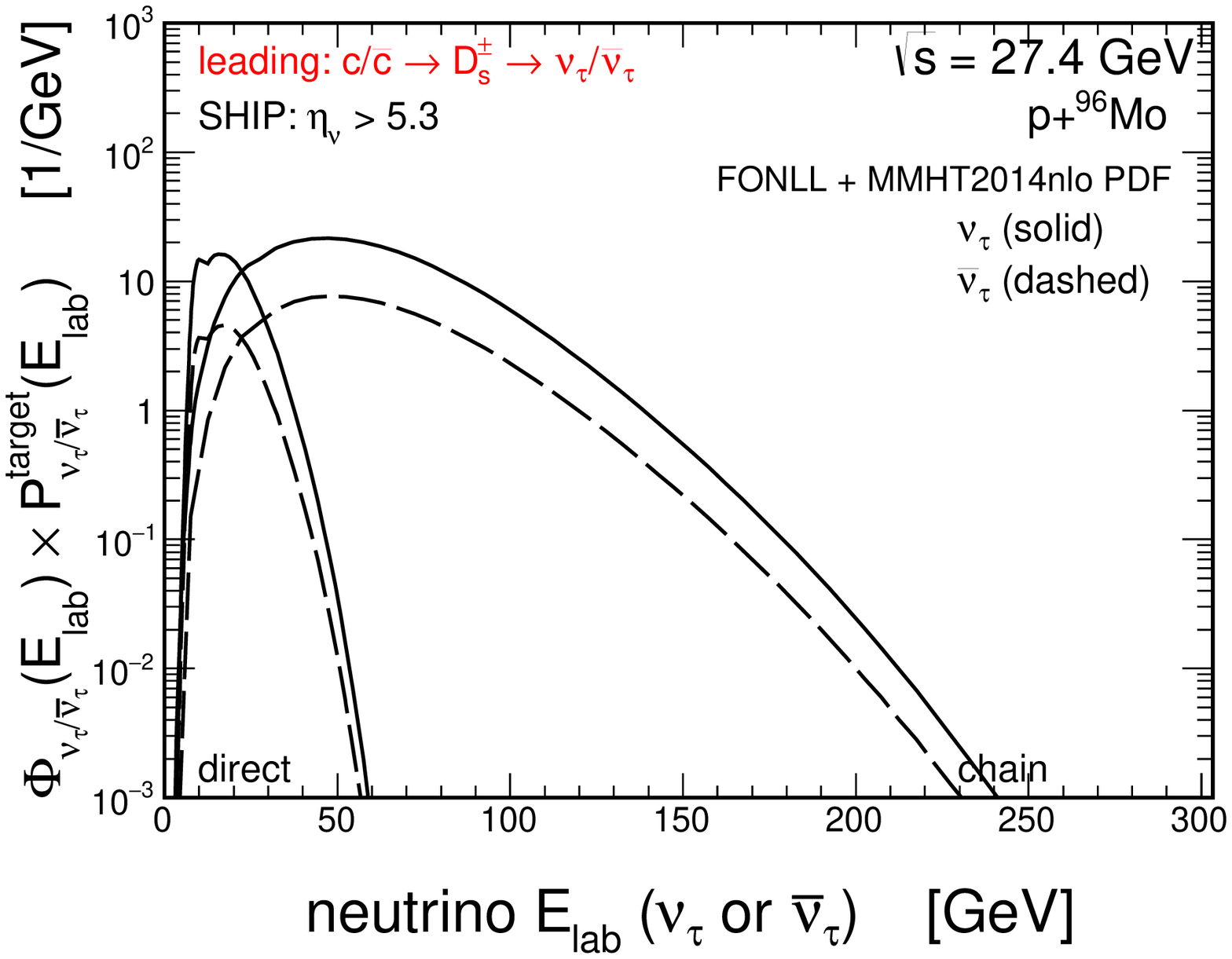}}
\end{minipage}
\begin{minipage}{0.47\textwidth}
  \centerline{\includegraphics[width=1.0\textwidth]{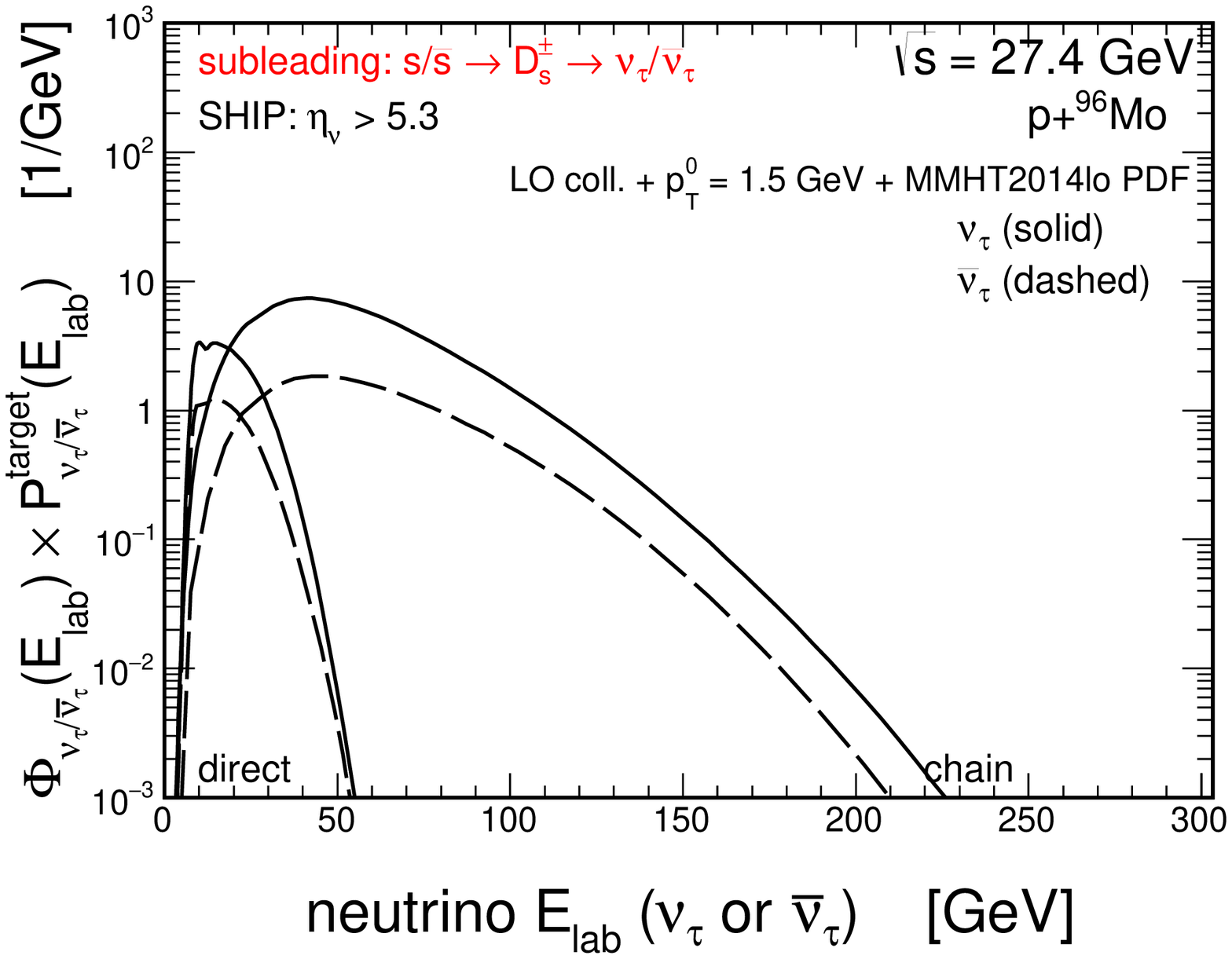}}
\end{minipage}\\
\begin{minipage}{0.47\textwidth}
  \centerline{\includegraphics[width=1.0\textwidth]{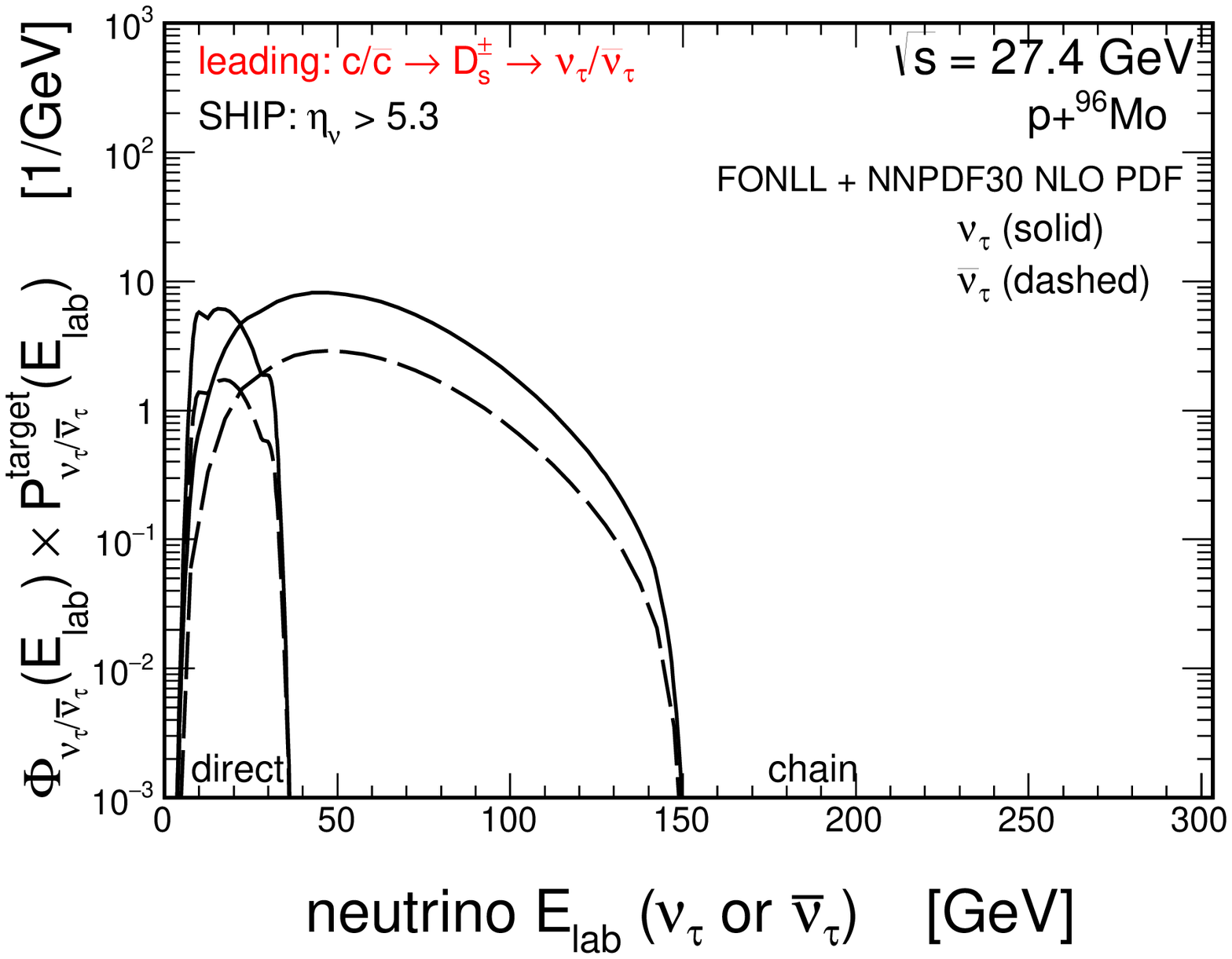}}
\end{minipage}
\begin{minipage}{0.47\textwidth}
  \centerline{\includegraphics[width=1.0\textwidth]{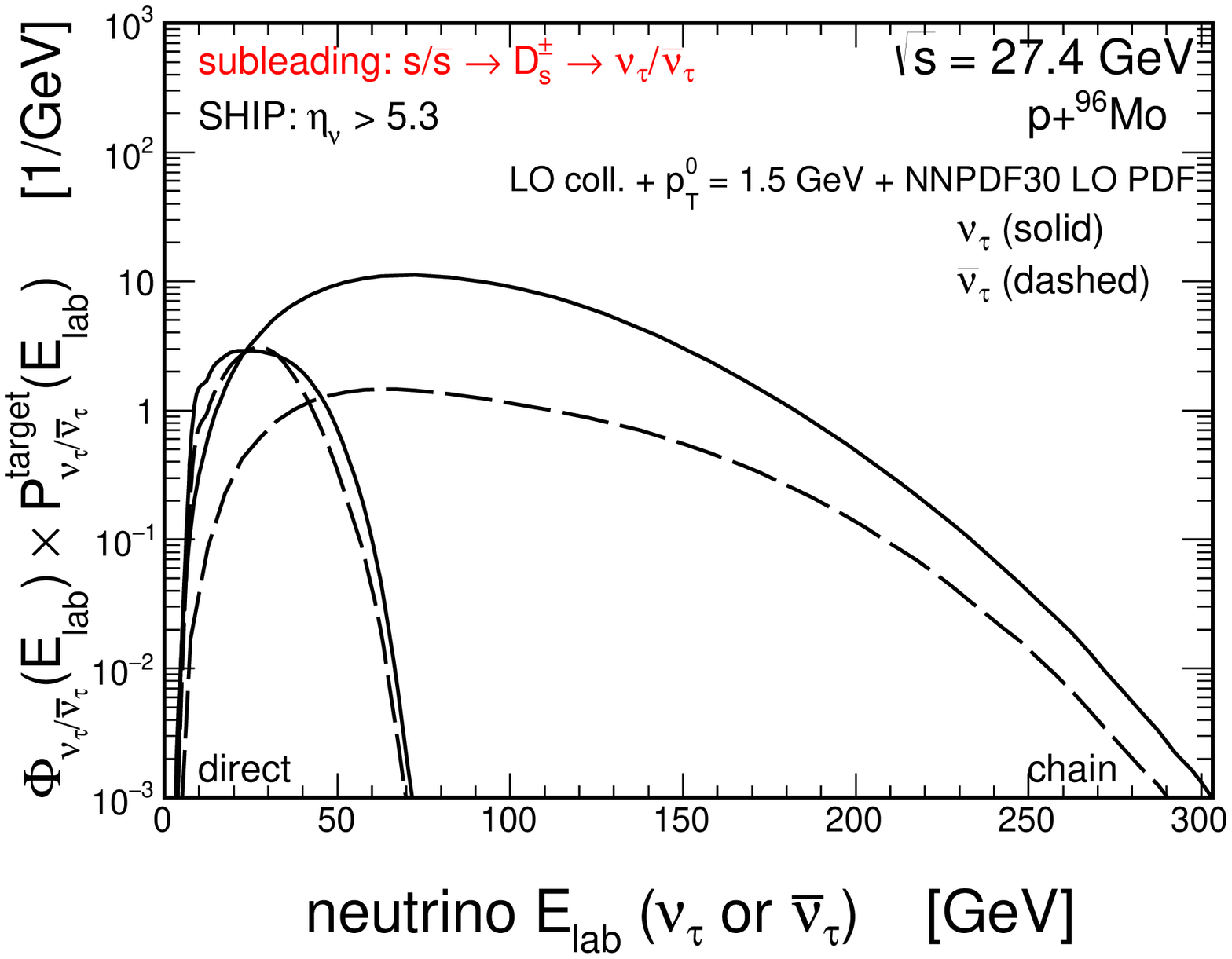}}
\end{minipage}
  \caption{
\small Integrand of Eq.(\ref{number_of_observed_neutrinos}) for the MMHT2014 (top) and the NNPDF30 (bottom)
sets of collinear PDF for the leading (left) and subleading (right) mechanisms of $D_s$ meson production.
We show results for $\nu_{\tau}$ (solid) and $\overline{\nu}_{\tau}$
(dashed) separately for both the direct and chain contributions.
}
\label{fig:integrand}
\end{figure}

In Fig.\ref{fig:integrand} we show the integrand of the integral in Eq.(\ref{number_of_observed_neutrinos}).
This can be interpreted as a number of produced neutrinos/antineutrinos
per interval of (laboratory) energy. As seen in the figure, the distributions
corresponding to the direct production are peaked at $E_{\mathrm{lab}} \approx$ 20 GeV.
For the chain neutrinos the situation strongly depends on the gluon distribution
for the leading contribution and on $s(x) / {\bar s}(x)$ distributions for the subleading contributions.
The latter ones are, however, much less certain and a better understanding
of the $s \to D_s$ transition is required. 
The maximum of the chain contributions is at $E_{\mathrm{lab}} \sim$ 50-100 GeV and depends on the details
of the model. The discussed measurement can be therefore used to verify
the existing parton distributions.
An extraction of gluon distributions seems, however, difficult.

After integrating the above integrands one gets numbers of neutrinos/antineutrinos collected in Table \ref{tab:number_nu}.
Quite different numbers are obtained for the different considered scenarios.
We get larger numbers than in Ref.~\cite{BR2018} but smaller than in Ref.~\cite{SHiP3}.
The chain contribution is significantly larger (about factor 7) than the direct one.
For the MMHT2014 distribution the contribution of the leading mechanism is much larger than
for the subleading one. For the NNPDF30 distributions the situation is reversed.
We predict large observation asymmetry (see the last column) for 
$\nu_{\tau}$ and $\overline{\nu}_{\tau}$. This asymmetry is bigger than
shown e.g. in Refs.~\cite{BR2018,SHiP3}. This is due to the subleading mechanism for $D_{s}^{\pm}$ meson production included in the present paper.
The observation asymmetry for the leading contribution which comes from the differences of the $\nu_{\tau}$ and $\overline{\nu}_{\tau}$ interactions with target are estimated at the level of 50\%. In the case of the subleading contribution the asymmetry increases to 60-70\%, depending on PDF model.

\begin{table}[tb]%
\caption{Number of observed $\nu_{\tau}$ and $\overline{\nu}_{\tau}$ for the SHiP experiment.}

\label{tab:number_nu}
\centering %
\begin{tabularx}{0.9\linewidth}{c c c c c c}
\\[-3.5ex] 
\toprule[0.1em] %

\multirow{2}{7.cm}{Framework/mechanism} & \multicolumn{5}{c}{\multirow{1}{6.cm}{Number of observed neutrinos}}  \\ [-0.ex]
\multirow{2}{7.cm}{}   & \multirow{1}{1.5cm}{flavour}  & \multirow{1}{1.5cm}{direct} & \multirow{1}{1.5cm}{chain} & \multirow{1}{1.5cm}{$\nu_{\tau} + \overline{\nu}_{\tau}$} & \multirow{1}{1.5cm}{$\frac{\nu_{\tau} - \overline{\nu}_{\tau}}{\nu_{\tau} + \overline{\nu}_{\tau}}$} \\ [+0.1ex]

\bottomrule[0.1em]

   \multirow{1}{7.cm}{FONLL + NNPDF30 NLO PDF}      &   \multirow{1}{1.5cm}{$\nu_{\tau}$}  &   \multirow{1}{1.5cm}{96}  &  \multirow{1}{1.5cm}{515}  & \multirow{2}{1.5cm}{818} & \multirow{2}{1.5cm}{0.49} \\[-1.0ex]
 \multirow{1}{7.cm}{$c/\bar{c} \to D_{s}^{\pm} \to \nu_{\tau}/\overline{\nu}_{\tau}$}      &   \multirow{1}{1.5cm}{$\overline{\nu}_{\tau}$}  &   \multirow{1}{1.5cm}{27}  &  \multirow{1}{1.5cm}{180}  & \multirow{2}{1.5cm}{} & \multirow{2}{1.5cm}{}  \\
 \hline
   \multirow{1}{7.cm}{LO coll. + NNPDF30 LO PDF}      &   \multirow{1}{1.5cm}{$\nu_{\tau}$}  &   \multirow{1}{1.5cm}{93}  &  \multirow{1}{1.5cm}{1092}  & \multirow{2}{1.5cm}{1416} & \multirow{2}{1.5cm}{0.67}   \\[-1.0ex]
 \multirow{1}{7.cm}{$s/\bar{s} \to D_{s}^{\pm} \to \nu_{\tau}/\overline{\nu}_{\tau}$}      &   \multirow{1}{1.5cm}{$\overline{\nu}_{\tau}$}  &   \multirow{1}{1.5cm}{75}  &  \multirow{1}{1.5cm}{156} & \multirow{2}{1.5cm}{}     \\
\bottomrule[0.1em]
 \multirow{1}{7.cm}{FONLL + MMHT2014nlo PDF}      &   \multirow{1}{1.5cm}{$\nu_{\tau}$}  &   \multirow{1}{1.5cm}{277}  &  \multirow{1}{1.5cm}{1427} & \multirow{2}{1.5cm}{2292}       & \multirow{2}{1.5cm}{0.49} \\[-1.0ex]
 \multirow{1}{7.cm}{$c/\bar{c} \to D_{s}^{\pm} \to \nu_{\tau}/\overline{\nu}_{\tau}$}      &   \multirow{1}{1.5cm}{$\overline{\nu}_{\tau}$}  &   \multirow{1}{1.5cm}{80}  &  \multirow{1}{1.5cm}{508} & \multirow{2}{1.5cm}{}  & \multirow{2}{1.5cm}{}    \\
 \hline
  \multirow{1}{7.cm}{LO coll. + MMHT2014lo PDF}      &   \multirow{1}{1.5cm}{$\nu_{\tau}$}  &   \multirow{1}{1.5cm}{59}  &  \multirow{1}{1.5cm}{435}  & \multirow{2}{1.5cm}{632}     & \multirow{2}{1.5cm}{0.56} \\[-1.0ex]
 \multirow{1}{7.cm}{$s/\bar{s} \to D_{s}^{\pm} \to \nu_{\tau}/\overline{\nu}_{\tau}$}      &   \multirow{1}{1.5cm}{$\overline{\nu}_{\tau}$}  &   \multirow{1}{1.5cm}{21}  &  \multirow{1}{1.5cm}{117} & \multirow{2}{1.5cm}{} & \multirow{2}{1.5cm}{}     \\

\bottomrule[0.1em]
\end{tabularx}
\end{table}

\begin{figure}[!h]
\begin{minipage}{0.47\textwidth}
  \centerline{\includegraphics[width=1.0\textwidth]{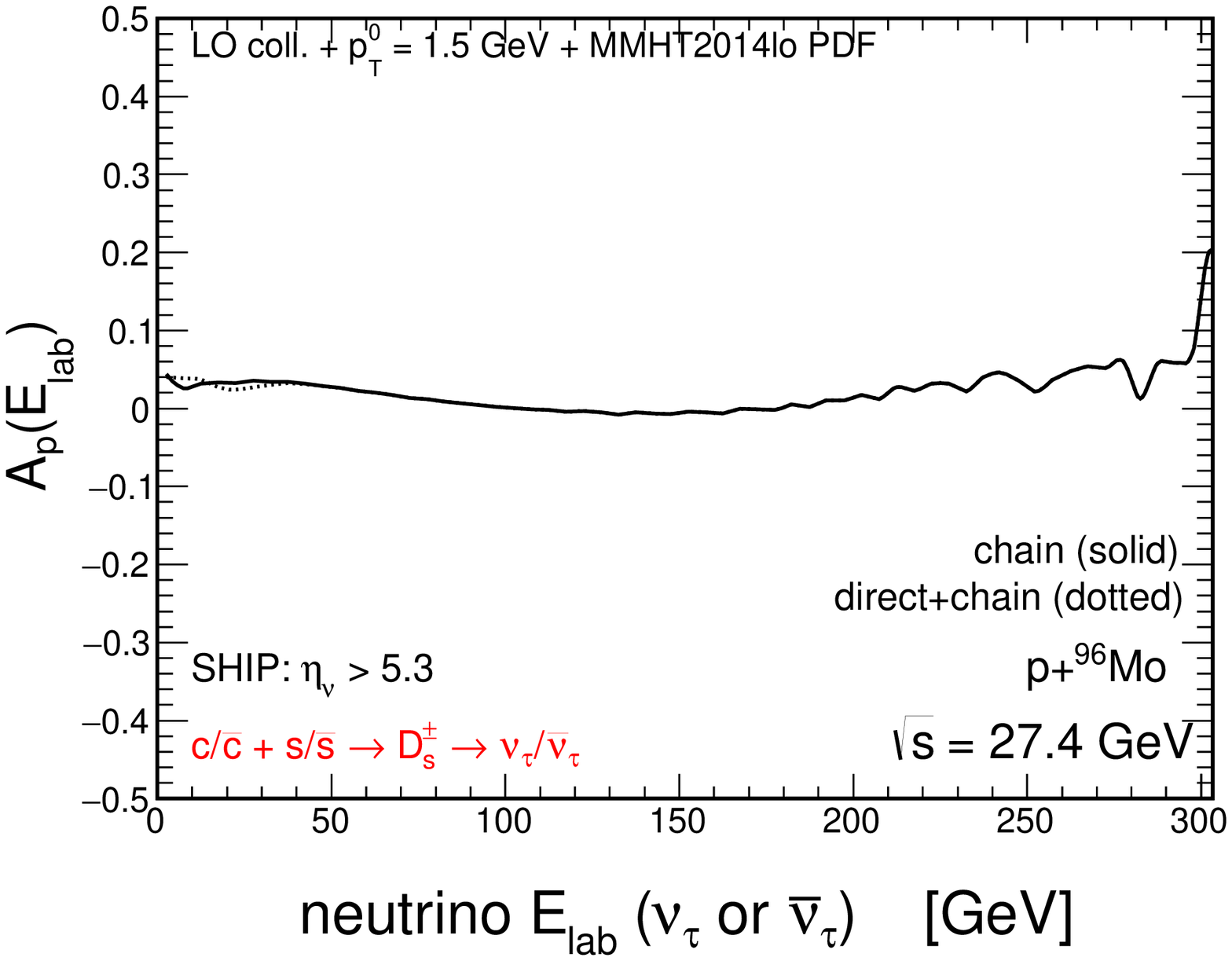}}
\end{minipage}
\begin{minipage}{0.47\textwidth}
  \centerline{\includegraphics[width=1.0\textwidth]{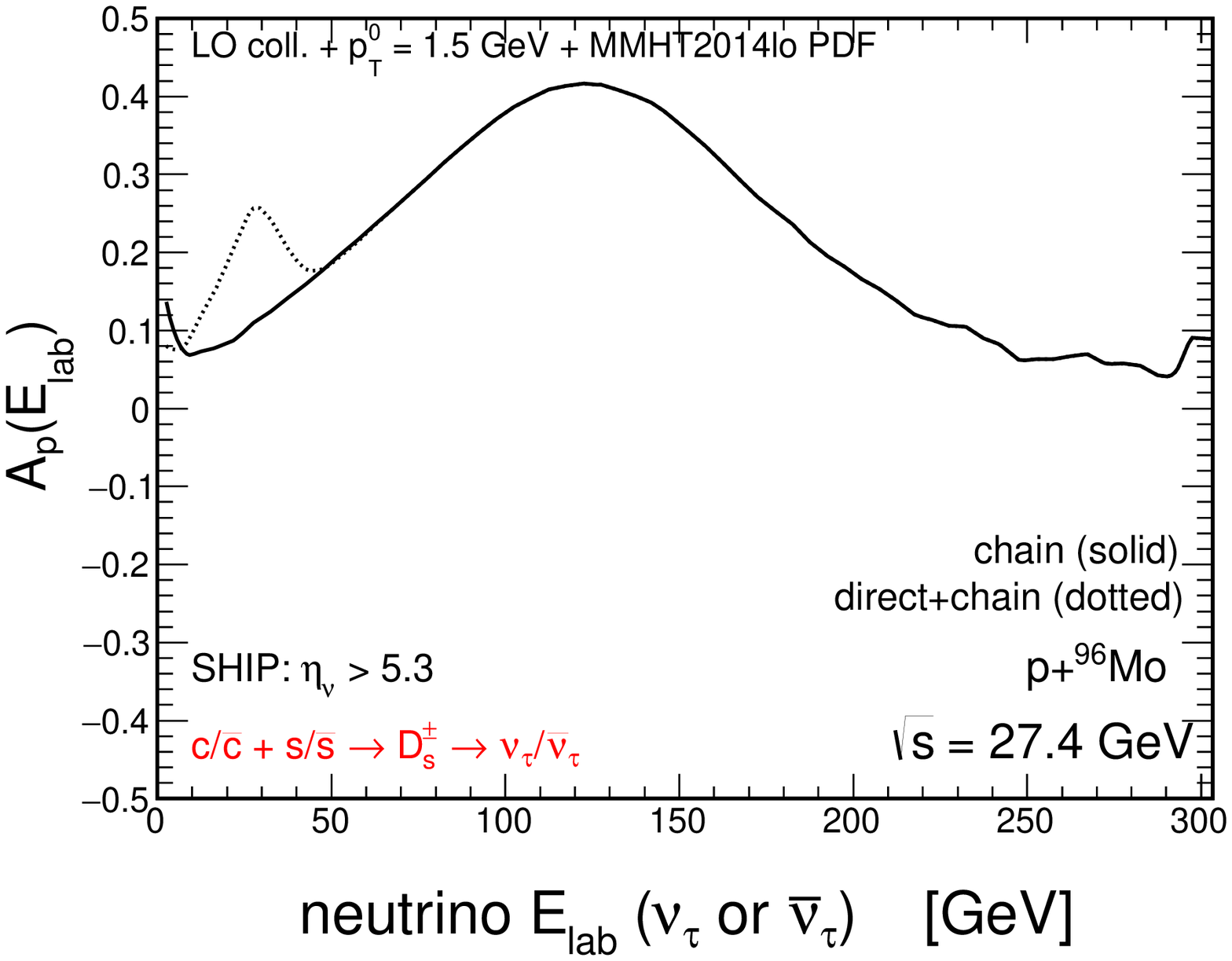}}
\end{minipage}
  \caption{
\small The production asymmetry $A(E_{\mathrm{lab}})$ as a function of 
neutrino/antineutrino laboratory frame energy for the MMHT2014 PDFs
(left panel) and the NNPDF30 PDFs (right panel) for the sum of the leading and
subleading mechanisms for $D_s$ meson production. Both the direct and
chain contributions are included.
}
\label{fig:asymmetry}
\end{figure}

Finally, in Fig.~\ref{fig:asymmetry} we show asymmetry in the production of $\nu_{\tau}$ and $\overline{\nu}_{\tau}$ defined as follows:
\begin{equation}
A(E_{\mathrm{lab}}) = 
\frac{d\sigma_{\nu_{\tau}}/dE_{\mathrm{lab}} - d\sigma_{\overline{\nu}_{\tau}}/dE_{\mathrm{lab}}}
     {d\sigma_{\nu_{\tau}}/dE_{\mathrm{lab}} + d\sigma_{\overline{\nu}_{\tau}}/dE_{\mathrm{lab}}}
\; ,
\end{equation}
for the sum of the leading and subleading production mechanisms. Here we
have included both the direct and chain contributions. Particularly
large positive asymmetry is observed in the case of the NNPDF30 PDF set
due to the relative large contribution of the subleading mechanism as
compared to the case of the MMHT2014 PDF. We conclude that the
asymmetry may strongly depend on the parton distributions used in the
calculations. Therefore we think that the SHiP experiment will be able 
to verify the latter.

\section{Conclusions}

In the present paper we have discussed the mechanism and cross sections for production
of $\nu_{\tau}$ and $\overline{\nu}_{\tau}$ in fixed target experiment
for $\sqrt{s_{NN}}$ = 27.4 GeV with 400 GeV proton beam and molybdenum target.
In the present analysis we have assumed that the neutrinos/antineutrinos
are produced exclusively from $D_s^{\pm}$ mesons. Other, probably small,
contributions (Drell-Yan, $\gamma \gamma$ fusion, $B$ decays, etc.) 
have been neglected here.

We include two different contributions of $D_s$ meson production: the leading fragmentation of $c$ and $\bar c$
and the subleading fragmentation of $s$ and $\bar s$. The cross section for $c / \bar c$ production has been obtained
either using the \textsc{Fonll} framework or in the $k_T$-factorization approach using the Kimber-Martin-Ryskin unintegrated parton distributions.
The $s$ and $\bar s$ production cross sections have been calculated here in the leading-order collinear factorization approach with on-shell initial state partons and with a special treatment of minijets at low transverse momenta, as adopted \textit{e.g.} in \textsc{Pythia}.

The neutrinos are produced then via the direct decay mode
$D_s^{\pm} \to \tau^{\pm} \nu_{\tau}/\overline{\nu}_{\tau}$ and via the chain decay of $\tau^{+}$ or $\tau^{-}$ leptons.
The direct production is very simple as pseudoscalar (spin zero) $D_s$ mesons decay isotropically in its rest frame.
The chain decay is more involved technically. In the present paper we have used \textsc{Tauola} package to generate
sequentional $\tau$ decays. All decay channels implemented in \textsc{Tauola} have been included in the calculation.
The four-momenta of
$\nu_{\tau}/\overline{\nu}_{\tau}$ in the $\tau$ rest frame have been transformed
event-by-event to the proton-nucleon center-of-mass system or laboratory frame using relevant Lorentz transformations and then 
corresponding distributions have been constructed.

The cross section for $p+^{96}\!\mathrm{Mo}$ was obtained from that for 
the proton-proton or proton-neutron collisions via a simple counting of
individual $pp$ and $pn$ collisions. We have taken the well known
probabilities of $c \to D_s$ fragmentation and branching fraction
for the $D_s \to \nu_{\tau} + \tau$ decay.

We have presented resulting distributions of neutrinos/antineutrinos
in transverse momentum and laboratory energy. Such distributions are
crucial to calculate interactions of $\nu_{\tau}$ and 
$\overline{\nu}_{\tau}$ with the lead target.
We have presented also production asymmetry for $\nu_{\tau}$ and
$\overline{\nu}_{\tau}$ as a function of neutrino/antineutrino
energy.


In the present paper we have included also subleading (unfavored) 
fragmentation ($s \to D_s^{-}$ or ${\bar s} \to D_s^{+}$). 
In principle, when $s(x) \ne {\bar s}(x)$ such a subleading mechanism 
could lead to different distributions of $D_s^+$ and $D_s^-$ and 
in the consequence different distributions of 
$\nu_{\tau}$ and $\overline{\nu}_{\tau}$.

The subleading fragmentation leads to asymmetry provided
$s$ and $\bar s$ distributions are different.
We have discussed a possible role of the subleading production of
$D_s$ mesons in the context of ``increasing'' the production of 
$\nu_{\tau}/\overline{\nu}_{\tau}$ neutrino/antineutrino at the SHiP experiment.
A similar effect for production of high-energy $\nu_{\tau}/\overline{\nu}_{\tau}$
neutrinos/antineutrinos was discussed very recently in Ref.~\cite{Goncalves:2018zzf}.
The subleading fragmentation may increase
the probability of observing $\nu_{\tau}/\overline{\nu}_{\tau}$ neutrinos/antineutrinos
by the planned SHiP fixed target experiment at CERN.
We have found that present knowledge of $s /{\bar s}$ parton 
distributions and especially $s /{\bar s}$ fragmentation to $D_s$ mesons does not allow for
precise estimations.
The SHiP experiment could be therefore useful to test $s/{\bar s}$ distributions.

\vskip+5mm
{\bf Acknowledgments}

This study was partially supported by the Polish National Science Center
grant UMO-2018/31/B/ST2/03537 and by the Center for Innovation and
Transfer of Natural Sciences and Engineering Knowledge in Rzesz{\'o}w.
We are indebted to Adam Kozela, Jacek Otwinowski, Jan Sobczyk 
and Agnieszka Zalewska for discussion on some issues 
presented in this paper. We are particularly indebted to Jan Sobczyk
for his help in using the NuWro computer program.


\end{document}